\documentclass{aa} 

\usepackage{amsmath}
\usepackage{graphicx}
\usepackage{siunitx}
\usepackage{breqn}
\usepackage{float}
\usepackage{txfonts}
\usepackage{bm}
\DeclareMathOperator{\e}{e}
\graphicspath{{fig/}}

\begin{document} 

   \title{Combining high-contrast imaging with high-resolution spectroscopy: Actual on-sky MIRI/MRS results compared to expectations}

   \author{S. Martos,
          \inst{1}
          A. Bidot,
          \inst{2}
          A. Carlotti
          \inst{1}
          \and
          D. Mouillet
          \inst{1}
          }

   \institute{\inst{1} Institute of Planetology and Astrophysics of Grenoble (IPAG), University Grenoble Alpes, CNRS, France\\ \inst{2} Space Telescope Science Institute (STScI), 3700 San Martin Drive, Baltimore, MD 21218, USA \\
              \email{steven.martos@univ-grenoble-alpes.fr}}

   \date{Received 11 December 2024 / Accepted 2 April 2025}

\abstract
{Combining high-contrast imaging with high-resolution spectroscopy represents a powerful approach to detecting and characterizing exoplanets around nearby stars, despite the challenges posed by their faintness. Instruments like VLT/SPHERE represent the state of the art in high-contrast imaging; however, their spectral resolution ($R\approx50$) limits them to basic characterization of close companions. These instruments can observe planets with masses as low as $5 \text{--} 10~\mathrm{M}_{\mathrm{Jup}}$ at distances of around $10~\mathrm{AU}$ from their stars. Detection limits are primarily constrained by speckle noise, which dominates over photon and detector noise at short separations around bright stars, even when advanced differential imaging techniques are used. Similarly, image stability also limits space-based high-contrast imaging capability. This speckle noise can, however, be largely mitigated by molecular mapping, a more recent method that leverages information from high-resolution spectroscopic data.}
{Our objective is to understand and predict the effective detection limits associated with spectro-imaging data after processing with molecular mapping. This involves analyzing the propagation of fundamental noise sources, such as photon and detector noise, and comparing these predictions to real instrument data to assess performance losses due to instrument-based factors. Our goal is to identify and propose potential mitigation strategies for these additional sources of noise. Another key aim is to compare the predictions made by our analytical approach with actual observational data to validate and refine the model’s accuracy where necessary.}
{We analyzed JWST/MIRI/MRS data using the recently developed semi-analytical and numerical tool, FastCurves, and compared the results with outputs from the end-to-end MIRI simulator. This simulator allows one to examine nonideal instrumental effects in detail. Additionally, we applied principal component analysis (PCA), a statistical method that identifies correlated patterns in the data, to help isolate systematic effects, both with and without molecular mapping.}
{Our analysis involves investigating the systematic effects introduced by the instrument, identifying their origins, and evaluating their impact on both noise and signal. We show that valuable insights are gained regarding the effects of straylight, fringes, and aliasing artifacts, each linked to different residual systematic noise terms in the data. The results are further supported by principal component analysis, which also demonstrates its effectiveness in mitigating these effects. Additionally, we explore the similarities and discrepancies between observed and modeled companion spectra from an astronomical perspective.}
{We modified FastCurves to account for systematic effects and improve its modeling of MIRI/MRS noise, with its signal-to-noise predictions validated against empirical data. In high-stellar-flux regimes, systematic noise imposes an ultimate contrast limit when using molecular mapping alone. Our methodology, demonstrated with MIRI/MRS data, could greatly benefit other instruments, aiding in the planning of observational programs. For future instruments like ELT/ANDES and ELT/PCS, it could also inform and guide their development.}

   
   \titlerunning{MIRI/MRS actual on-sky results compared to expectations}
   \authorrunning{S. Martos et al.}
   \maketitle

\section{Introduction}

The detection of exoplanets has revolutionized our understanding of the Universe and raised many fascinating questions. Their exhaustive characterization has three important underlying objectives: (1) to elucidate the complexities of their formation and early evolution by measuring the relative abundances of various molecules, which serve as a proxy for determining the physical processes at play \citep[][]{Oberg_2016, Molliere_2022}; (2) to explore the rich diversity of planetary systems and configurations \citep{Lissauer_2011}, allowing for comparisons within and between planetary systems, particularly in relation to our own Solar System; and (3) to determine the habitability of Earth-like planets, and search for potential indicators of extraterrestrial biological activity \citep[][]{Turbet_2018, J_Wang_2018}. 

The spectral characterization of the atmosphere of these planets is essential to achieve these objectives. Acquiring data with a sufficiently high signal-to-noise ratio (S/N) is challenging due to the observational properties of these planetary systems, the specifications of our telescopes and instruments, and, for ground-based observations, the presence of Earth’s atmosphere, which spectrally filters incoming light and introduces short-lived wavefront aberrations.

The minimum planet-to-star flux ratio ranges from $10^{-4}$ to $10^{-6}$ for young giant planets and decreases to $10^{-8}$ to $10^{-10}$ (or even smaller) for older, smaller planets. In the first case, the photons received from the young planets are emitted during their cooling process, while in the second case, they mostly result from the star's light reflecting off the planet's surface. These low flux ratios would not be an issue if the angular separation between the planet and its host star were sufficiently large, considering our telescopes' size, and the wavelengths at which observations are made. Unfortunately, even for nearby stars, the angular separation is typically a fraction of an arcsecond, translating into a few units of wavelength over diameter.

While some observations do not require the star and planet to be angularly resolved, there are several advantages to resolving them. Unresolved observations at a high spectral resolution can leverage the relative radial velocity of the star and the planet to spectrally resolve the planet. However, the small planet-to-star flux ratio induces photon noise that strongly limits this type of observation. In the more favorable case of transiting planets, the change in the spectroscopic signal over time (before, during, and after the transit) can help in isolating the planetary contribution. However, transits occur over short durations, which limits the performance of this technique and underscores the need for instrument stability when using data obtained over multiple orbits. On the other hand, high-angular-resolution observations significantly reduce photon noise, albeit at the cost of increased instrument complexity: an extreme adaptive optics system is required to achieve diffraction-limited observations with ground-based telescopes, and a coronagraph, or other similar techniques such as dark-hole generation, to further lower photon noise.

High-contrast imaging, which spatially isolates most of the light of a host star from its planetary companions, has advanced significantly over the past two decades. Instruments such as the Spectro-Polarimetric High-contrast Exoplanet REsearch \citep[SPHERE;][]{Beuzit_2019} at the Very Large Telescope (VLT), the Gemini Planet Imager \citep[GPI;][]{Macintosh_2014} at the Gemini South Telescope, and the Subaru Coronagraphic Extreme Adaptive Optics system \citep[SCExAO;][]{Jovanovic_2015} at the Subaru Telescope have enabled the direct imaging of young (<100 Myr) planets with unprecedented spatial resolution. Despite these achievements, detecting exoplanets through high-contrast imaging remains challenging due to speckles (stellar light aberrations) that can mask faint planetary signals, particularly at close angular separations. Although various differential imaging techniques have been developed to extract companion signals from speckle noise, the latter still establishes the detection and characterization limits when searching for companions around bright stars at short separations.

To address this challenge, a powerful detection and characterization technique, called molecular mapping, has been developed. This pioneering method, first applied by \cite{Hoeijmakers_2018}, uses high-resolution spectroscopy to identify the distinct molecular signatures of a companion with a cooler atmosphere than its host star. As a result, the faint signals emitted by planets can be distinguished from their much brighter stars through cross-correlation techniques \citep{Konopacky_2013}, even in the presence of speckles. This approach has already demonstrated its effectiveness with several instruments, including the Keck Planet Imager and Characterizer \citep[KPIC;][]{Delorme_2021} at the Keck Observatory, the High-Resolution Imaging and Spectroscopy of Exoplanets \citep[HiRISE;][]{Vigan_2023} at the VLT, and the Multi-Unit Spectroscopic Explorer \citep[MUSE;][]{Jorquera_2024} also at the VLT.

The founding paper for the combination of high-contrast imaging and high spectral resolution by \cite{Snellen2015} introduced the idea that if direct imaging could achieve a contrast of $10^{-3}$ and spectroscopy a contrast of $10^{-4}$, then their combination should theoretically offer a contrast of $10^{-7}$. The propagation of noise and the useful signal resulting from this combination, considering post-processing techniques such as molecular mapping, has already been investigated by \cite{Landman_2023a} and \cite{Bidot_2024}. It was demonstrated that speckles are effectively filtered by this method, with detection becoming potentially limited by fundamental noise sources. However, this analysis has not yet been extended to real data to confirm whether fundamental noise indeed sets the limit. In this study, we test this assumption and investigate whether additional systematic effects further constrain detection. To achieve this, we employ FastCurves \citep{Bidot_2024}, an analytical and numerical tool designed to predict integral field spectrograph (IFS) performance, which is used to estimate both fundamental and systematic noise contributions.

Currently, only a few IFSs provide diffraction-limited imaging on large telescopes with a high spectral resolution. In this regard, the medium-resolution spectroscopy mode of the Mid-IR Instrument (MIRI/MRS) on the \textit{James Webb} Space Telescope (JWST) is particularly interesting, as it is not affected by adaptive optics residuals, making it simpler to study than ground-based instruments such as the Enhanced Resolution Imager and Spectrograph \citep[ERIS;][]{Kravchenko_2022} on the VLT or the OH-Suppressing Infra-Red Integral-field Spectrograph \citep[OSIRIS;][]{Petit_dit_de_la_Roche_2018} at the Keck Observatory. The proposed use of MIRI/MRS data, along with the MIRISim end-to-end simulator \citep{Klaassen_2020}, offers a promising approach to understanding this data.

Firstly, we present the principles and formalism of molecular mapping, incorporating a new systematic term, along with the FastCurves tool, in Sect. \ref{section2}. Then, we introduce MIRI/MRS, detailing its specifications and the method used for estimating systematic errors through end-to-end simulations (Sect. \ref{section3}). Next, we detail an analysis of on-sky data, highlight the presence of systematic errors, and confirm the FastCurves noise model (Sect. \ref{section4}). Following this, we identify the origins of these systematic errors, and we demonstrate how they impact the detection by altering noise statistics and by diminishing the planetary signal (Sect. \ref{section5}). We also show how FastCurves can accurately estimate the planetary signal under these conditions and predict the potential mismatch between the observed planet spectrum and the considered model (Sect. \ref{section6}). Finally, we show that principal component analysis (PCA) can be applied as a last resort to push the detection limits imposed by systematics, albeit with some reduction in signal (Sect. \ref{section7}).

\section{Methods and tools: Formalism and FastCurves} \label{section2}

\begin{figure*}
    \centering
    \includegraphics[width=16cm]{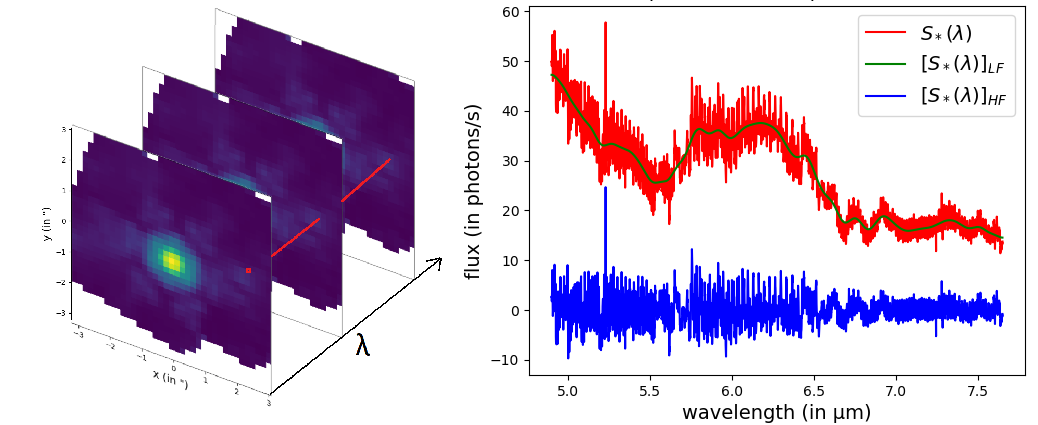}
    \caption{\label{modulations speckles cube} Left: MIRI/MRS data cube representation of a single star. Right: Spectrum of the star at a given point (x, y) in red, with its modulated continuum in green and with high-pass filtering in blue.}
\end{figure*}

\subsection{Molecular mapping principle} \label{subsection2_1}

Speckles are intensity aberrations in starlight caused by wavefront errors that can evolve over time. When analyzing only photometric data, speckles can easily be mistaken for planetary signals and establish the detection limits for high-contrast imaging at close separations. Originating from diffraction effects \citep{Perrin_2003}, speckles have a minimum size set by the wavelength-to-diameter ratio and a pattern that scales with wavelength at first order. As a result, the wavelength dependency of random speckle occurrences folds into the spectral dimension at a specific position, resulting in smooth (low-frequency) modulations within the spectra (see Fig. \ref{modulations speckles cube}). High-pass filtering can be applied to remove these uncontrollable low-frequency modulations, which act as noise. At sufficiently high resolution ($R>100$), the spectral diversity between the star and the planet can be leveraged to distinguish and isolate the planetary flux.

The formalism that we present in this paper is based on \cite{Bidot_2024}. This framework quantifies (i) the high-pass filtered planetary signal relevant for detection (after accounting for its distinction from the stellar contribution, including potential self-subtraction when the stellar and planetary spectra share similarities) and (ii) the noise level due to fundamental sources. Observational data from spectro-imagers is represented as a 3D flux (data cube), which is a function of wavelength $\lambda$ and spaxel (spatial pixel) position $(x, y)$. We let $S$ denote a data cube (after reduction and calibration) corresponding to the measured flux (in electrons) integrated per pixel. It can be expressed as

\begin{equation}
     {\underbrace{S(\lambda,x,y)}_{\text{data}}  = \underbrace{M(\lambda,x,y) \gamma(\lambda) S_{\mathrm{*}}(\lambda)}_{\text{stellar halo}}
     + \underbrace{M_{\mathrm{p}}(\lambda,x,y) \gamma(\lambda) S_{\mathrm{p}}(\lambda)}_{\text{planet}}
     + \underbrace{n(\lambda,x,y)}_{\text{noise}}}
     \label{eq_cube_data}
,\end{equation}

where $\gamma$ is the total system throughput, $S_*$ is the stellar spectrum, and $M(\lambda, x, y)$ represents the stellar modulation function within the field of view (FoV), including the PSF and speckle patterns. The planetary signal is similarly modulated by the planetary modulation function $M_{\mathrm{p}}$, which accounts for both the spatial distribution of the planetary signal and nonideal instrumental effects impacting the planetary spectrum $S_{\mathrm{p}}$. Lastly, $n$ represents fundamental noise sources (photon noise, detector noise, etc.).

Although the planetary modulation function $M_{\mathrm{p}}$ is introduced here for completeness, its impact is not explicitly considered in Sects. \ref{section2} to \ref{section4}, as its effect on the signal remains relatively minor ($<10\%$) as shown in Sect. \ref{section5} (see Fig. \ref{Systematic effect on correlation}). This simplification allows for a more straightforward analysis of the dominant noise and systematics affecting detection.

We let denote $[S]_{\mathrm{LF}}$ the low-frequency content of $S$ (with a certain cutoff resolution, $R_{c}$) and $[S]_{\mathrm{HF}} = S - [S]_{\mathrm{LF}}$ its high-frequency content (see Appendix \ref{appendixA}). In the following, we consider a cutoff resolution of $R_c=100$ (unless otherwise stated), since \cite{Bidot_2024} have shown that this is the optimum cutoff resolution in order to suppress speckles efficiently. We will therefore assume that the speckle signature is negligible beyond this cutoff resolution. We define the Hermitian scalar product in the spectral dimension as $\langle S(\lambda), S'(\lambda) \rangle = \sum_{i} S^{*}(\lambda_i) \times S'(\lambda_i) $. The norm is then given by $\|S(\lambda)\| = \sqrt{\langle S(\lambda), S(\lambda) \rangle} = \sqrt{\sum_{i} |S(\lambda_i)|^2} $.

\begin{figure*}
    \centering
    \includegraphics[width=16cm]{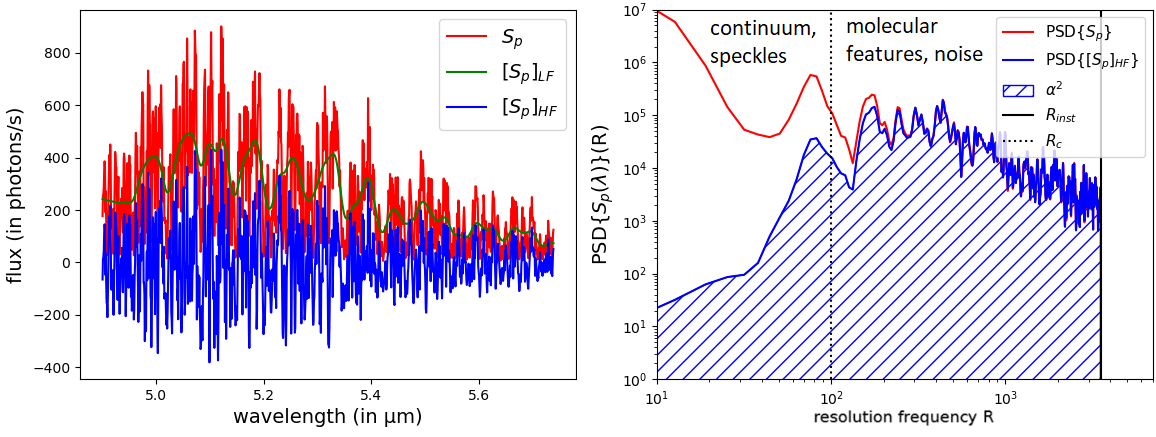}
    \caption{\label{PSD filtering planet spectrum} Left: Thermal emission spectrum of a planet at 500K with the Exo-REM atmospheric model \citep{Charnay_2018} downgraded to 1SHORT band resolution ($R\approx3500$) of MIRI/MRS. Right: PSD of the same spectrum with (blue) and without (red) high-pass filtering. The PSD's limits are defined by the cutoff resolution, $R_c$, and the instrumental resolution, $R_{inst}$. Due to the Gaussian filter, some spectral content may remain below $R_c$, while no lines exist beyond $R_{inst}$ (see Appendix A).}
\end{figure*}

Initially, the data cube is filtered ($S_{\mathrm{res}}$) by estimating and subtracting low-frequency spectral variations along with the stellar component. Then, molecular mapping consists in applying cross-correlations within the spectral dimension in each spaxel between the residual signal, $S_{\mathrm{res}}$, and templates of planets spectra, $\hat{t}$, which depend on various properties (temperature, C/O ratio, metallicity, radial velocity, etc.). This process aims to discern the planet's faint signal from the noise: with an appropriate model (and matching radial velocity), a correlation peak at the planet's location can be found. Detailed calculations and mathematical formalism are provided in Appendix \ref{appendixB}. Thus, at a given radial velocity, the 2D cross-correlation function (CCF) is expressed as
\begin{dmath}
      \mathrm{CCF}(x,y) = \langle S_{\mathrm{res}}(\lambda,x,y) , \hat t(\lambda) \rangle 
      = \underbrace{\alpha(x,y) \cos \theta_{\mathrm{p}} - \beta(x,y)}_{\text{measured signal}} 
      \newline{+ \underbrace{\langle n(\lambda,x,y)+n_{\mathrm{syst}}(\lambda,x,y) , \hat t(\lambda) \rangle}_{\text{noise}}}
\end{dmath}
This is the equation that governs the analysis here. The first term, $\alpha \cos \theta_{\mathrm{p}}$, represents the projection of the template onto the observed planetary spectrum, quantifying its spectral content. Where $\alpha = \| \gamma [M_{\mathrm{p}} S_{\mathrm{p}}]_{\mathrm{HF}} \|$ and $\cos \theta_{\mathrm{p}}$ denotes the similarity factor (correlation) between the template and the observed planetary spectrum (after high-pass filtering), describing the intrinsic mismatch between model and reality. The signal of interest in molecular mapping, $\alpha$, is easily predicted and quantified in the Fourier domain as the high-frequency part of the planetary spectrum, lying between the low cutoff frequency and the maximum frequency determined by the instrument's spectral resolution (see Fig. \ref{PSD filtering planet spectrum})\footnote{This follows from the Parseval-Plancherel theorem, which ensures the conservation of energy in the Fourier domain.}: 
\begin{equation}
     \\{ \alpha(x,y) = \sqrt{ \sum_i \mathrm{PSD} \lbrace \gamma(\lambda) [M_{\mathrm{p}}(\lambda, x, y) S_{\mathrm{p}}(\lambda)]_{\mathrm{HF}} \rbrace (R_i) }}
     \label{alphaPSD}
.\end{equation}
Fig. \ref{PSD filtering planet spectrum} illustrates that the higher the cutoff resolution, the lower the spectral richness. This is the purpose of molecular mapping: part of the signal is sacrificed in order to get rid of speckles (or other low-frequency noise). Lastly, $\alpha$ depends solely on the flux and properties of the planet, particularly its temperature: higher planetary temperatures mean fewer atmospheric lines (i.e., $\alpha$ will be smaller for the same flux).

The second term, $\beta = \| [ {M_{\mathrm{p}} S_{\mathrm{p}}}/{{S}_*} ]_{\mathrm{LF}} \gamma [S_*]_{\mathrm{HF}} \| \cos \theta_*$, represents the projection of the template onto the lines of the residual stellar spectrum and indicates the amount of signal of interest that is lost. $\cos \theta_*$ denotes the similarity factor between the template and the observed stellar spectrum. Indeed, $\beta$ is a self-subtraction term representing the similarity degree between the star and the planet that cannot be used to discriminate the planet from the star (since subtracting the stellar component also eliminates common lines). This term is minimized when the planetary spectrum significantly differs from the stellar spectrum (e.g., when their temperatures are sufficiently different). It is also minimized if the Doppler shift between the planet and the star and/or the instrumental resolution is sufficiently important.

The last term represents the noise projection induced by the correlation. Unlike previous work \citep{Bidot_2024}, where it was assumed that all modulations were low-frequency ($M \approx [M]_{\mathrm{LF}}$), we assume here that the modulation may contain high-frequency components. As a result, we introduce a new term, $n_{\mathrm{syst}}(\lambda,x,y) = [M(\lambda,x,y)]_{\mathrm{HF}} \gamma(\lambda) S_*(\lambda)$\footnote{If only speckles were present as systematic effects, $M \approx [M]_{\mathrm{LF}}$ and $n_{\mathrm{syst}}$ would be null. However, some systematic effects introduce high-frequency modulations in the stellar halo, making this term non-zero and generating systematic noise.}. Instrumental effects, as well as signal extraction and calibration processes that introduce high-frequency artifacts, distort the original spectral information, projecting onto the template and creating a non-uniform CCF signal that effectively acts as a noise source. It is therefore crucial to consider this new term, as we show that it is the primary source of systematic noise. Similar to speckle noise, this systematic noise term is directly proportional to the integrated stellar flux, thereby setting new detection limits.

\subsection{Signal and noise calculation} \label{subsection2_2}

First, the S/N of the CCF is simply given by
\begin{equation}
     \\{ \textrm{S/N}(x,y) = \frac{E[\mathrm{CCF}(x,y)]}{\sqrt{\mathrm{Var}[\mathrm{CCF}(x,y)]}} = \frac{\alpha(x,y)\cos \theta_{\mathrm{p}} - \beta(x,y)}{\sigma_{\mathrm{CCF}}(x,y)} }
.\end{equation}
To go a step further, a description of the various fundamental noise sources and their propagation in the CCF through correlation is needed:
\begin{equation}
     \\{ n(\lambda,x,y) = n_{\mathrm{halo}}(\lambda,x,y) + n_{\mathrm{bkgd}}(\lambda) + n_{\mathrm{dc}} + n_{\mathrm{RON}} }
,\end{equation}
where $n_{\mathrm{halo}}$ is the photon noise associated with the stellar halo, $n_{\mathrm{bkgd}}$ is the photon noise from the background emission, $n_{\mathrm{dc}}$ is the detector dark current noise, and $n_{\mathrm{RON}}$ is the readout noise. The latter three components are assumed to be spatially homogeneous across the FoV.

If there is a spatial dependence of the noise, we assume that its statistics is axisymmetric, meaning that the noise variance only depends spatially on the separation $\rho = \sqrt{x^2+y^2}$ from the star. We also assume that the various noise sources are independent of each other, and that there is no spectral covariance for fundamental noises (although spatial covariance is permitted). Deviations from these assumptions could introduce slight errors in the subsequent estimates. Using template's normalization ($\sum_i \hat t^2(\lambda_i) = 1$) and the spectral non-covariance assumption for fundamental noises, we can write:
\begin{dmath}
     \sigma_{\mathrm{CCF}}^2(\rho) = \mathrm{Var}[\langle n(\lambda,x,y) , \hat t(\lambda) \rangle] + \mathrm{Var}[\langle n_{\mathrm{syst}}(\lambda,x,y) , \hat t(\lambda) \rangle]
     = \mathrm{Var}[\langle n_{\mathrm{halo}}(\lambda,x,y) + n_{\mathrm{bkgd}}(\lambda) + n_{\mathrm{dc}} + n_{\mathrm{RON}}, \hat t(\lambda) \rangle] + \sigma_{\mathrm{syst}}^{\prime 2}(\rho)
     = \sum_i \hat t^2(\lambda_i) \times (\mathrm{Var}[n_{\mathrm{halo}}(\lambda_i,x,y)] + \mathrm{Var}[n_{\mathrm{bkgd}}(\lambda_i)] + \mathrm{Var}[n_{\mathrm{dc}}] + \mathrm{Var}[n_{\mathrm{RON}}]) + \sigma_{\mathrm{syst}}^{\prime 2}(\rho)
     = \sum_i \hat t^2(\lambda_i) \times (\sigma_{\mathrm{halo}}^2(\lambda_i,\rho) + \sigma_{\mathrm{bkgd}}^2(\lambda_i)) + \sigma_{\mathrm{dc}}^2 + \sigma_{\mathrm{RON}}^2 + \sigma_{\mathrm{syst}}^{\prime 2}(\rho)
\end{dmath}
with
\begin{equation}
    \\{\begin{split}
        \sigma_{\mathrm{syst}}^{\prime 2}(\rho) &= \mathrm{Var}[\langle n_{\mathrm{syst}}(\lambda,x,y) , \hat t(\lambda) \rangle] \\
        &= \mathrm{Var}[\langle [M(\lambda,x,y)]_{\mathrm{HF}} \gamma(\lambda) S_*(\lambda) , \hat t(\lambda) \rangle]
    \end{split}}
    \label{sigma_syst}
.\end{equation}
To simplify the explanation, we note
\begin{equation}
     \\{\sigma_{\mathrm{halo}}^{\prime 2}(\rho) = \sum_i \hat t^2(\lambda_i) \times \sigma_{\mathrm{halo}}^2(\lambda_i,\rho)
     \label{eq_sigma_halo_prime}}
\end{equation}
and
\begin{equation}
     \\{\sigma_{\mathrm{bkgd}}^{\prime 2} = \sum_i \hat t^2(\lambda_i) \times \sigma_{\mathrm{bkgd}}^2(\lambda_i)
     \label{eq_sigma_bkgd_prime}}
.\end{equation}
$\sigma^2$ then denotes the variances per spectral channel and $\sigma^{\prime 2}$ the variances projected into the CCF (no distinction is made for the readout noise and dark current components). 

In order to obtain a Poisson statistic for photon noises, we consider all quantities in electron per detector integration time (DIT) and per pixel. In addition, since the planet's signal is spatially distributed across multiple spaxels, we conduct a spatial integration over a box of $A_{\mathrm{fwhm}}$ spaxels (equivalent to the size of the PSF's full width at half maximum (FWHM)) to optimize the S/N \citep{Ruffio_2019}. For $N_{\mathrm{int}}$ integrations, noise components combine quadratically (assuming independence), while the signal combines linearly. However, as the systematic noise term $n_{\mathrm{syst}}$ is directly proportional to the integrated stellar flux, its variance scales with $N_{\mathrm{int}}^2$. If the data are dithered, the noise statistics will be modified and it is needed to multiply by a corrective factor $R_{\mathrm{corr}}$ (see Appendix \ref{appendixC}), otherwise $R_{\mathrm{corr}} = 1$. The S/N is thus expressed as
\begin{equation}
      \tiny \\ {\textrm{S/N}(\rho) = \frac{\sqrt{N_{\mathrm{int}}}(\alpha_{\mathrm{fwhm}}\cos \theta_{\mathrm{p}} - \beta_{\mathrm{fwhm}})}{\sqrt{A_{\mathrm{fwhm}} R_{\mathrm{corr}} (\sigma_{\mathrm{halo}}^{\prime 2}(\rho) + \sigma_{\mathrm{bkgd}}^{\prime 2} + \sigma_{\mathrm{dc}}^2 + \sigma_{\mathrm{RON}}^2 + N_{\mathrm{int}}\sigma_{\mathrm{syst}}^{\prime 2}(\rho)) }} }
     \label{eq_SNR}
,\end{equation}
where $\alpha_{\mathrm{fwhm}}$ and $\beta_{\mathrm{fwhm}}$ are the same quantities but integrated over the FWHM (so they no longer have any spatial dependency). Since the quantity $\alpha_{\mathrm{fwhm}}\cos \theta_{\mathrm{p}} - \beta_{\mathrm{fwhm}}$ is proportional to the integrated flux of the planet $F_{\mathrm{p}}$ , $\alpha_0 = (\alpha_{\mathrm{fwhm}}\cos \theta_{\mathrm{p}} - \beta_{\mathrm{fwhm}}) \frac{F_*}{F_{\mathrm{p}}}$ can be defined and corresponds to the signal the planet would have if it had the same integrated flux as the star $F_*$. The contrast limit (ensuring $5\sigma$ detection if the contrast between the planet and the star $F_{\mathrm{p}}/F_*$ is greater) is then defined by
\begin{equation}
     \tiny \\ { \mathrm{C}(\rho) = \frac{5 \sqrt{A_{\mathrm{fwhm}} R_{\mathrm{corr}} (\sigma_{\mathrm{halo}}^{\prime 2}(\rho) + \sigma_{\mathrm{bkgd}}^{\prime 2} + \sigma_{\mathrm{dc}}^2 + \sigma_{\mathrm{RON}}^2 + N_{\mathrm{int}}\sigma_{\mathrm{syst}}^{\prime 2}(\rho)) }}{\sqrt{N_{\mathrm{int}}} \alpha_0} }
     \label{eq_contrast}
.\end{equation}

Finally, for long integration times (i.e., $N_{\mathrm{int}}$ large) and/or high stellar fluxes (i.e., $\sigma_{\mathrm{syst}}^{\prime 2}$ large), the systematic term will dominate and set the detection limit:
\begin{equation}
     \\{ \textrm{S/N}(\rho) = \frac{\alpha_{\mathrm{fwhm}}\cos \theta_{\mathrm{p}} - \beta_{\mathrm{fwhm}}}{{ \sigma_{\mathrm{syst}}^{\prime}(\rho) }} \; \text{ and } \; \mathrm{C}(\rho) = \frac{5 \sigma_{\mathrm{syst}}^{\prime}(\rho)}{\alpha_0}}
     \label{eq_SNR_limit}
.\end{equation}

\subsection{FastCurves} \label{subsection2_3}

FastCurves\footnote{Initial version: \url{https://github.com/ABidot/FastCurves} \\ Updated version: \url{https://github.com/StevMartos/FastYield}} is a numerical and analytical tool initially developed by \cite{Bidot_2024} to compute the S/N (Eq. \ref{eq_SNR}) and the contrast limit (Eq. \ref{eq_contrast}). In this study, we have extended FastCurves to incorporate systematic effects that impact both the noise and the planetary signal. Specifically, the updated version now models systematic noise through the stellar modulation function $M$ and accounts for distortions in the planetary signal via the planetary modulation function $M_{\mathrm{p}}$. Additionally, FastCurves can now estimate performance when considering PCA in the post-processing of data, as discussed in Sect. \ref{section7}.

First, we compute the planetary signal component with $\alpha_{\mathrm{fwhm}} = f_{\mathrm{fwhm}} \| \gamma [M_{\mathrm{p}} \mathbb{S}_{\mathrm{p}}]_{\mathrm{HF}} \|$ where $f_{\mathrm{fwhm}}$ represents the fraction of flux within the PSF’s FWHM (the computation of $M_{\mathrm{p}}$ is discussed below). The planetary spectrum model $\mathbb{S}_{\mathrm{p}}$ can be selected from various atmospheric models such as BT-Settl \citep{Allard_2012}, Exo-REM \citep{Charnay_2018}, SONORA \citep{Marley_2021}, Morley \citep{Morley_2012} or PICASO \citep{Batalha_2019}. In FastCurves, we assume that the observed planetary spectrum $S_{\mathrm{p}}$ and the template $\hat t$ are identical ($\cos \theta_{\mathrm{p}} = 1$). This implies that no prior assumptions regarding the degree of mismatch are considered, allowing the detection limits to be determined by the instrument specifications rather than our limited understanding of planetary models. In reality, it should be acknowledged that this mismatch could potentially constitute a limitation to molecular mapping (since the $1 - \cos \theta_{\mathrm{p}}$ fraction of the signal will be lost), and disparities between the template and the observed planetary spectrum are likely to persist. $\beta_{\mathrm{fwhm}}$ can also be computed by considering a BT-NextGen \citep{Allard_2012} stellar spectrum based on its properties. However, both planetary and stellar spectra need to be degraded to the instrumental resolution, adjusted to their respective magnitudes, and Doppler-shifted to their respective radial velocities to calculate these two quantities. Regarding contrast calculation, a magnitude for the planet is unnecessary (only that of the star), as we renormalize $\alpha_0$ with the stellar flux.

Next, $\sigma_{\mathrm{halo}}^{2}$ (and then $\sigma_{\mathrm{halo}}^{\prime 2}$ according to Eq. \ref{eq_sigma_halo_prime}) can be calculated by considering a stellar spectrum and PSF profiles derived from calibration data, $\sigma_{\mathrm{bkgd}}^{2}$ (and then $\sigma_{\mathrm{bkgd}}^{\prime 2}$ according to Eq. \ref{eq_sigma_bkgd_prime}) by considering a background model \citep{Rigby_2023}, $\sigma_{\mathrm{dc}}$ and $\sigma_{\mathrm{RON}}$ by considering detector characteristics \citep{Rieke_2015}.

A key advancement in this version of FastCurves is the inclusion of systematic noise estimation. The calculation of $\sigma_{\mathrm{syst}}^{\prime 2}$ requires an estimation of the high-frequency residual modulations in the stellar halo $[M]_{\mathrm{HF}}$ (see Eq. \ref{sigma_syst}) and this can be achieved through two approaches:
\begin{itemize}
    \item Direct measurement from single-star data with high S/N per spectral channel, where systematic noise dominates over fundamental noises ($M_{\mathrm{data}}$).
    \item End-to-end simulations (e.g., with MIRISim) to derive a model of the stellar (or planetary) modulation function $M_{\mathrm{sim}}$ (see Sect. \ref{subsection3_2}).
\end{itemize}
With these estimates, we computed the systematic noise component as
\begin{equation}
      \sigma_{\mathrm{syst}}^{\prime 2}(\rho) = \mathrm{Var}[\langle [M_{\mathrm{sim, data}}(\lambda,x,y)]_{\mathrm{HF}} \gamma(\lambda) S_*(\lambda)  , \hat t(\lambda) \rangle]
     \label{eq_sigma_syst_prime_est}
.\end{equation}
By directly computing $\sigma_{\mathrm{syst}}^{\prime 2}$ in this manner, no assumptions regarding spectral covariance are necessary. Incorporating these new features into the framework ensures that both noise and signal distortions are accurately accounted for when estimating molecular mapping detection limits. These enhancements make FastCurves a more robust tool for evaluating the impact of systematics on molecular mapping and optimizing observational strategies for exoplanet detection.

\section{MIRI/MRS case} \label{section3}

\subsection{Presentation of MIRI/MRS} \label{subsection3_1}

\begin{figure*}
    \centering
    \includegraphics[width=16cm]{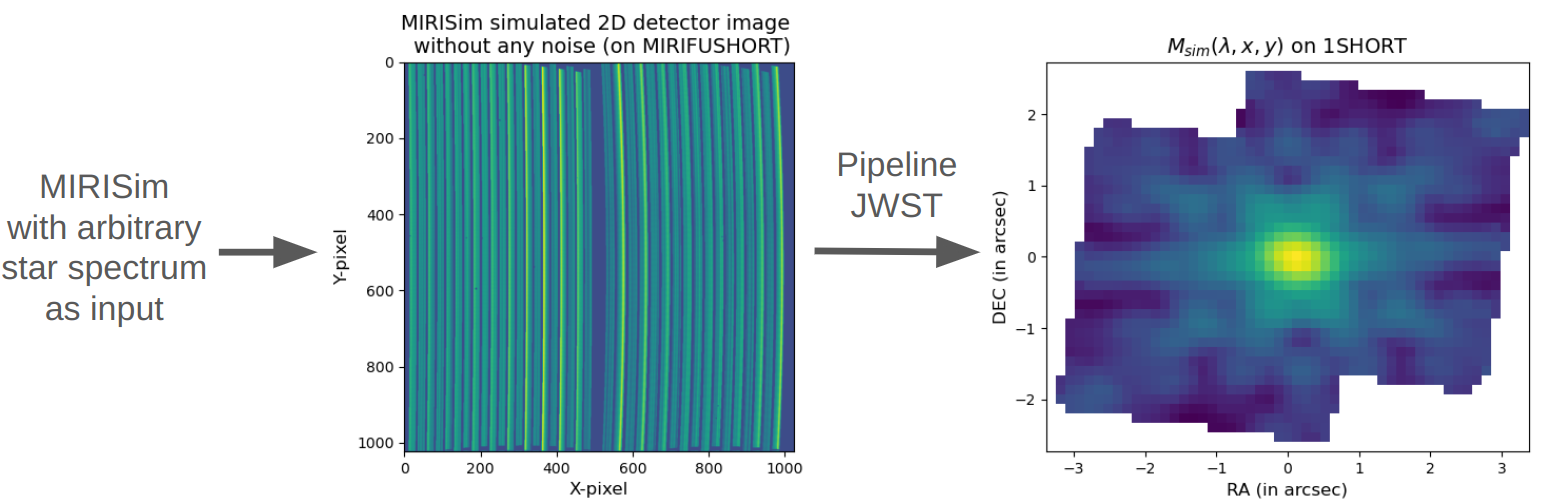}
    \caption{\label{MIRISim estimating modulations} Left: Noiseless MIRI/MRS 2D detector image simulated with MIRISim with an arbitrary stellar spectrum as input. The x axis correspond to the spatial position in the FoV and y axis to the spectral dispersion. Right: Cube reconstructed with the JWST pipeline from the noiseless 2D detector image simulated and divided by the input stellar spectrum.}
\end{figure*}

\begin{figure}
    \centering
    \includegraphics[width=8cm]{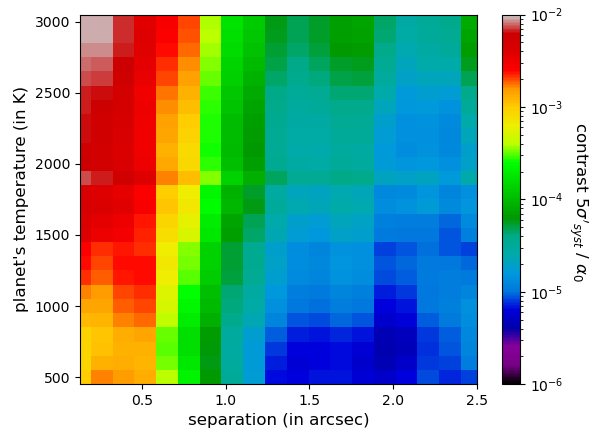}
    \caption{\label{contrast MIRIMRS limits on 1SHORT} MIRI/MRS molecular mapping detection limits set by systematics estimated with FastCurves and MIRISim on 1SHORT with BT-Settl models and $R_c = 100$.}
\end{figure}

\begin{figure*}
    \centering
    \includegraphics[width=17cm]{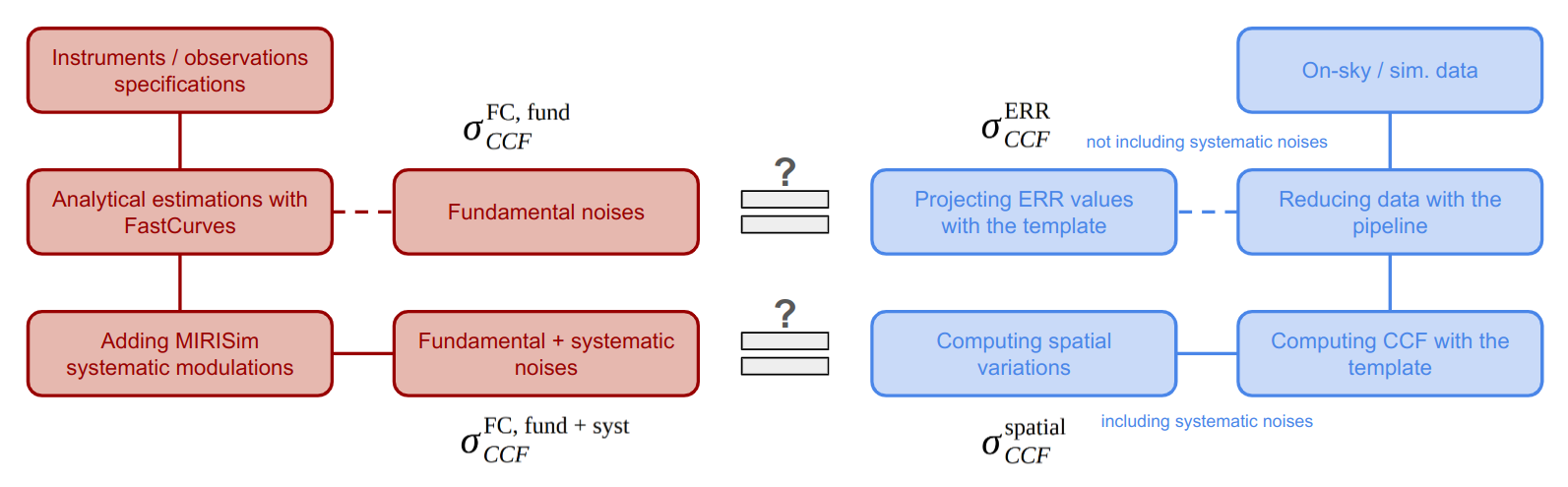}
    \caption{\label{Summary of the methodology for comparing analytical and empirical noise estimates} Methodology for comparing analytical (red) and empirical (blue) noise estimates. Empirical errors are obtained from on-sky data cubes by projecting the ERR extension values onto the template, as well as by calculating the spatial standard deviation of the CCF. These empirical values are then compared to the analytical errors derived from FastCurves, both with and without consideration of the systematic noise component (estimated with MIRISim), in addition to the fundamental noise sources.}
\end{figure*}

For this analysis, we selected MIRI/MRS due to its key role in probing exoplanet atmospheres. MIRI/MRS operates in a spectral domain that offers significant advantages. This particular wavelength range is relatively unexplored, presenting numerous opportunities for new discoveries. There is a notable interest in this range, as reflected by the number of studies targeting it \citep[][]{Deming_2024, Henning_2024, Pontoppidan_2024, Miles_2023, Worthen_2024}. Furthermore, because MIRI/MRS operates from space, it avoids atmospheric interference, leading to high-quality observations. The instrument operates in a regime where, despite the inherent challenges of large wavelength-to-diameter ratios, it maintains high performance at close angular separations, creating favorable conditions for exoplanet detection and characterization. MIRI/MRS is also particularly convenient to study due to the availability of the MIRISim end-to-end simulator \citep[version 2.4.2;][]{Klaassen_2020} and of the JWST pipeline \citep[version 1.13.4;][]{Labiano_Ortega_2016}. 

As the only mid-infrared instrument aboard JWST, MIRI’s coverage spans from $4.9$ to $27.9 ~ \mu\mathrm{m}$ \citep{Wells_2015}. It offers four operational modes: imaging, coronagraphy, low-resolution spectroscopy (LRS) with $R\approx100$, and medium-resolution spectroscopy (MRS) with $R\approx1500-3500$. MIRI/MRS functions as an IFS, delivering diffraction-limited spectroscopy. The wavelength range is divided into four spectral channels, each subdivided into three bands denoted as 1SHORT ($4.9 \text{--} 5.74~\mu\mathrm{m}$ at $R\approx3515$), 1MEDIUM ($5.66 \text{--} 6.63~\mu\mathrm{m}$ at $R\approx3470$), 1LONG ($6.53 \text{--} 7.65~\mu\mathrm{m}$ at $R\approx3355$), 2SHORT ($7.51 \text{--} 8.77~\mu\mathrm{m}$ at $R\approx3050$), etc. \citep[see Table 1 of][]{Argyriou_2023}. A single exposure allows one to observe one band of each channel at the same time, meaning that three exposures are required to cover the entire wavelength range. Only the first two channels are considered, since channel 4 has lower sensitivity and exhibits significant photometric errors. Consequently, channel 3, which shares the same detector as channel 4, is also excluded from consideration. By design, the MRS is spatially and spectrally undersampled in channels 1 and 2. For this reason, point source observations are always made using dithered exposures to improve spatial sampling. This involves placing the point source at a different location on the detector on different exposures and combining them by drizzle \citep[][]{Fruchter_2002, Law_2023}. In this way, all MIRI/MRS observations are considered with four-point dithering, providing an effective angular resolution of 0.13 arcsec/pixel for channel 1 and 0.17 arcsec/pixel for channel 2.

Even in space, Reference Differential Imaging (RDI) is ultimately limited by PSF stability and pointing reproducibility \citep[][]{Ruffio_2024, Malin_2024, Boccaletti_2024}. This limitation leaves room for molecular mapping approaches, which remain valuable for both detection and characterization, despite requiring the sacrifice of the continuum. This motivates the need to quantify the detection capabilities of molecular mapping in various observational scenarios, allowing for a discussion of its potential complementarity to RDI.

\subsection{MIRISim + JWST pipeline as a systematic estimator} \label{subsection3_2}

The JWST pipeline \citep{Labiano_Ortega_2016} is designed to handle the processing and calibration of raw data acquired from the JWST. MIRISim  \citep{Klaassen_2020}, an end-to-end simulator, is specifically crafted to produce realistic MIRI data, incorporating various effects such as fringing and distortion. MIRISim outputs are uncalibrated 2D detector images, which are JWST pipeline-compatible inputs. The scene to be observed and the observation parameters can be arbitrarily defined.

Having an end-to-end simulator enables us to estimate systematic effects through the following approach: we generate a 2D detector image using MIRISim without any fundamental noise while injecting an arbitrary stellar spectrum. We then use the JWST pipeline to reconstruct and calibrate a data cube, which we subsequently normalize by the input stellar spectrum to derive an estimate of the stellar modulation function $M_{\mathrm{sim}}$ (see Fig. \ref{MIRISim estimating modulations}). Similarly, we can estimate the planetary modulation function $M_{\mathrm{p}}$ using the same approach as $M_{\mathrm{sim}}$, but by injecting a planetary spectrum into MIRISim instead of a stellar spectrum. Although the renormalization step removes the direct influence of the input spectrum, residual dependencies between different injected spectra can still arise due to the nonlinearity of the signal projection by the instrument and its subsequent extraction by the pipeline. These effects become noticeable when strong spectral features are present (particularly in $M_{\mathrm{p}}$ estimations), introducing minor variations in the derived modulation functions (see Fig. \ref{PSD Mp HF Tp}). However, these differences remain negligible in practice, making a single simulation sufficient for estimating systematic noise through $M_{\mathrm{sim}}$. Additionally, the modulation function is shaped by signal extraction and calibration procedures during cube reconstruction, as well as the object's position within the FoV. Thus, $M_{\mathrm{sim}}$ serves as a valuable tool for approximating spatial systematic noise in the CCF using Eq. \ref{eq_sigma_syst_prime_est}.

Nevertheless, it is crucial to bear in mind that the level of contrast induced by systematic effects represents a detection limit (Eq. \ref{eq_SNR_limit}). Indeed, $5 \sigma_{\mathrm{syst}}^{\prime} / \alpha_0 $ remains independent of exposure time and the star's magnitude, given that both $\sigma_{\mathrm{syst}}^{\prime}$ and $\alpha_0$ are directly proportional to the integrated stellar flux. It is therefore possible to calculate the contrast limit for different planet temperatures, showing the detection limits that MIRI/MRS can reach with molecular mapping according to systematics estimated with MIRISim (see Fig. \ref{contrast MIRIMRS limits on 1SHORT}). Notably, lower temperatures of the planet yield better contrast, as the spectral content ($\alpha_0$) increases with decreasing temperature, while $\sigma_{\mathrm{syst}}^{\prime}$ does not vary significantly with temperature. This raises the important question of whether this systematic noise floor actually appears in the data.

\section{Identification of systematics in the noise budget} \label{section4}

To validate the accuracy of noise level estimates obtained with FastCurves and MIRISim, we compare our results with both on-sky and simulated data. First of all, we analyze a particular case in detail (Sect. \ref{subsection4_1}), highlighting the presence of systematics and validating the fundamental and systematic noise estimates with FastCurves. Next, we extend the validation of fundamental and systematic noise estimates to a diverse set of observational scenarios (Sect. \ref{subsection4_2}), demonstrating the reliability of the FastCurves noise model. Finally, in the last two subsections, we show how to improve detection performance through the optimization of cutoff resolution selection (Sect. \ref{subsection4_3}), and how to optimize observation time in the presence of systematics (Sect. \ref{subsection4_4}).

\subsection{Particular case: CT Cha b} \label{subsection4_1}

\begin{figure*}
    \centering
    \includegraphics[width=16cm]{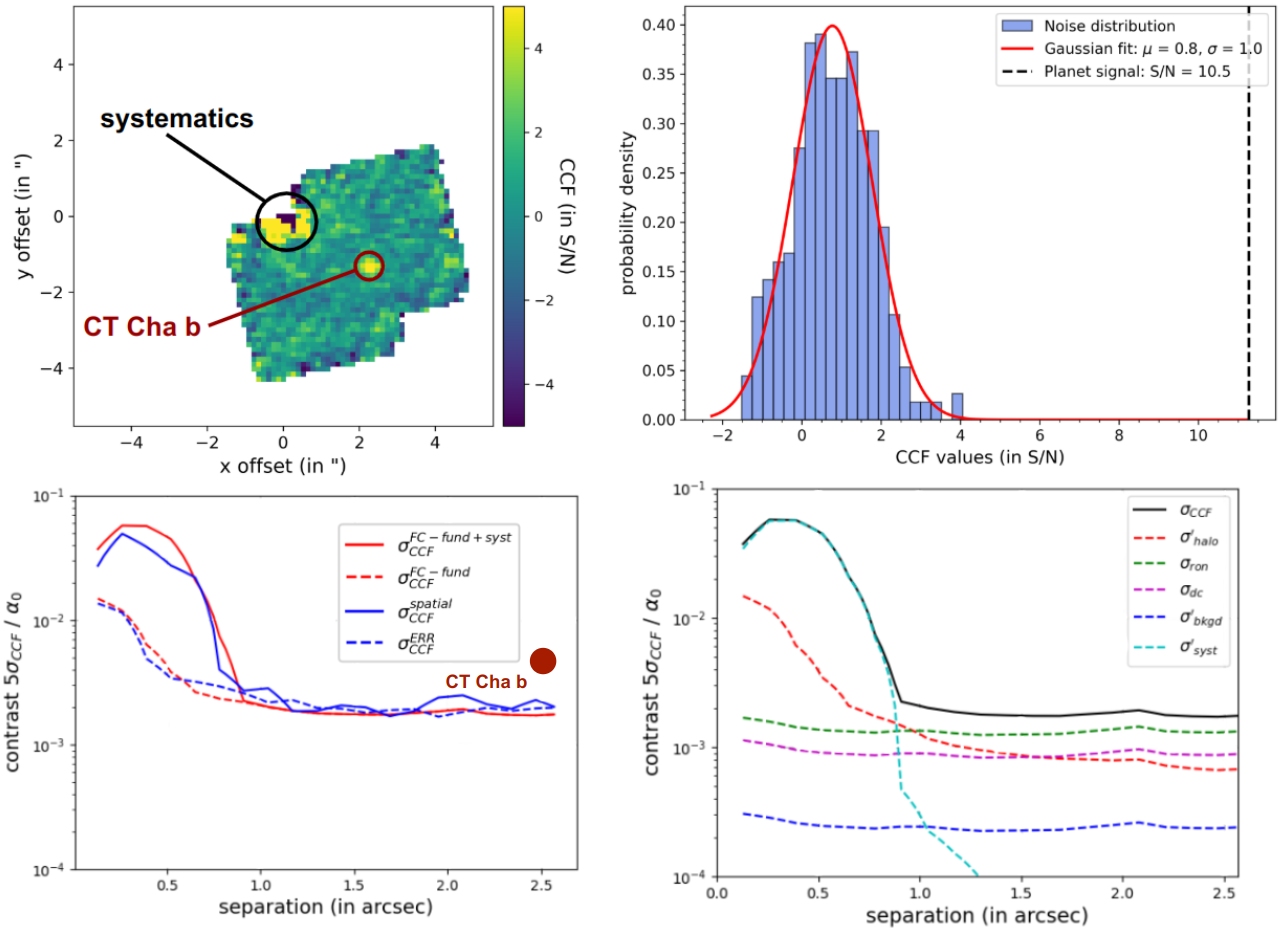}
    \caption{\label{CT Cha perf 1SHORT} Top left: CCF of CT Cha b on-sky data on 1SHORT, calculated with a BT-Settl template at $2600~\mathrm{K}$ Doppler shifted with $13.5 ~ \mathrm{km/s}$ and $R_c=100$ (the S/N color scale is cropped to $\pm5$). Top right: Histogram of the CCF. Bottom left: Contrast curves calculated analytically using FastCurves are shown in red, assuming $t_{exp}=56~\mathrm{min}$, $K_*=8.7$, and the same planetary spectrum used to compute the CCF. The solid red curve represents the result including both fundamental and systematic noise ($\sigma_{\mathrm{CCF}}^{\mathrm{FC, fund+syst}}$), while the dashed red curve represents only fundamental noise ($\sigma_{\mathrm{CCF}}^{\mathrm{FC, fund}}$). Empirical contrast curves derived from on-sky data are shown in blue, where the solid blue curve corresponds to the spatial standard deviation of the CCF ($\sigma_{\mathrm{CCF}}^{\mathrm{spatial}}$), and the dashed blue curve corresponds to the errors estimated from the ERR extension values ($\sigma_{\mathrm{CCF}}^{\mathrm{ERR}}$). Bottom right: Analytical contrast contributions.}
\end{figure*}

To begin with, we consider the on-sky data of CT Cha b with MIRI/MRS, which has been the subject of an observing program for the spectroscopy of its circumplanetary by Rab et al.\footnote{PI: C. Rab, https://www.stsci.edu/jwst/phase2-public/1958.pdf}. Originally discovered and characterized with the IFU VLT/SINFONI by \cite{Schmidt_2008}, CT Cha b exhibits a magnitude of $K_s = 14.9$, an effective temperature of $T_{\mathrm{eff}} = 2600\pm250~\mathrm{K}$, and a mass of $M = 17\pm6~\mathrm{M_{jup}}$, classifying it as a sub-stellar companion (brown dwarf). Positioned at an angular separation of $\rho\approx2.5~"$ from its host star, we propose a new basic characterization of CT Cha b with molecular mapping in Sect. \ref{section6}, supplementing previous characterizations proposed post-discovery by \cite{Patience_2012}, \cite{Bonnefoy_2014} and \cite{Wu_2015}. 

On these data cubes, a CCF can be computed using an arbitrary template, as described earlier. This allows us to estimate an empirical spatial standard deviation for each separation $\sigma_{\mathrm{CCF}}^{\mathrm{spatial}}$. This standard deviation must include all fundamental noises, but especially systematic noise, if any. This way, we compare the empirical contrast limit, derived from the spatial standard deviation of the CCF ($\sigma_{\mathrm{CCF}}^{\mathrm{spatial}}$), with the analytical contrast limit from FastCurves, taking systematic noise into account ($\sigma_{\mathrm{CCF}}^{\mathrm{FC, fund+syst}}$). Secondly, the standard deviations of fundamental noises (i.e., Poisson and detector noises) are estimated by the pipeline during ramp fitting, propagated through different pipeline stages, and stored in the ERR extension of the data cubes \footnote{The error values in the ERR extensions may be incorrectly estimated by the pipeline, sometimes differing by factors as large as 10-50. Therefore, the ERR values should be taken with caution.}. While this provides an empirical standard deviation $\sigma_{\lambda}^{\mathrm{ERR}}$ at each point in the cube, it does not take into account spatial variation and therefore systematic noise. We project the mean standard deviation for each separation $\sigma_{\lambda}^{\mathrm{ERR}}$ into the CCF $\sigma_{\mathrm{CCF}}^{\mathrm{ERR}}$ (likewise Eq. \ref{eq_sigma_halo_prime} or \ref{eq_sigma_bkgd_prime}), and we compare it to the analytical contrast limit given by $\sigma_{\mathrm{CCF}}^{\mathrm{FC, fund}}$ calculated with FastCurves, without taking systematics into account (see Fig. \ref{Summary of the methodology for comparing analytical and empirical noise estimates}).

The template that we used to calculate the CCF is the one yielding the best correlation on 1SHORT (see Sect. \ref{section6}); that is, a BT-Settl spectrum at $2600K$. On 1SHORT, we observe a favorable agreement between the analytical approach (FastCurves) and the on-sky data (see bottom left of Fig. \ref{CT Cha perf 1SHORT}). The analytical noise level with systematics corresponds closely to the empirical noise level derived from spatial variance. Similarly, the analytical noise level without systematics also aligns with the empirical noise level computed from the ERR extension, as expected. The slight discrepancies between the different noise levels may be due to the fact that the simulated noiseless cube used to estimate $M_{\mathrm{sim}}$ has not been calibrated and reconstructed by the pipeline in the exact same way as the data cube (inducing different systematics). They may also be due to the fact that the noise induced by cosmic rays (or other outliers) is not taken into account and from the different assumptions used.

\begin{figure}
    \centering
    \includegraphics[width=8cm]{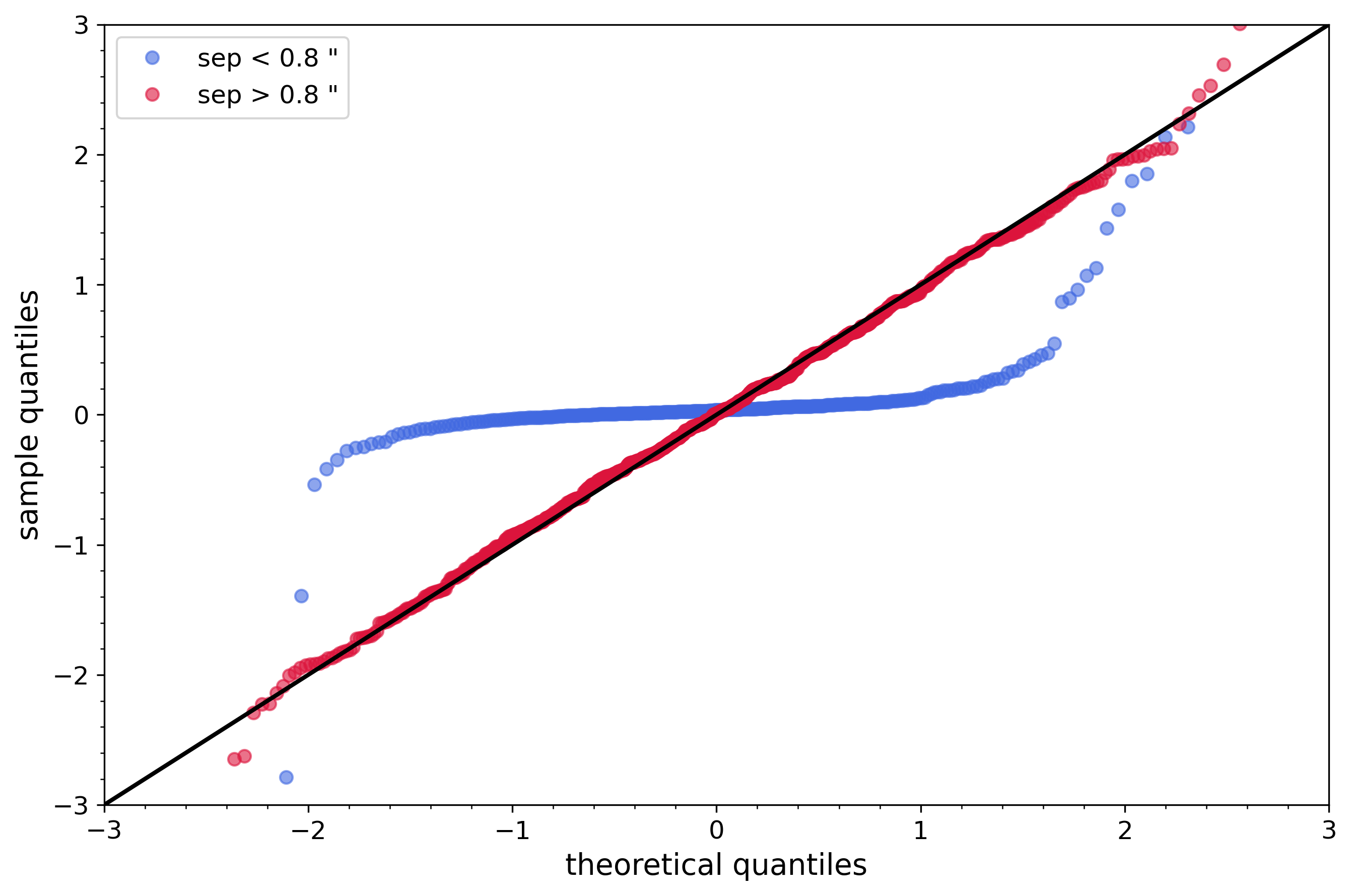}
    \caption{\label{CT Cha b QQ plots 1SHORT} Q-Q plots of the CCF of CT Cha b data on 1SHORT for a separation greater (in red) or lesser (in blue) than $0.8~"$.}
\end{figure}

The noise contributions plot (bottom right panel of Fig. \ref{CT Cha perf 1SHORT}) indicates that detection is limited by systematic noises at short separations ($<0.8~"$), which is expected due to the high stellar flux in this area. At larger separations, the detection can be limited by other noises if the stellar flux and/or exposure time are not sufficiently high. These two regions are expected to exhibit different statistics: fundamental noises should follow a Gaussian distribution whereas systematic noises may not. This is confirmed through quantile-quantile plots (see Fig. \ref{CT Cha b QQ plots 1SHORT}), which compare the quantiles of the CCF distribution against the theoretical quantiles of a normal distribution (a Gaussian distribution would approximately lie on the identity line $y = x$). This implies that $5\sigma$ detection may not carry the same significance in regions dominated by systematic noises \citep{Garvin_2024}. Moreover, these Q-Q plots serve as additional validation of the predictions made by FastCurves.

\begin{figure}
    \centering
    \includegraphics[width=9cm]{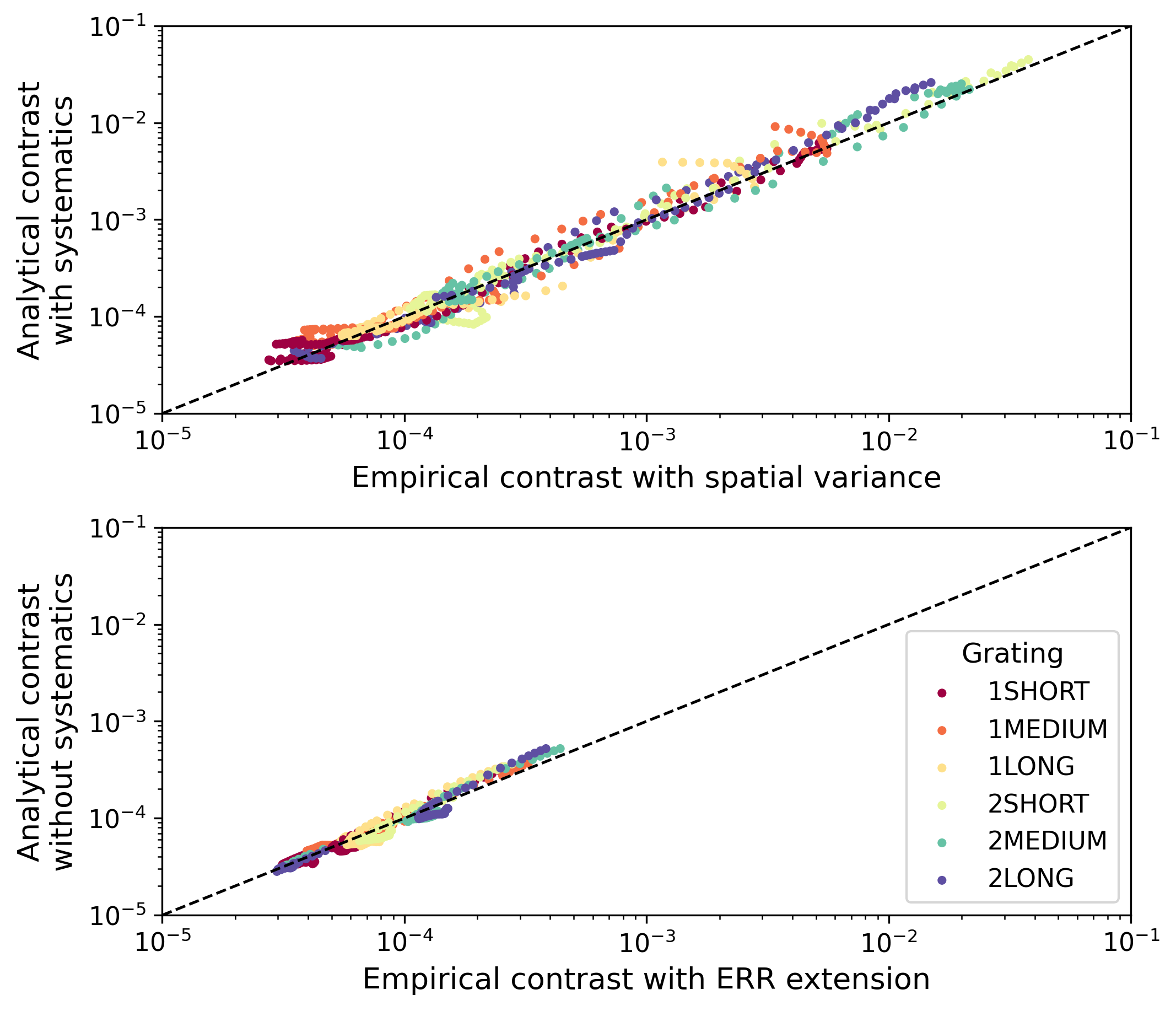}
    \caption{\label{sim data noise level comparison} Comparison of analytical noise levels (y axis) with systematics $\sigma_{\mathrm{CCF}}^{\mathrm{FC, fund+syst}}$ (top) and without systematics $\sigma_{\mathrm{CCF}}^{\mathrm{FC, fund}}$ (bottom), against empirical noise levels in simulated data (x axis) from spatial standard deviations $\sigma_{\mathrm{CCF}}^{\mathrm{spatial}}$ (top) and ERR extensions $\sigma_{\mathrm{CCF}}^{\mathrm{ERR}}$ (bottom). Each point is a given separation, for a given band and data set. The simulated data considered are those used in \cite{Malin_2023} for HR8799 (with $K_*=5.24$ and $t_{\mathrm{exp}}=171~\mathrm{min}$ per band) and $\beta$ Pictoris (with $K_*=3.48$ and $t_{\mathrm{exp}}=93~\mathrm{min}$ per band).}
\end{figure}

\begin{figure}
    \centering
    \includegraphics[width=9cm]{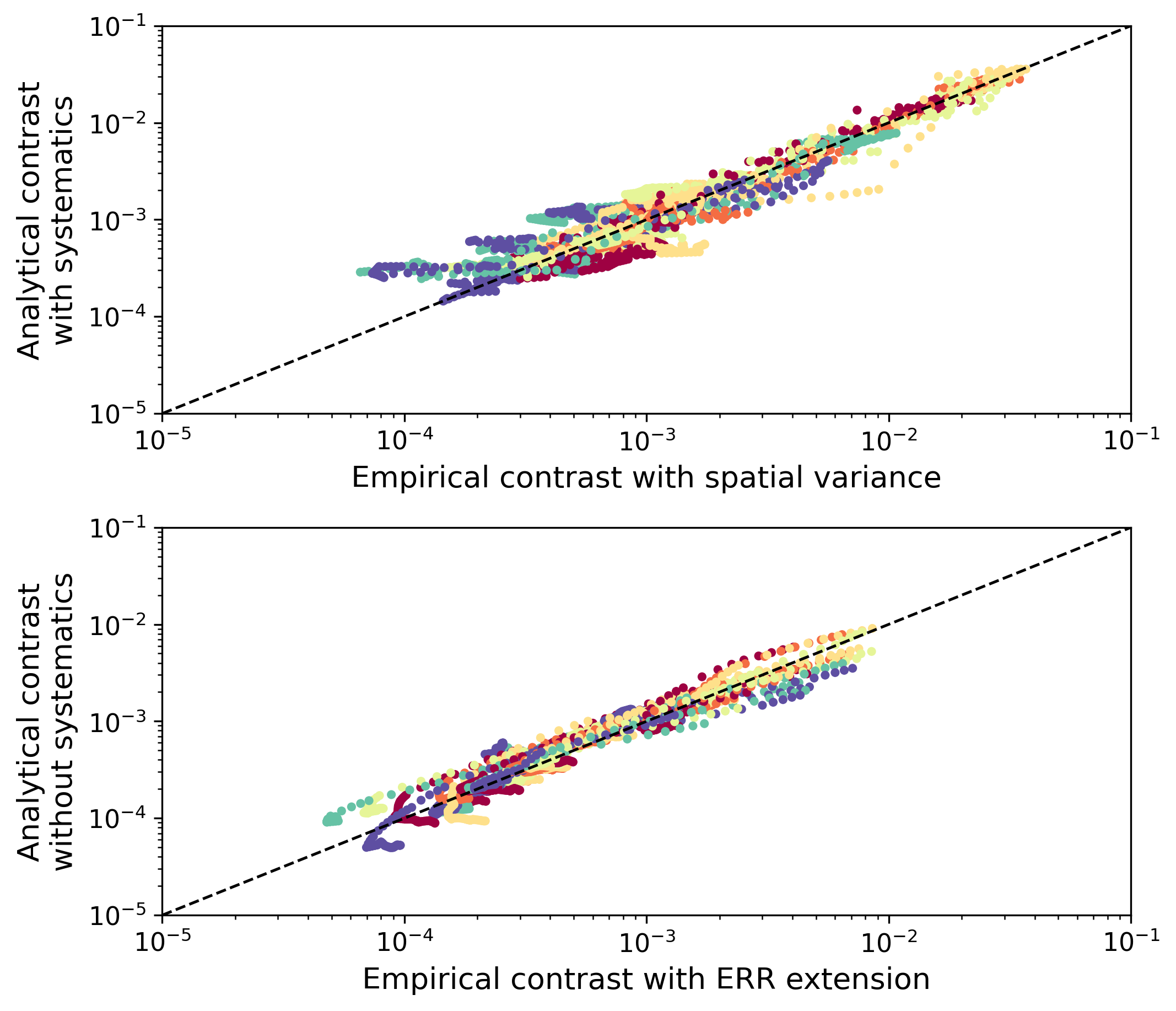}
    \caption{\label{on-sky data noise level comparison} Same as Fig. \ref{sim data noise level comparison} but with on-sky data listed in Table \ref{tab:observed_targets}. To recover the agreement between FastCurves and on-sky data for the top panel, systematics (i.e., $M_{\mathrm{sim}}$) for which fringes and straylight had been incorrectly subtracted by the pipeline had to be considered (see Sect. \ref{section5}).}
\end{figure}

\subsection{Validation of the noise model on various stellar cases} \label{subsection4_2}

It is essential to verify whether FastCurves accurately estimates noise levels across various observational cases and spectral bands. To achieve this, we compare the analytical contrasts (with and without systematics) with empirical contrasts derived from both spatial variance and ERR extension for each separation, band, and various simulated and on-sky data cases. We observe a high level of agreement  between the FastCurves estimates and the simulated data used in \cite{Malin_2023} across all bands (see Fig. \ref{sim data noise level comparison}). This alignment is expected for systematic noise, given its estimation from MIRISim simulated data as well.

\begin{table}[h]
    \centering
    \caption{Considered targets, including stellar magnitude, exposure time, program number, and principal investigator (PI) name.}
    \label{tab:observed_targets}
    \begin{tabular}{lcccc}
        \hline
        {Target} & {${K}$} & {${t_{\mathrm{exp}}}$ (min)} & {Program} & {PI} \\
        \hline
        HD159222    & 5.00  & 18  & 1050  & B. Vandenbussche  \\
        del UMi     & 4.26  & 7   & 1524  & D. Law  \\
        HD2811      & 7.06  & 24  & 6604  & K. Gordon  \\
        V* GO Tau   & 9.33  & 39  & 1640  & A. Banzatti  \\
        V* TW Cha   & 8.62  & 40  & 1549  & K. Pontoppidan  \\
        HD163296    & 4.78  & 7   & 2025  & K. Oberg  \\
        \hline
    \end{tabular}
\end{table}

In order to have a full comparison of the noise levels estimated with FastCurves and on-sky data, we consider the cases listed in Table \ref{tab:observed_targets}. Although the agreement between noise levels estimated with FastCurves and on-sky data is relatively good, it is not as good as with simulated data (since the dispersion is wider, see Fig. \ref{on-sky data noise level comparison}). For noise levels including systematics, it suggests that the systematics simulated by MIRISim may not perfectly mirror those encountered in on-sky observations (as detailed in Sect. \ref{section5}). Moreover, it seems that in all cases the behavior of fundamental noises is correctly estimated. In conclusion, it can be seen that the noise model used in FastCurves seems to be appropriate and sufficient to recover the noise statistics in the data.

Consequently, the initial challenge remains: although the combination of high-contrast imaging and high-resolution spectroscopy with molecular mapping effectively removes speckles that limit detection at short separations, another type of systematic noise continues to constrain detection in this region. One potential strategy to mitigate this systematic noise is by opting for a higher cutoff resolution, but a trade-off must be found to avoid filtering out too much signal (see Sect. \ref{subsection4_3}). However, we shall also demonstrate later (see Sect. \ref{section7}) that a more efficient method of reducing systematic noise is to apply a PCA.

\subsection{Impact of systematics on optimal cutoff resolution} \label{subsection4_3}

\begin{figure}
    \centering
    \includegraphics[width=8cm]{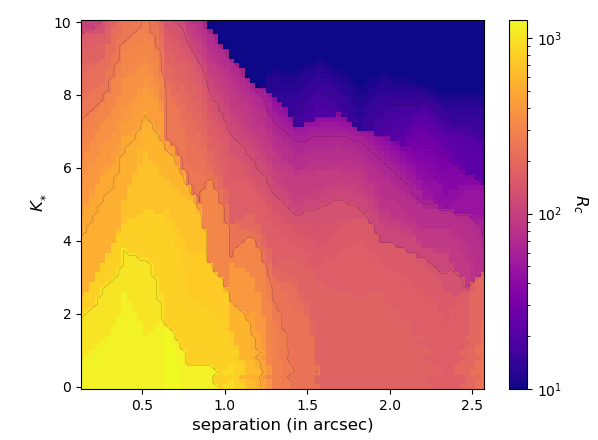}
    \caption{\label{optimum cut-off 1SHORT} Optimum cutoff resolution as a function of stellar magnitude with a BT-Settl planetary spectrum at $1000~\mathrm{K}$ and a total exposure time of one hour.}
\end{figure}

To determine the optimum cutoff resolution for each specific case, we calculated the S/N according to Eq. \ref{eq_SNR} for different cutoff resolutions and star magnitudes. Consequently, we could determine the optimum cutoff resolution (yielding the best S/N) as a function of the star's magnitude and separation (see Fig. \ref{optimum cut-off 1SHORT}). This map delineates the most favorable trade-off in cutoff resolution between signal retention and systematic noise suppression. It reveals that the optimum cutoff resolution decreases with increasing separation, reflecting the decrease in systematics attributable to the declining stellar flux. Additionally, for brighter stars, a higher optimal cutoff resolution is preferable for each separation. While estimating the optimal cutoff resolution may not be inherently useful, given that it can be empirically determined when applying molecular mapping to the data, it is noteworthy that in scenarios dominated by systematics, a higher cutoff resolution is required compared to the one needed for eliminating speckles ($R_c\approx100$).

\subsection{Impact of systematics on total exposure time} \label{subsection4_4}

Selecting an appropriate exposure time is crucial, as the S/N does not increase indefinitely with the square root of exposure time in the presence of systematics. Molecular mapping is employed to remove speckles, assuming they are the primary source of systematic noise. Beyond a cutoff resolution of $R_c = 100$, speckles become negligible, implying that no residual systematic noise should remain. The key question is to determine the exposure time at which this assumption no longer holds, as additional systematics start to dominate and degrade detection performance. Identifying this threshold helps establish the exposure time limit beyond which molecular mapping alone becomes less effective.

\begin{figure}
    \centering
    \includegraphics[width=8cm]{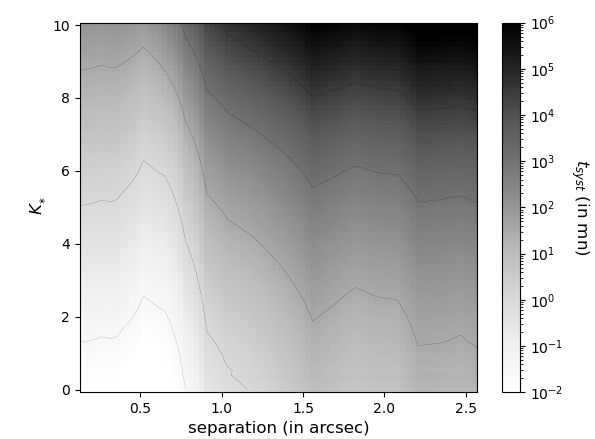}
    \caption{\label{optimum exposure time 1SHORT} Optimum exposure time as a function of the stellar magnitude on 1SHORT with $R_c = 100$.}
\end{figure}

To quantify this limit, we defined the exposure time, $t_{\mathrm{syst}}$, at which systematic noise becomes comparable to fundamental noise, marking the point where systematics become predominant:
\begin{dmath}
     {{\sigma_{\mathrm{syst}}^{\mathrm{total}}} (\rho) = \sigma_{\mathrm{fund}}^{\mathrm{total}} (\rho) }
     \newline{\rightarrow N_{\mathrm{int}}^2 \sigma_{\mathrm{syst}}^{\prime 2} (\rho) = N_{\mathrm{int}} (\sigma_{\mathrm{halo}}^{\prime 2} (\rho)+\sigma_{\mathrm{bkgd}}^{\prime 2}+\sigma_{\mathrm{dc}}^2+\sigma_{\mathrm{RON}}^2) }
     \newline{\rightarrow N_{\mathrm{int}}  = \frac{\sigma_{\mathrm{halo}}^{\prime 2} (\rho)+\sigma_{\mathrm{bkgd}}^{\prime 2}+\sigma_{\mathrm{dc}}^2+\sigma_{\mathrm{RON}}^2}{\sigma_{\mathrm{syst}}^{\prime 2} (\rho)}}
     \newline{\rightarrow  t_{\mathrm{syst}}(\rho) = \mathrm{DIT} \frac{\sigma_{\mathrm{halo}}^{\prime 2} (\rho)+\sigma_{\mathrm{bkgd}}^{\prime 2}+\sigma_{\mathrm{dc}}^2+\sigma_{\mathrm{RON}}^2}{\sigma_{\mathrm{syst}}^{\prime 2} (\rho)}}
     \label{eq_t_syst}
,\end{dmath}
where $\mathrm{DIT}$ is the exposure time per integration. $t_{\mathrm{syst}}$ represents the exposure time at which the S/N  begins to reach the detection limit (Eq. \ref{eq_SNR_limit}), indicating the point at which further observation becomes less beneficial. Figure \ref{optimum exposure time 1SHORT} highlights that the time required to reach the detection limit imposed by systematics increases with a lower stellar flux (due to greater magnitude and/or separation). While this outcome is not surprising, it underscores the feasibility of estimating the optimal exposure time for a given observational case and highlights the utility of FastCurves as an exposure time calculator, particularly for optimizing telescope time allocation, in addition to its performance estimation capabilities.

\section{Contributors to systematics} \label{section5}

Since we have highlighted the dominant role of systematics in limiting molecular mapping detection capabilities at short separations for bright targets, we further explore in this section their origin, behavior, and impact. As was previously stated, these systematics (Eq. \ref{eq_sigma_syst_prime_est}) arise from any instrument-based imperfections in the signal that include high-frequency content and subsequently project onto the template in the CCF. 

We begin this section by listing and describing the potential effects contributing to the observed systematics (Sect. \ref{subsection5_1}). Next, we discuss the impact of different effects and correction procedures on the structure and spatial variations in modulations (Sect. \ref{subsection5_2}). We then explore how these spatial variations affect noise statistics (Sect. \ref{subsection5_3}) and examine how the modulation structure influences the signal (Sect. \ref{subsection5_4}).

\subsection{Possible sources of systematic effects} \label{subsection5_1}

Errors in photometric and wavelength calibration (due to errors in the calibration data, imperfections in re-interpolation operations, cube reconstruction, slice processing, etc.) are likely to induce residuals proportional to the initial flux. A detailed study of the various known effects and in-flight performances of MIRI/MRS is given in \cite{Argyriou_2023}. Here is a summary of the ones that may impact our study:

\begin{itemize}
\item{{Wavelength calibration:} An absolute error in the wavelength solution would not impact detection with molecular mapping, but would simply induce an error in the radial velocity retrieval. If there are relative errors between spaxels for the wavelength solution, this can be more troublesome. In particular, since the planet's signal is integrated over several spaxels, any wavelength shift in the spectra of the different spaxels would result in spectral line broadening, thereby diminishing correlation (see Sect. \ref{subsection5_4}). Such shifts could also introduce high-frequency modulations in the stellar halo spectrum, potentially contributing to systematic noise (see Sect. \ref{subsection5_3}). The current solution (FLT-6) provides calibration accuracy of a few kilometers per second in channels 1 and 2. }
\newline{}
\item{{Photometric calibration:} Similar to wavelength calibration, an absolute error in photometry would not impact the actual S/N found. However, if photometric calibration errors introduce high-frequency modulations, this could result in diminished correlation (see Sect. \ref{subsection5_4}) and systematic noise (see Sect. \ref{subsection5_3}).  All the effects listed below are likely to induce multiplicative errors (modulations) on the flux of each spaxel, and are therefore considered as photometric systematics.}
\newline{}
\item{{Straylight (detector internal scattering):} Scattered light in MIRI/MRS arises from the scattering of photons within the detector substrate and from light diffraction at the narrow gaps between pixels. This results in a broader PSF and an overlap between the detector's diffraction pattern and the instrument's PSF, inducing horizontal band and small spike features in the 3D reconstructed cube \citep[see Fig. 4 of][]{Argyriou_2023}. Those features may be either over- or under-subtracted at the percent level by the pipeline.}
\newline{}
\item{{Fringing:} Similar to many infrared spectrometers, MIRI/MRS exhibits notable spectral fringing attributed to Fabry-Perot interference within the detectors \citep{Argyriou_2020}. The amplitude ($10\text{--}30\%$ of the flux) and phase of the fringes depend on the source extension (whether the source is point-like, non-resolved, or extended, resolved) and wavelength. Presently, the JWST pipeline incorporates two dedicated steps for fringe removal. The first correction involves dividing the 2D detector images by a static 2D flat of fringes derived from spatially extended sources, effectively eliminating fringing from the spectra of such sources. But for point sources, this static correction unavoidably leaves residual fringes, as they generate distinct fringe patterns on the detectors compared to those accounted for by the flat. For this reason, the second step is a residual fringe elimination, which iteratively finds and removes the remaining periodic features in the spectrum, reducing the fringe contrast down to less than $6\%$. Nonetheless, a key question regarding this second correction step remains open for point sources with high spectral richness: either the lines can be identified as fringes by the correction algorithm (and are suppressed), or they can prevent the algorithm from identifying fringes \citep[preventing their suppression;][]{Gasman_2023}. Consequently, the latter step is not considered here.}
\newline{}
\item{{Resampling noise:} This effect is due to an interplay between the cube reconstruction algorithm (re-interpolation of the native detector pixel data into a regular cube grid), spatial undersampling and the curvature of the spectral traces on the detector \citep{Smith_2007}. The resampling of unresolved point sources results in aliasing artifacts, manifesting in the data as relatively low-frequency sinusoidal modulations \citep[see Fig. 9 of][]{Law_2023}. The amplitude of these artifacts is estimated at around $10\%$ for channels 1 and 2 with four-point dithering \citep[see Fig. 10 of][]{Law_2023}.} 
\end{itemize}

\subsection{Structure and spatial variations in the modulations} \label{subsection5_2}

\begin{figure}
    \centering
    \includegraphics[width=9cm]{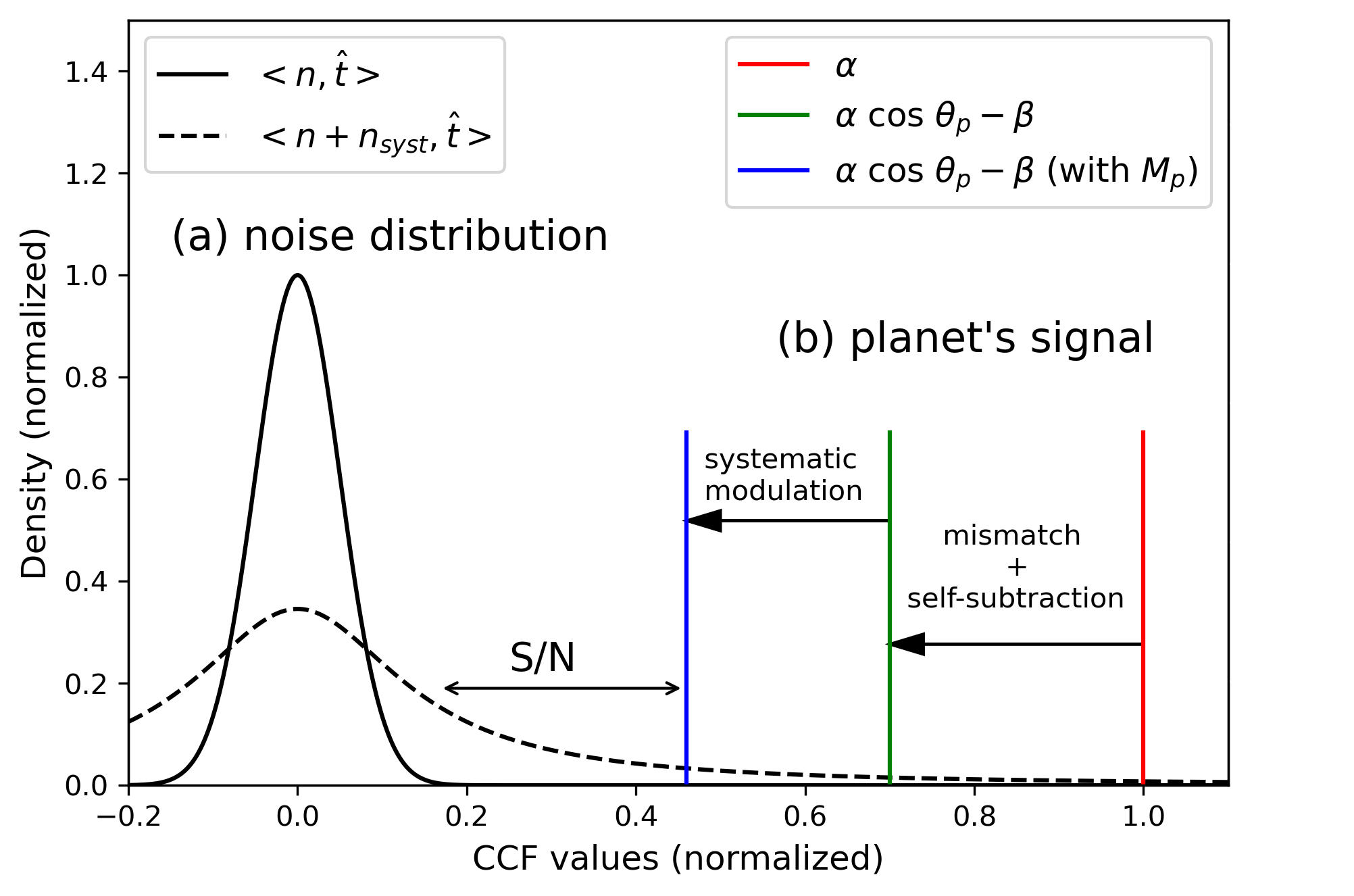}
    \caption{\label{Schema CCF distribution} Schematic diagram of the effect of systematics on noise statistics (a) and on the planet's signal (b). It is important to note that, as previously mentioned, the noise distribution in the presence of systematics will not be Gaussian, making the S/N ratio an unsuitable metric \citep{Garvin_2024}.}
\end{figure}

All of the effects outlined in Sect. \ref{subsection5_1}, along with potentially unidentified factors, contribute to high-frequency modulations in the data, resulting in systematic noise and a reduction in correlation strength. On the one hand, if the residual high-frequency modulations of the stellar halo, $[M]_{\mathrm{HF}}$, vary spatially, the projection values onto the template in the CCF will also exhibit spatial variations, broadening the noise distribution. Conversely, if these modulations were spatially uniform, they would not introduce systematic noise. On the other hand, the planetary modulation function, $M_{\mathrm{p}}$, describes how the planetary signal is projected by the instrument and subsequently extracted by the pipeline. These nonideal modulations degrade the alignment with the template, thereby reducing correlation effectiveness. In summary, the spatial variations in the stellar modulation function, $M$, impact the noise statistics (i.e., the noise distribution (a) in Fig. \ref{Schema CCF distribution}), while the structure of the planetary modulation function $M_{\mathrm{p}}$ over the planet's FWHM impacts the correlation strength (i.e., the planet's signal (b) in Fig. \ref{Schema CCF distribution}).

In Appendix \ref{appendixD}, we analyze the impact of different instrumental corrections (fringe, straylight, outlier subtraction) and reconstruction methods (drizzle vs. EMSM) on the simulated modulation function $M_{\mathrm{sim}}$. The structure and spatial variations in the resulting high-frequency modulations are evaluated using PSD calculations. Our results indicate that, even after applying all pipeline corrections, residual modulations persist, primarily due to aliasing artifacts. Furthermore, uncorrected fringes and straylight significantly affect the PSD of the modulations. We also observe discrepancies between the modulations in simulated and real data, with the latter showing additional modulations likely attributed to uncorrected fringe and straylight effects.

\subsection{Systematic effects on noise} \label{subsection5_3}

\begin{figure}
    \centering
    \includegraphics[width=9cm]{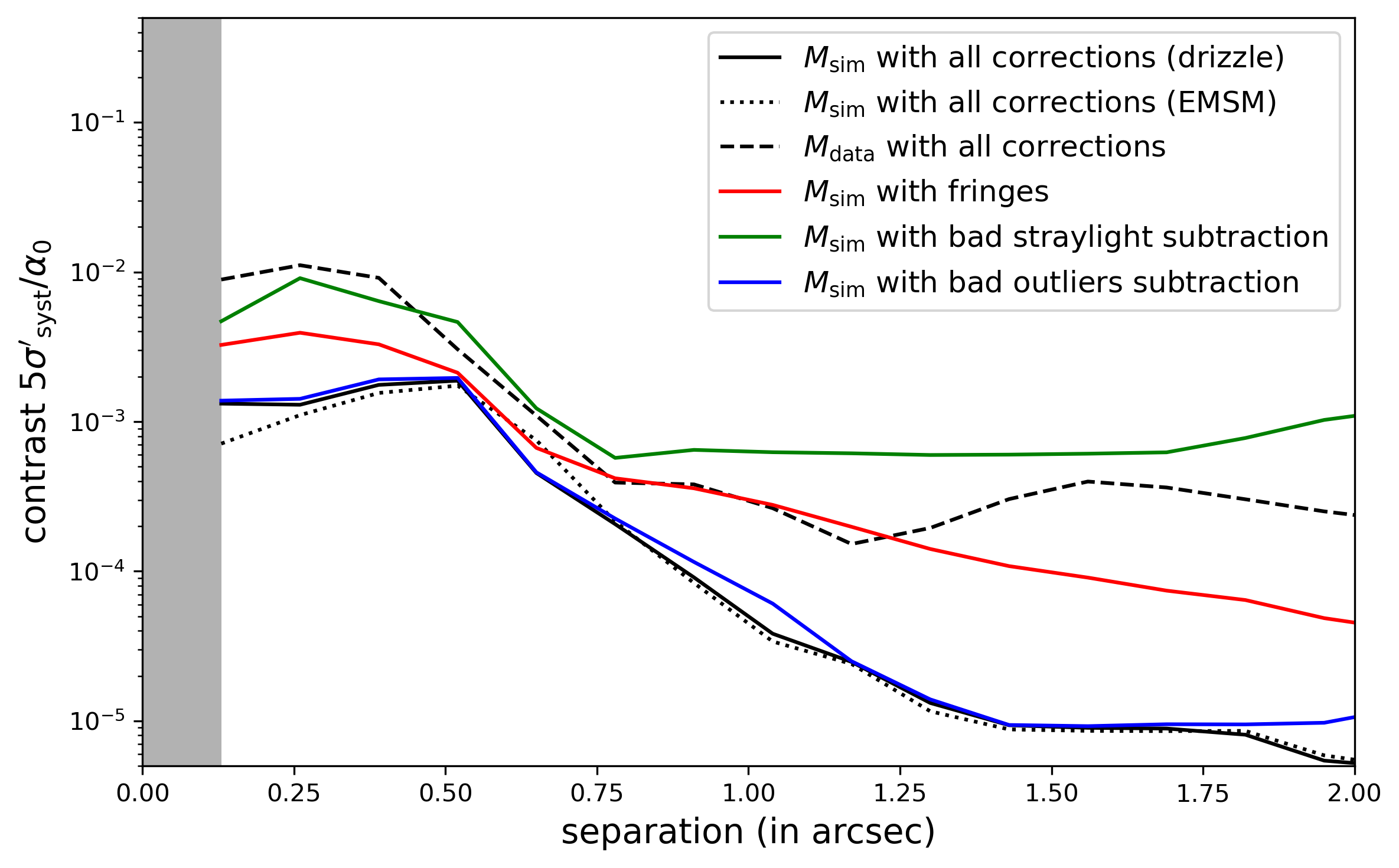}
    \caption{\label{contrast level due to different effects 1SHORT} Contrast set by systematics according to various pipeline corrections on 1SHORT with an arbitrary BT-Settl template at $1000~\mathrm{K}$ and $R_c = 100$.}
\end{figure}

To assess the impact of different effects on modulation-induced noise (i.e., on the noise distribution (a) in Fig. \ref{Schema CCF distribution}), we calculated systematic noise levels using Eq. \ref{eq_sigma_syst_prime_est} with an arbitrary template. We depict the corresponding contrasts in Fig. \ref{contrast level due to different effects 1SHORT}. The results show that straylight is the most critical effect requiring accurate correction, followed by fringing. Poor outlier subtraction or different reconstruction methods appear to have minimal impact. The contrast derived from the observed modulations in on-sky data falls between the levels associated with bad straylight and fringing correction, suggesting that both effects currently limit detection performance in on-sky data.\footnote{This result aligns with the necessity of accounting for bad subtraction of straylight and fringes in the estimation of systematics with MIRISim to achieve the agreement seen in Fig. \ref{on-sky data noise level comparison} between FastCurves and the on-sky data.}

\begin{figure}
    \centering
    \includegraphics[width=9cm]{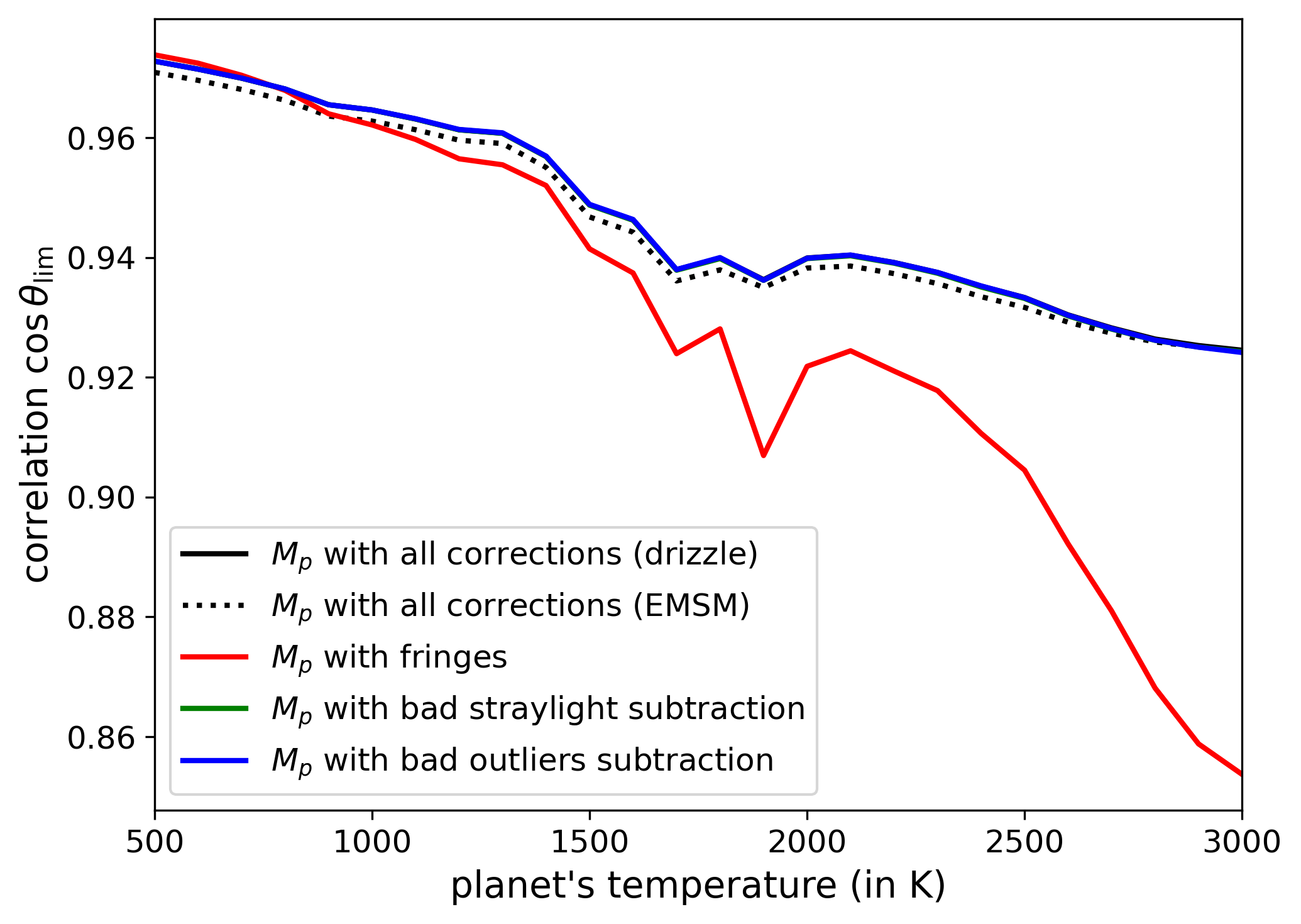}
    \caption{\label{Systematic effect on correlation} MIRI/MRS correlation limits set by MIRISim systematics according to various pipeline corrections on 1SHORT with $R_c=100$. Blue, green, and black curves are overlaid.}
\end{figure}

A question that remains when straylight and fringes are properly corrected is what the reason is for the remaining contrast level (solid black line in Fig. \ref{contrast level due to different effects 1SHORT}). In line with the previous discussion, it seems reasonable to speculate that this residual contrast stems from residual aliasing artifacts post high-pass filtering. To test this hypothesis, we simulated photometric calibration errors of the same nature as resampling noise (see Appendix \ref{appendixE_1}). We find that when the simulated systematics have the same amplitude as that measured for the aliasing artifact in the data ($\approx7\%$), we recover the order of magnitude of the contrast limit obtained with $M_{\mathrm{sim}}$ (with all corrections, i.e., solid black line in Fig. \ref{contrast level due to different effects 1SHORT}). This suggests that resampling noise is likely the primary systematic effect when straylight and fringes are properly corrected.

At the same time, we simulated the impact of wavelength calibration errors on contrast (see Appendix \ref{appendixE_2}). In terms of magnitude, the effects of wavelength calibration errors (of a few kilometers per second) on noise appear to be largely negligible. In fact, a calibration error of $100\text{--}1000~\mathrm{km/s}$ would be required to account for this systematic noise floor (solid black line in Fig. \ref{contrast level due to different effects 1SHORT}).

In conclusion, the systematic noise level appears to be primarily influenced by straylight, fringing, and resampling noise. To alleviate the contribution of the first two, corrective measures are essential during both the correction stages and the subsequent signal extraction and cube reconstruction processes, as mishandling at any stage can exacerbate these effects \citep{Gasman_2024}. 

Secondly, despite the benefits of dithering in reducing aliasing artifacts through improved spatial sampling, their presence remains significant. To mitigate the resampling noise effect mentioned in Sect. \ref{subsection5_1}), it may be wise to avoid the interpolation steps that are introduced in the cube reconstruction step. One potential approach involves directly using 2D calibrated detector images and implementing an alternative molecular mapping technique, similar to the forward modeling method proposed in \cite{Ruffio_2024} for JWST/NIRSpec/IFU. This approach also benefits from the use of spline filtering as a high-pass filter, which allows bad pixels to be ignored, unlike other convolution filters that propagate interpolation errors from bad pixels. Although this technique has shown great improvement for high-contrast spectroscopy with NIRSpec/IFU, adapting it to MIRI/MRS presents additional challenges, particularly regarding fringing, which must first be addressed appropriately. This forward modeling approach is currently under investigation (Bidot et al. in prep.), incorporating recent calibration improvements for point source reduction, as described in \cite{Argyriou_2020, Gasman_2023, Gasman_2024}.

\subsection{Systematic effects on correlation} \label{subsection5_4}

\begin{figure*}
    \centering
    \includegraphics[width=16cm]{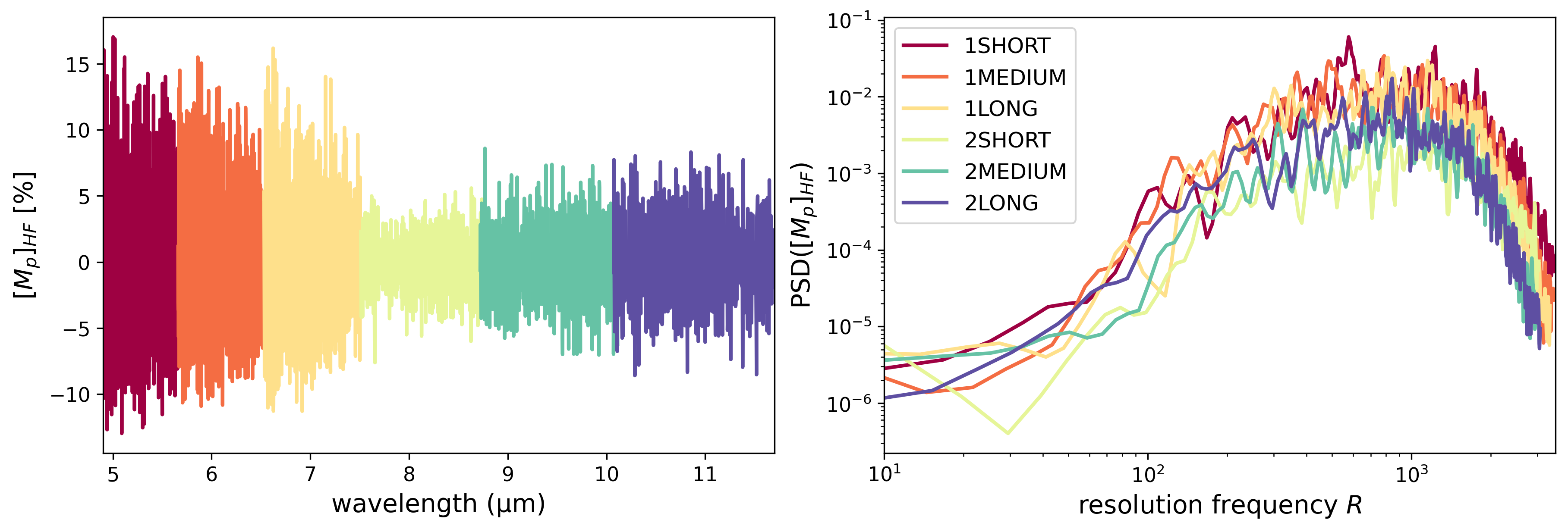}
    \caption{\label{Overall systematic modulation} Left: Overall high-filtered systematic modulations estimation while injecting a BT-Settl spectrum at 1000K in MIRISim with $R_c=100$. Right: PSDs of these modulations.}
\end{figure*}

To assess the influence of systematics on correlation (i.e., the planet's signal (b) in Fig. \ref{Schema CCF distribution}), we reiterated the method previously employed: we simulated noiseless cubes using MIRISim and processed them through the pipeline by injecting different planetary spectra (at different temperatures). From these cubes, we integrated the planetary flux over the FWHM and estimated the drop in correlation according to
\begin{equation}
      \\{ \cos \theta_{\mathrm{lim}} = \frac{\langle \gamma(\lambda)[M_{\mathrm{p}}(\lambda)\mathbb{S}_{\mathrm{p}}(\lambda)]_{\mathrm{HF}},\gamma(\lambda)[\mathbb{S}_{\mathrm{p}}(\lambda)]_{\mathrm{HF}} \rangle}{\| \gamma(\lambda)[M_{\mathrm{p}}(\lambda)\mathbb{S}_{\mathrm{p}}(\lambda)]_{\mathrm{HF}}\| \times \|\gamma(\lambda)[\mathbb{S}_{\mathrm{p}}(\lambda)]_{\mathrm{HF}} \|} }
      \label{eq_systematic_effect_on_correlation}
,\end{equation}
where $\mathbb{S}_{\mathrm{p}}(\lambda)$ is the injected planetary spectrum model. This means that the reduction in correlation arises not only from spectral mismatches between the observed spectrum and the template used but also from systematic effects, as quantified here. Thus, we assume that the effective correlation loss from both effects is simply given by the product $\cos \theta_{\mathrm{lim}} \times \cos \theta_{\mathrm{p}}$ \footnote{This assumption holds strictly when the systematic deviations of the observed planetary spectrum are entirely uncorrelated with the template used.}. $\cos \theta_{\mathrm{lim}}$ represents the maximal correlation achievable given the systematic distortions (see Fig. \ref{Systematic effect on correlation}). We find that the most important effect to correct properly for the correlation is the fringing, with a negligible impact observed from other effects. Also, the greater the temperature, the more the correlation is diminished by the modulations. This is because the lower the temperature, the greater the spectral content and the more it will dominate high-frequency systematic modulations. Finally, the same question arises again of why, when the fringes are perfectly corrected, there is a remaining drop in correlation (solid black line in Fig. \ref{Systematic effect on correlation}).

Similarly to the approach described earlier, we estimated the reduction in correlation caused by simulated photometric effects akin to resampling noise (see Appendix \ref{appendixF_1}). This revealed that the systematic decline in correlation observed in the data cannot be attributed to resampling noise, given that a photometric error of $20\text{--}50\%$ would be required to explain it. This is partly due to the fact that the modulations of the aliasing artifacts are relatively low-frequency compared to the spectral characteristics of the planetary spectra.

In the same way, it is possible to estimate the reduction in correlation caused by simulated wavelength calibration errors (see Appendix \ref{appendixF_2}). It suggests that the expected wavelength calibration errors also do not appear to be the dominant source of the systematic drop in correlation (given that it would take an error of around $60~\mathrm{km/s}$ to explain it). 

\begin{figure}
    \centering
    \includegraphics[width=8cm]{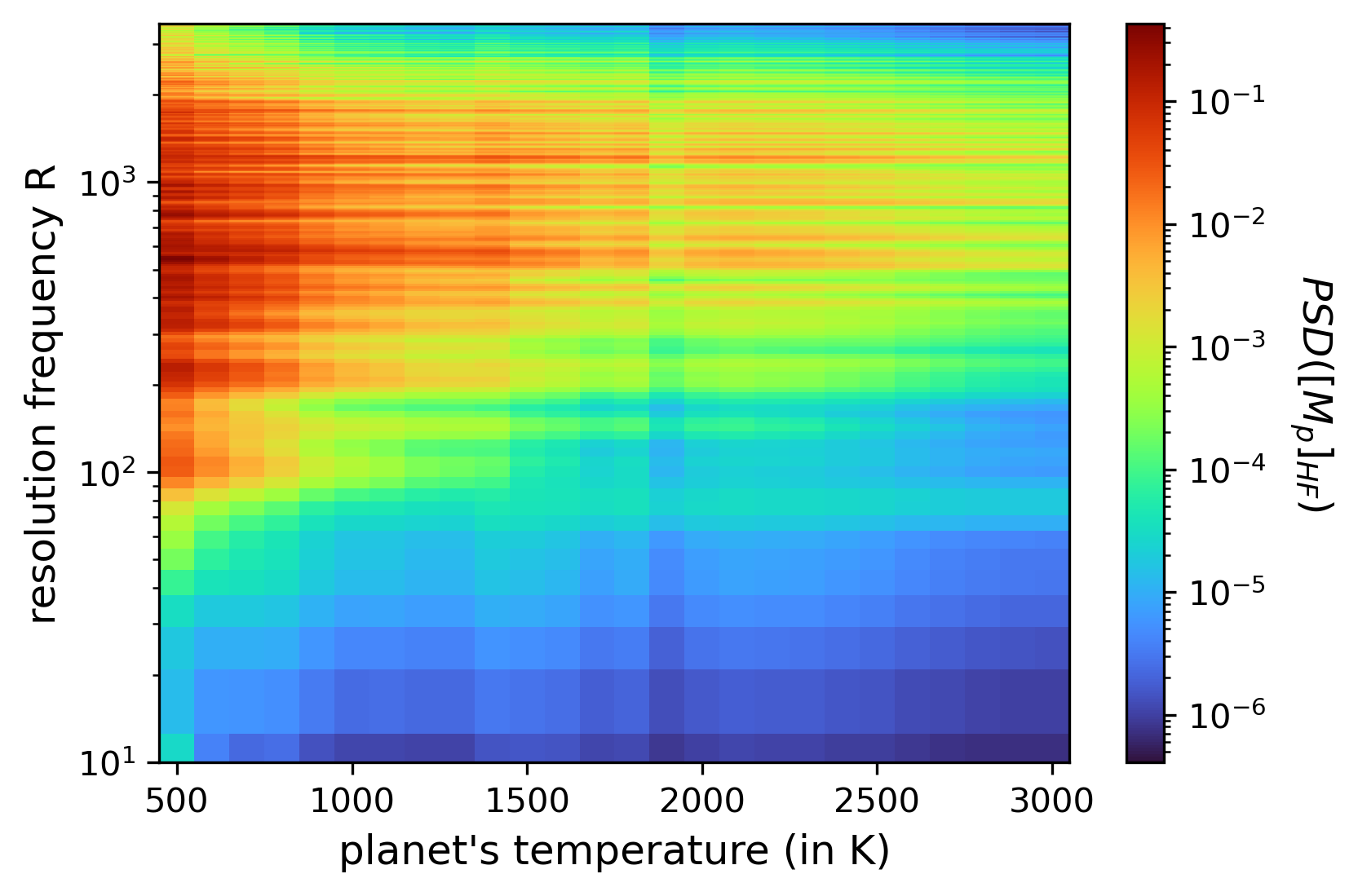}
    \caption{\label{PSD Mp HF Tp} PSDs of high-filtered modulations of the planetary spectrum as a function of its temperature on 1SHORT with $R_c = 100$ (same as the right panel of Fig. \ref{Overall systematic modulation} but for different temperatures).}
\end{figure}

In summary, the observed drop in correlation when fringes are corrected does not appear to be due to resampling noise or wavelength calibration errors. Instead, this reduction in correlation is attributed to very high-frequency ($R\approx1000$) modulations common to all spaxels, rather than relative modulations between spaxels (see Fig. \ref{Overall systematic modulation}). This overall modulation is strongly influenced by the injected spectral content: the greater the content (i.e., the lower the temperature), the more pronounced the modulations (see Fig. \ref{PSD Mp HF Tp}) \footnote{This explains why this modulation dominates over aliasing artifacts for the planetary modulation function over the FWHM, whereas it is the opposite for the stellar modulation function (since this modulation is not clearly visible in the left panel of Fig. \ref{mean and std PSD modulations with differents effects}). Even if this modulation dominated the stellar modulation function, it would not introduce noise, as it is spatially homogeneous.}. The origin of this overall modulation remains unclear, but it is reasonable to speculate that it arises from common photometric calibration errors and/or interpolation procedures. Finally, given that this systematic effect operates at high frequency, increasing the cutoff resolution would not improve correlation. However, even in the worst-case scenario -- at high temperatures and with poor fringe calibration -- the signal lost due to systematics remains below $15\%$. This implies that systematics will mainly limit detections by impacting spatial noise statistics, meaning they have a greater impact on (a) than on (b) in Fig. \ref{Schema CCF distribution}.

\section{Signal and correlation estimations} \label{section6}

\begin{figure}
    \centering
    \includegraphics[width=8cm]{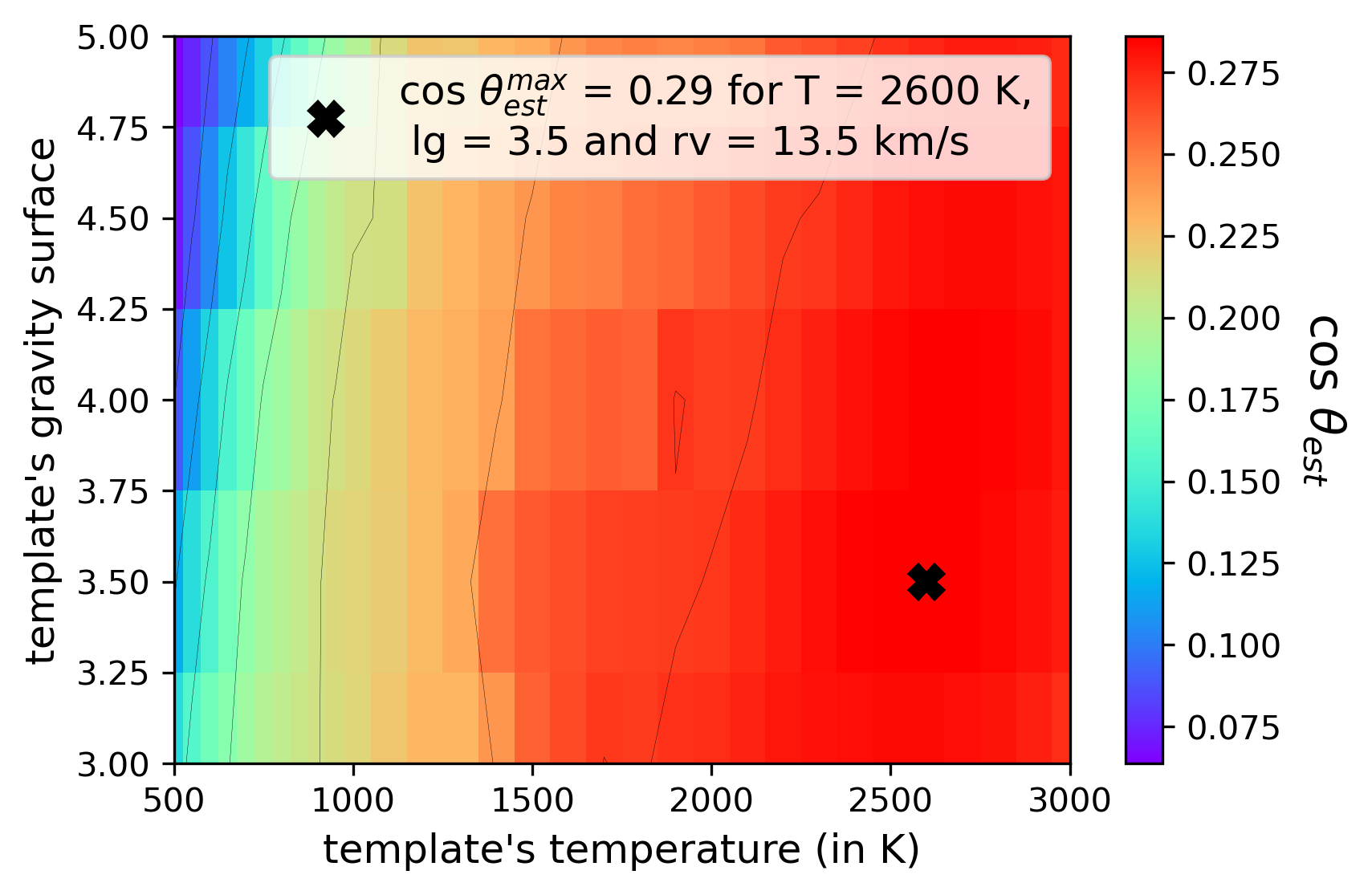}
    \caption{\label{CT Cha b correlation estimation 1SHORT} Estimated correlation between CT Cha b high-filtered data spectrum and BT-Settl spectra on 1SHORT with $R_c = 100$. }
\end{figure}
\begin{figure}
    \centering
    \includegraphics[width=8cm]{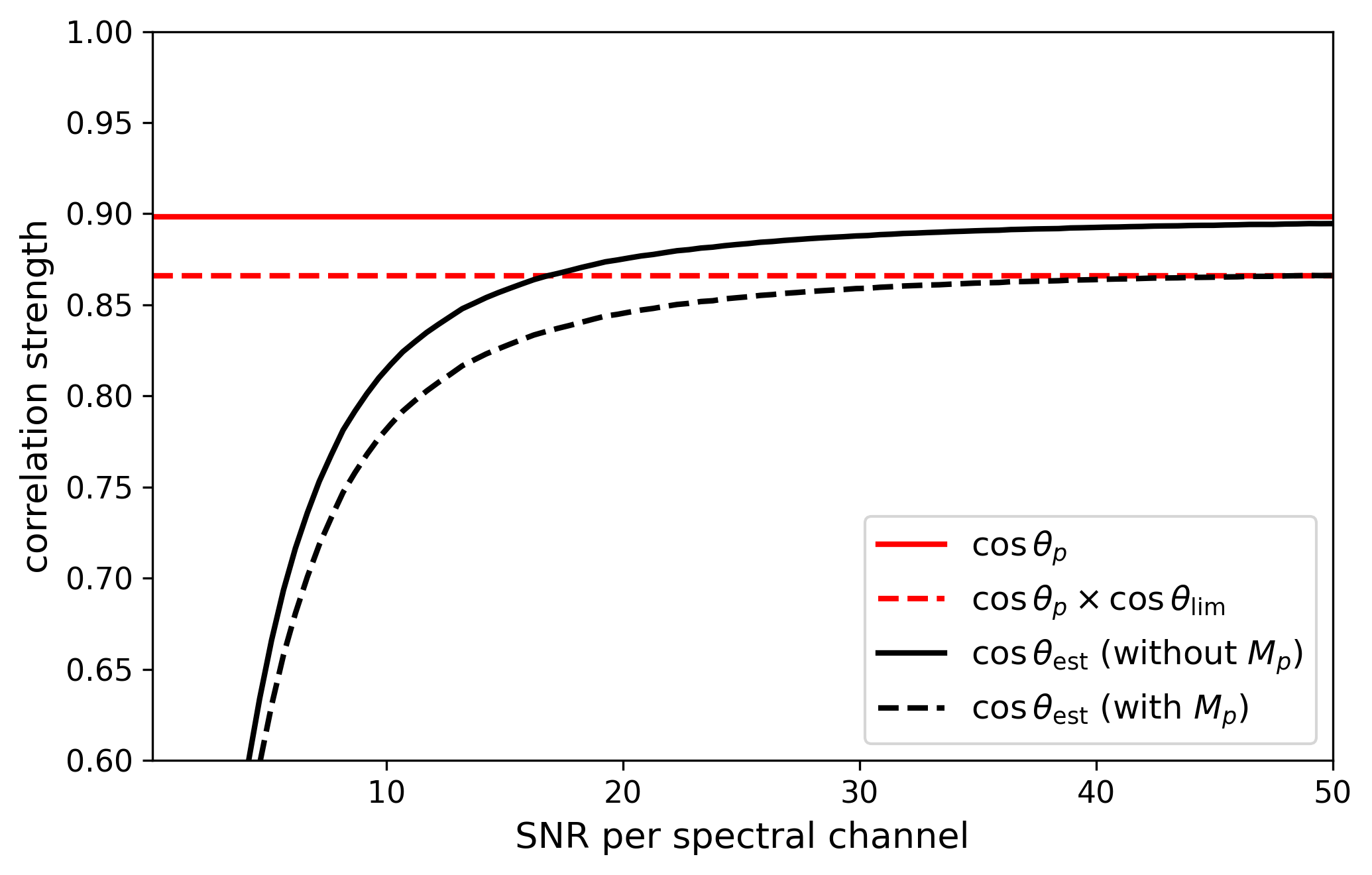}
    \caption{\label{Impact of noise on correlation estimation 1SHORT} Impact of noise on the estimation of the correlation on 1SHORT with $T_{\mathrm{p}} = 1000~\mathrm{K}$ and $R_c = 100$. A BT-Settl spectrum is associated with the observed spectrum and a SONORA spectrum with the template to create an artificial mismatch ($\cos \theta_{\mathrm{p}} \approx 0.9$). For each S/N per spectral channel, Gaussian noise is added to the observed spectrum to simulate the noise. }
\end{figure}
While FastCurves has proven reliable for estimating noise levels, its accuracy in modeling the signal itself remains to be assessed. To address this, we again considered the case of CT Cha b. First, we estimated the planetary spectrum model that best correlates with the observed data spectrum using
\begin{equation}
      \\{ \cos \theta_{\mathrm{est}} = \frac{\langle \hat d (\lambda) , \hat t (\lambda) \rangle}{ \| \hat d (\lambda) \| } }
,\end{equation}
where $\hat d$ is the signal of the planet in the high-filtered cube, $S_{\mathrm{res}}$, integrated over the FWHM. The spectrum that appears to be most similar to the CT Cha b spectrum is a BT-Settl spectrum at $2600~\mathrm{K}$ with a surface gravity of $3.5$ and a Doppler shift of around $13.5~\mathrm{km/s}$ (achieving a correlation of around $0.29$, see Fig. \ref{CT Cha b correlation estimation 1SHORT}). \footnote{Minimizing a $\chi^2$ in the same way as in \cite{Malin_2023} yields to the same parameters.}

Now, the idea is to deduce the intrinsic mismatch, $\cos \theta_{\mathrm{p}}$, from the estimated correlation, $\cos \theta_{\mathrm{est}}$. Neglecting self-subtraction, it is possible to write (using the fact that $\langle n + n_{\mathrm{syst}} , \hat t \rangle \approx 0 $ on average): 
\begin{dmath}
     { \hat d(\lambda) =  \gamma(\lambda) [M_{\mathrm{p}}(\lambda) S_{\mathrm{p}}(\lambda)]_{\mathrm{HF}} + n(\lambda) + n_{\mathrm{syst}}(\lambda)}
     \newline{\rightarrow \cos \theta_{\mathrm{est}} = \frac{\alpha }{\| \hat d (\lambda) \|} \times \cos \theta_{\mathrm{lim}} \times \cos \theta_{\mathrm{p}}}
     \newline{\rightarrow \underbrace{\cos \theta_{\mathrm{est}}}_{\substack{\text{estimated} \\ \text{correlation}}} = \underbrace{\cos \theta_{\mathrm{n}}}_{\substack{\text{noise-induced} \\ \text{correlation loss}}} \times \underbrace{\cos \theta_{\mathrm{lim}}}_{\substack{\text{systematic-induced} \\ \text{correlation loss}}} \times \underbrace{\cos \theta_{\mathrm{p}}}_{\substack{\text{model-induced} \\ \text{correlation loss}}}}
     \label{eq_cos_theta_est}
,\end{dmath}
defining $ \cos \theta_{\mathrm{n}} = \alpha / \| \hat d \|$, representing the drop in the estimated correlation due to fundamental noises. Therefore, the estimated correlation ($ \cos \theta_{\mathrm{est}} $) may not accurately reflect the intrinsic discrepancy between the observed spectrum and the template ($ \cos \theta_{\mathrm{p}} $). We estimate the quantity in Eq. \ref{eq_cos_theta_est} for different S/Ns per spectral channel and plot it in Fig. \ref{Impact of noise on correlation estimation 1SHORT}. It illustrates that the greater the S/N per spectral channel, the closer the estimated correlation will be to the mismatch $\cos \theta_{\mathrm{p}}$ (or to $\cos \theta_{\mathrm{lim}} \times \cos \theta_{\mathrm{p}}$ if there are systematic modulations). It is also observed that the greater the planet's temperature, the greater the S/N per spectral channel required to retrieved the mismatch. This is simply because the higher the temperature, the lower the spectral content and the less it will be distinguishable from the noise. 

\begin{figure}
    \centering
    \includegraphics[width=9cm]{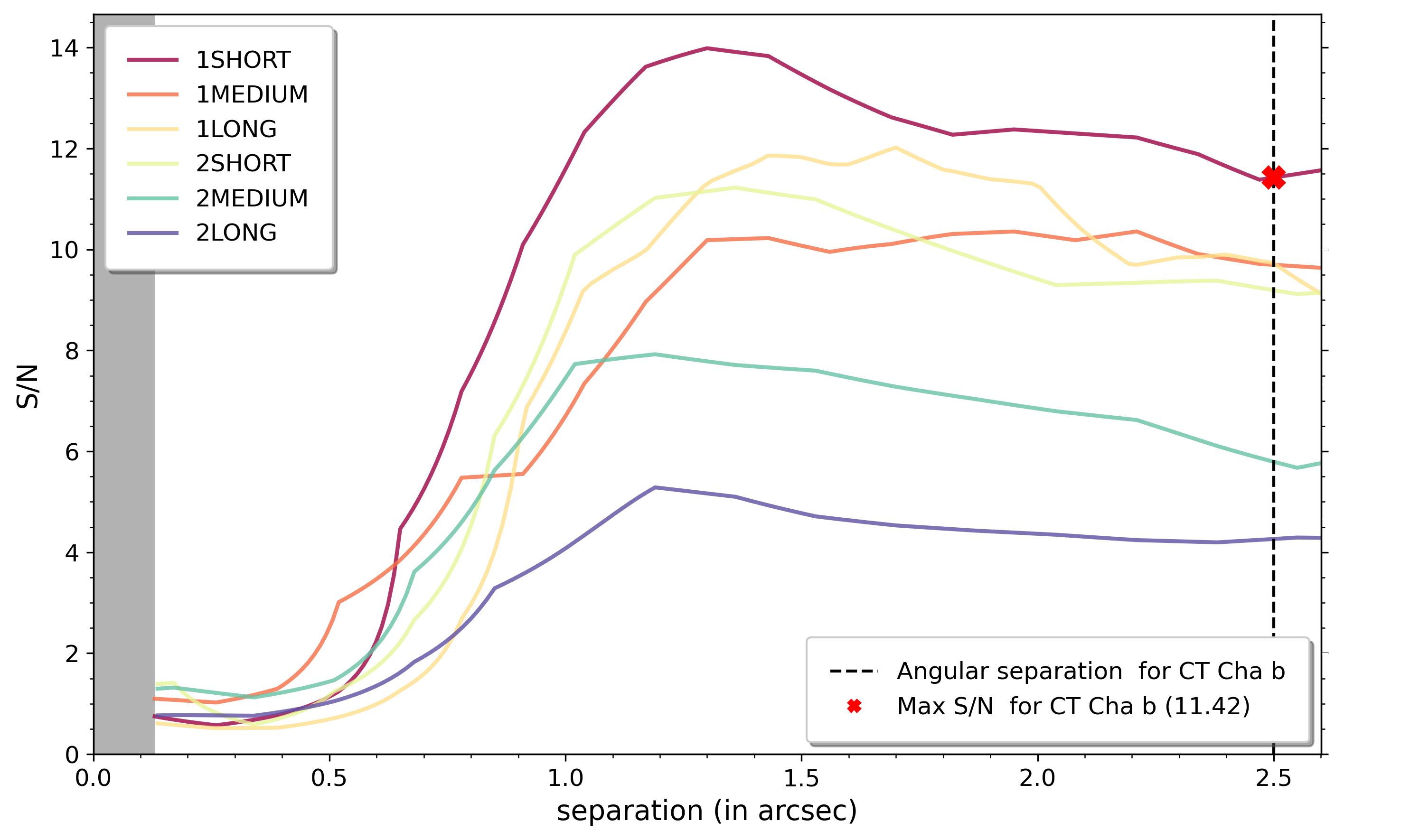}
    \caption{\label{CT Cha b SNR estimation} S/N curves estimated with FastCurves for the CT Cha b observation with MIRI/MRS (with $t_{exp}=56~\mathrm{min}$, $K_*=8.7$, $K_p=14.9$, and with the BT-Settl spectrum at $2600~\mathrm{K}$).}
\end{figure}

If self-subtraction is now considered, it is written:
\begin{dmath}
     {\cos \theta_{\mathrm{est}} = \cos \theta_{n} \times \cos \theta_{\mathrm{lim}} \times \cos \theta_{\mathrm{p}} - \frac{\beta}{\| \hat d (\lambda) \|}}
     \newline{\rightarrow \cos \theta_{\mathrm{p}} = \frac{\cos \theta_{\mathrm{est}}/\cos \theta_{n} + \beta/\alpha}{\cos \theta_{\mathrm{lim}}} }
     \label{eq_cos_theta_p}
,\end{dmath}
where $\beta/\alpha$ is the fraction of the signal that is self-subtracted. Hence, a straightforward and linear relationship exists between the estimated correlation and the mismatch. The best that can be done now is to assume that, even if the spectrum model differs from the observed one, the quantities $\cos \theta_n$, $\cos \theta_{\mathrm{lim}}$, and $\beta/\alpha$ are accurately estimated with FastCurves (knowing that the noise level is well estimated). This way, the estimated correlation of around $0.29$ actually implies a mismatch of around 0.82 with the template used (with $\cos \theta_{est} \approx 0.29$, $\cos \theta_{n} \approx 0.54$, $\cos \theta_{\mathrm{lim}} \approx 0.94$, and $\beta/\alpha \approx 0.23$ at the planet separation). It is noteworthy that if the template was exactly the observed spectrum ($\cos \theta_{\mathrm{p}} = 1$), a correlation of about $0.38$  would have been measured. The model used therefore seems to be much more similar to the observed spectrum than it initially appeared. Lastly, the S/N curves for the case of CT Cha b were calculated by introducing a mismatch of $0.82$ (see Fig. \ref{CT Cha b SNR estimation}), revealing that FastCurves estimates a S/N of around 11.4 on 1SHORT, which closely aligns with the observed S/N ($10.5$, see upper right panel of Fig. \ref{CT Cha perf 1SHORT}). Similar observations hold true for the other bands of channel 1. This underscores the capability of FastCurves to reliably estimate the signal as well.

\section{Enhancing detection limits with PCA} \label{section7}

\begin{figure}
    \centering
    \includegraphics[width=8.5cm]{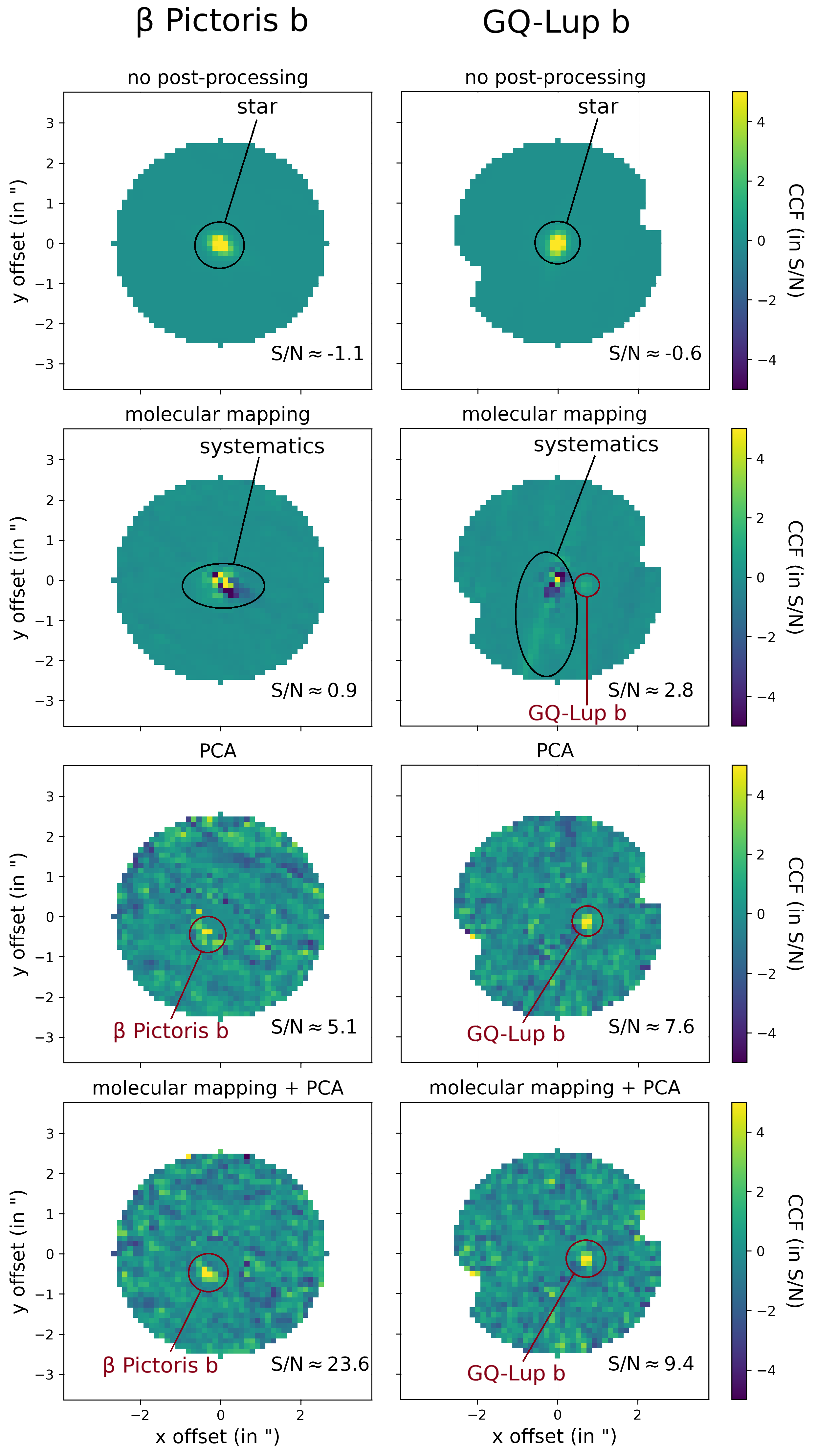}
    \caption{\label{bet pic GQ-Lup molecular mapping PCA comparison} CCF S/N maps on 1SHORT for $\beta$ Pictoris b data (with $t_{exp}=50~\mathrm{min}$, $K_*=3.5$, $K_p\approx12.7$ and $T_{\mathrm{eff}}\approx1600~\mathrm{K}$) and GQ-Lup b (with $t_{exp}=30~\mathrm{min}$, $K_*=7.1$, $K_p\approx13.5$ and $T_{\mathrm{eff}}\approx2700~\mathrm{K}$) using various post-processing techniques (the S/N color scales are cropped to $\pm5$). Molecular mapping was applied with a cutoff resolution of 100, and PCA was used to subtract 100 principal components.}
\end{figure}

Given the presence of systematics, it is worth attempting to mitigate their impact by applying a PCA to the spectral dimension of the data. Although PCA is typically used with temporal diversity, here it leverages the spectral diversity of spaxels within the data cube to subtract principal modes associated with spatially varying systematic spectral modulations. This application of PCA to the spectral dimension has been shown to be effective for exoplanet detection, as has been demonstrated by \cite{Ruffio_2024}, \cite{Parker_2024}, and \cite{Landman_2023b}. We applied PCA to the filtered cube, $ S_{\mathrm{res}} $ (see Eq. \ref{eq_S_res}), and then performed cross-correlation, making both $\beta$ Pictoris b\footnote{PI: C. Chen, https://www.stsci.edu/jwst/phase2-public/1294.pdf} and GQ-Lup b\footnote{PI: A. Banzatti, https://www.stsci.edu/jwst/phase2-public/1640.pdf} detectable (see Fig. \ref{bet pic GQ-Lup molecular mapping PCA comparison}). Detailed analyses of these detections are provided in \cite{Worthen_2024} and \cite{Cugno_2024}, respectively.

\begin{figure}
    \centering
    \includegraphics[width=9cm]{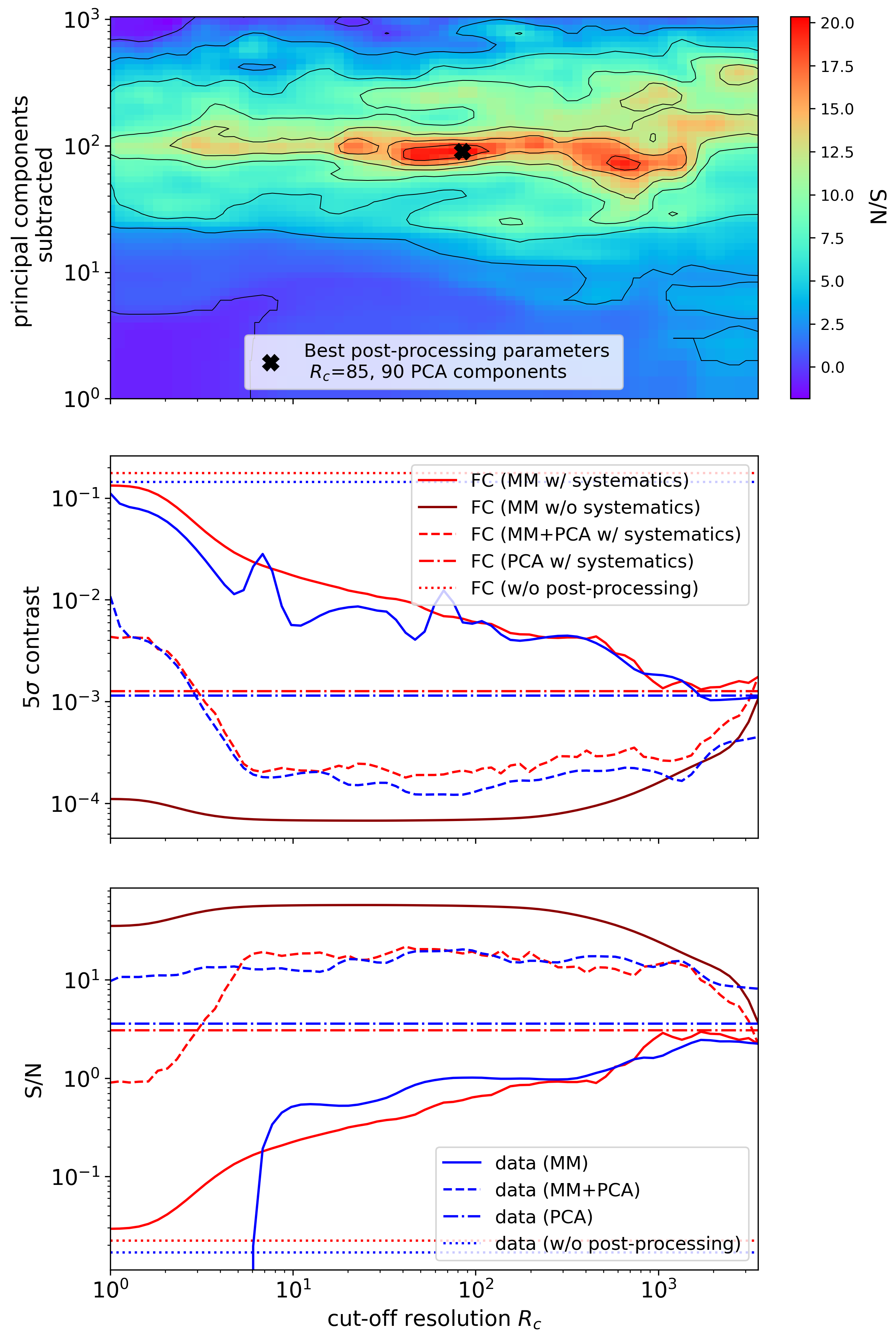}
    \caption{\label{bet pic b best post-processing parameters} Top: S/N fluctuation map as a function of post-processing parameters on 1SHORT for $\beta$ Pictoris b on-sky data. Middle: Contrast curves calculated analytically with FastCurves (in red) and empirically with $\beta$ Pictoris b on-sky data (in blue) on 1SHORT for various post-processing techniques. Bottom: S/N curves calculated similarly. The plots are given at the planet's separation ($\rho \approx 0.5~"$) and for 90 principal components subtracted when the PCA is applied. For FastCurves estimations, as in Fig. \ref{on-sky data noise level comparison}, systematics ($M_{\mathrm{sim}}$) where fringes and straylight were incorrectly subtracted by the pipeline had to be considered (see Sect. \ref{section5}).}
\end{figure}

\begin{figure}
    \centering
    \includegraphics[width=9cm]{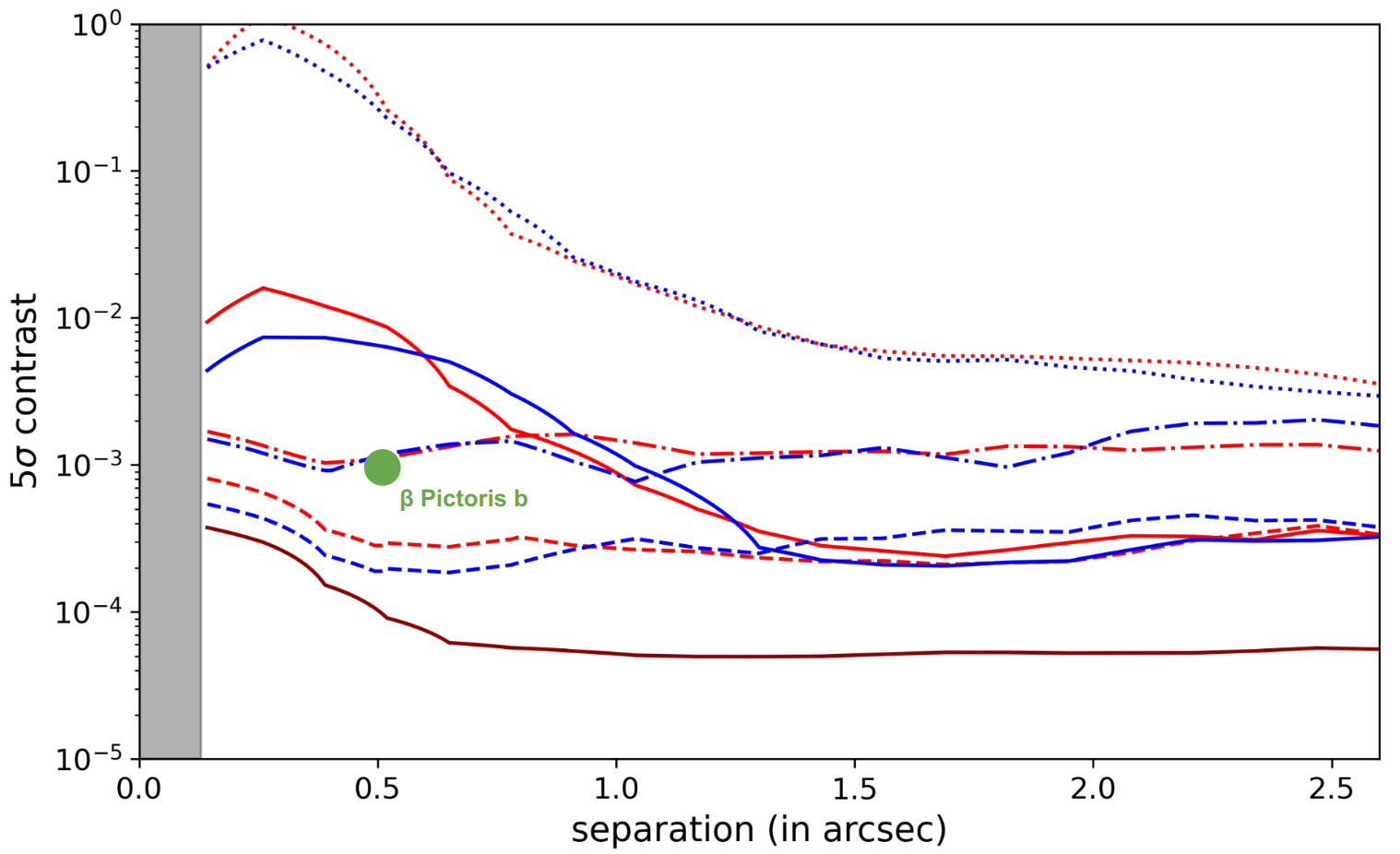}
    \caption{\label{bet pic b best post-processing parameters CONTRAST} Same as the middle panel of Fig. \ref{bet pic b best post-processing parameters} (using the same color scheme) but presented as a function of separation instead of cutoff resolution. The plots correspond to the optimal post-processing parameters for $\beta$ Pictoris b on-sky data in 1SHORT, with a cutoff resolution of 85 and the subtraction of 90 principal components.}
\end{figure}

Focusing on $\beta$ Pictoris b, we calculated the S/N for various post-processing parameters, such as varying cutoff resolutions and the number of subtracted principal components (see Fig. \ref{bet pic b best post-processing parameters} and Fig. \ref{bet pic b best post-processing parameters CONTRAST}). Additionally, we incorporated PCA into FastCurves predictions by applying it to the simulated noiseless cubes, which we used to assess the level of systematics and estimate the corresponding signal drop. In the middle panel of Fig. \ref{bet pic b best post-processing parameters}, we show the contrasts estimated from the data that were obtained using the $\alpha_0$ calculated by FastCurves, allowing for a comparison of the different noise levels in the first instance. It is observed that the noise levels estimated by FastCurves still align well with those in the data for the same post-processing methods. Similarly, the estimated and measured signal levels are consistent (bottom panel of Fig. \ref{bet pic b best post-processing parameters}). In this case, PCA alone outperforms molecular mapping alone, but the best detection performance is achieved by combining PCA with molecular mapping. This combined approach closely approaches the performance set by fundamental noise limits, though it does not completely reach that level due to the signal self-subtraction induced by PCA. The positive impact of PCA is further highlighted in Fig. \ref{bet pic b best post-processing parameters CONTRAST}, which shows its critical role in detecting close companions in systematics-dominated regimes.

\begin{figure}
    \centering
    \includegraphics[width=8.5cm]{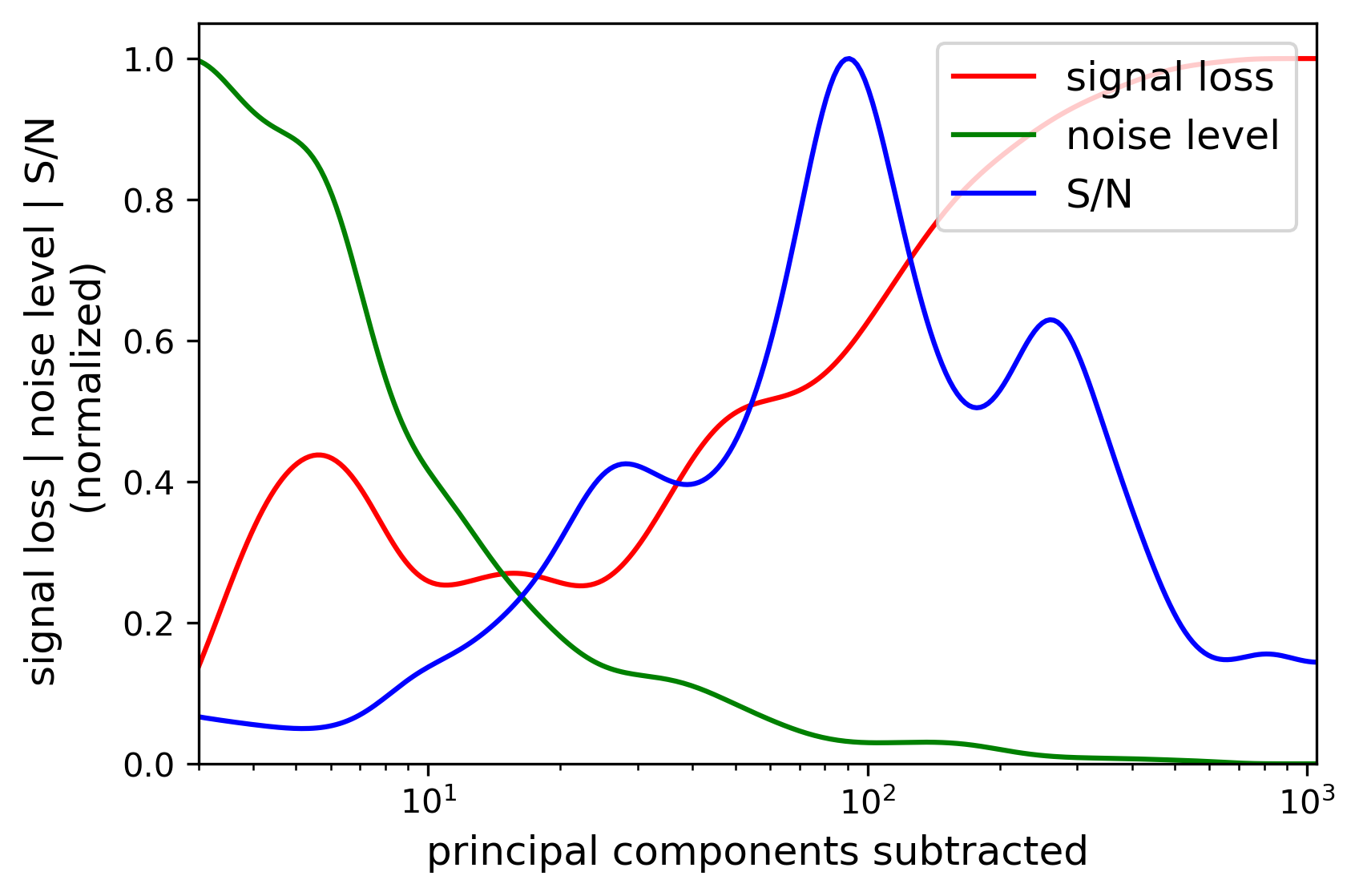}
    \caption{\label{bet pic b PCA trade-off} Trade-off on the optimal number of principal components subtracted for $\beta$ Pictoris b on 1SHORT when combining molecular mapping (with $R_c=100$) and PCA.}
\end{figure}

The optimal choice of the number of subtracted principal components is a balance between reducing systematic noise and minimizing signal loss (see Fig. \ref{bet pic b PCA trade-off}). This optimal number primarily depends on the planet's separation and the contrast between the planet and the star, which determines how much of the planet's spectrum enters the principal components, and thus the degree of self-subtraction of the signal. For $\beta$ Pictoris b, this optimal number is around 100 components, resulting in a signal reduction of about $60\%$ but a systematic noise reduction of approximately $95\%$.

\begin{figure*}
    \centering
    \includegraphics[width=17cm]{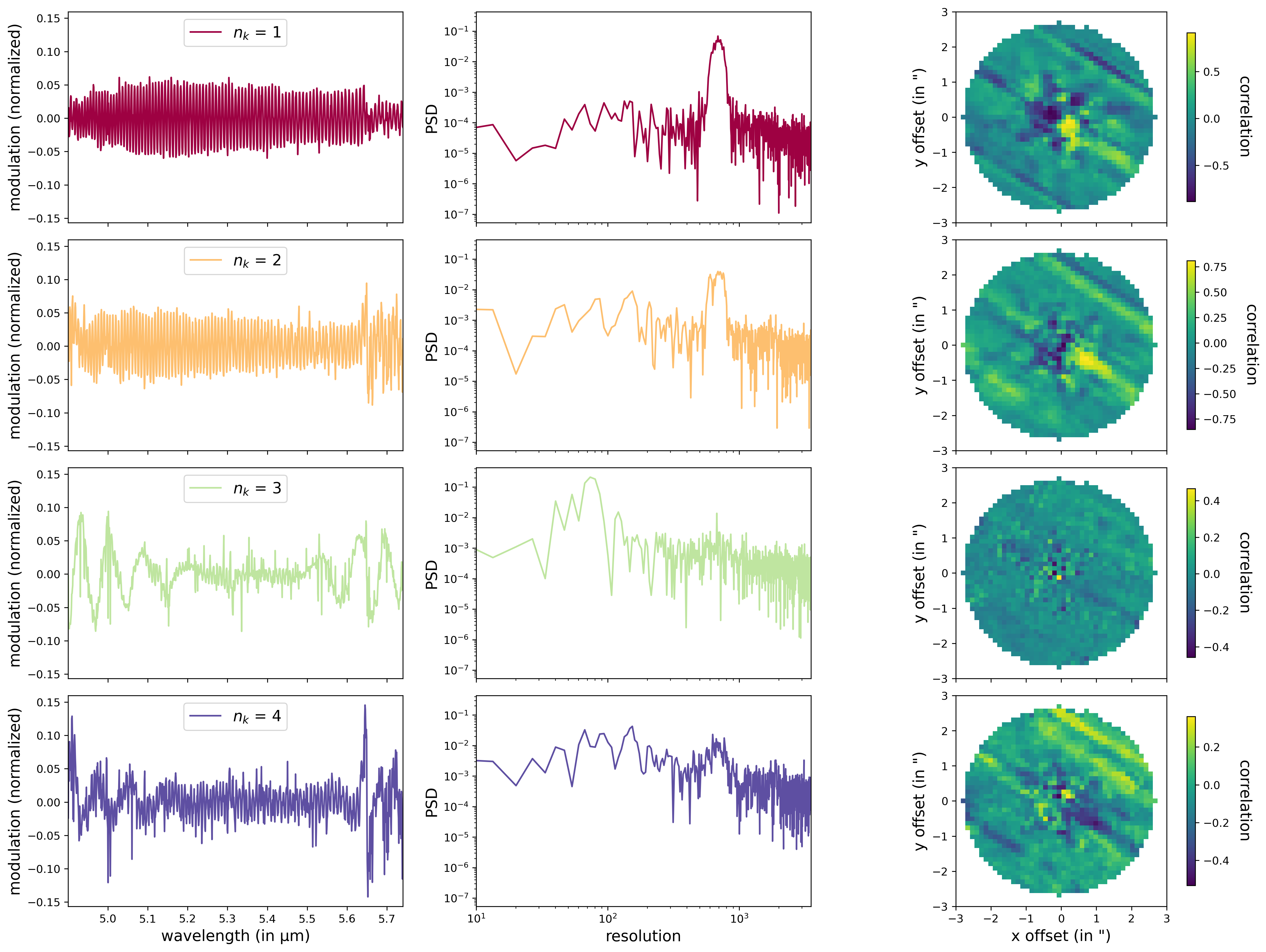}
    \caption{\label{bet pic b PCA components visual} First four modes of the PCA performed on the filtered cube $S_{\mathrm{res}}$ of the $\beta$ Pictoris b data on 1SHORT (left), their PSD (middle), and their correlation projection onto $S_{\mathrm{res}}$ (right).}
\end{figure*}

Finally, examining the modulation patterns of the first principal components estimated by the PCA on the filtered cube ($S_{\mathrm{res}}$), along with the frequencies they influence and their projection by correlation on $S_{\mathrm{res}}$, provides valuable insights (see Fig. \ref{bet pic b PCA components visual}). It is observed that the first two modes are predominantly affected by fringing (evident from their frequency peaks in the PSDs), while subsequent modes begin to reflect modulations caused by aliasing artifacts (with peaks at lower frequencies around $R \approx 100$), consistent with previous observations. Furthermore, the correlation projection of these modes on $S_{\mathrm{res}}$ reveals significant correlation intensity (up to approximately 0.8) at the stellar position, along with diagonal and parallel structures typical of straylight effects. Essentially, straylight would not be as limiting without the presence of high-frequency modulations induced by fringing and aliasing. However, straylight scatters stellar flux away from the star, resulting in straight, parallel structures that enhance the effects of fringe- and aliasing-induced modulations.

It is important to note that PCA appears effective in both cases ($\beta$ Pictoris b and GQ-Lup b), as detection is otherwise limited by systematics. However, when systematics are not dominant, such as in short exposure times, low stellar flux, or high separation, PCA will not be relevant.

\section{Conclusions}\label{conclusion}

In this study, we compare the effective detection limits for close companions, derived from real-life IFS data processed with molecular mapping, to the limits obtained from the semi-analytical FastCurves algorithm. Molecular mapping is theoretically capable of eliminating the detection limits imposed by speckle noise, potentially yielding photon noise-limited data unless other noise sources, related to nonideal instrument effects, become dominant. We selected JWST/MIRI/MRS IFU data for this comparison due to their sufficiently high spectral resolution, the absence of AO residuals, and the ability to compare these results with predictions from MIRISim, the MIRI/MRS simulation tool.

\subsection{Understanding systematic effects and their impact}

We have accurately characterized the fundamental noises of MIRI/MRS with the analytical tool, as the FastCurves predictions align with the data ERR extension. Noise propagation through the CCF is also well captured, with agreement between analytical predictions from FastCurves and empirical spatial variance on CCF maps. Since the noise in the CCF cannot be explained by fundamental noise sources alone, it becomes essential to incorporate a systematic noise component proportional to the stellar flux. This component is well characterized through tests on MIRISim and the pipeline, aligning closely with real data and attributed to nonideal instrument effects and signal projection and retrieval on the detector, setting an ultimate contrast limit that cannot be overcome solely with molecular mapping.

We have identified and explored the detection limits of MIRI/MRS and their underlying sources. Various origins of systematic noise were investigated, excluding factors such as bad outlier subtraction, different 3D reconstruction algorithms, or wavelength calibration errors. The study highlights the significant impact of straylight, fringing, and resampling noise on spatial noise statistics. Systematic effects during signal extraction can also decrease the signal correlation from observed planets. Adjusting the cutoff resolution enhances performance under systematic conditions, and exposure durations must be meticulously selected to optimize observation efficiency given these effects.

\subsection{Implications of noise propagation and template assumptions}

We have discussed the implications of noise propagation and template assumptions in interpreting planetary spectrum correlations with assumed models, ensuring accurate astronomical interpretations from correlation intensity. In particular, systematic noise not only impacts detection sensitivity but can also introduce biases in inferred planetary properties if not properly accounted for. Understanding the interplay between noise characteristics, template accuracy, and signal retrieval methods is essential to derive robust planetary spectra cross-correlation techniques.

\subsection{PCA as a key tool for systematic noise mitigation}

A major result of this study is the demonstrated effectiveness of PCA in mitigating systematic noise. By leveraging spectral diversity, PCA removes dominant systematic components and enhances detection sensitivity. In the case of $\beta$ Pictoris b, PCA increases the S/N by a factor of $\sim25$, making it a crucial technique for extracting planetary signals. However, PCA also induces signal loss, requiring careful tuning of the number of subtracted components to optimize detection performance.

\subsection{Recommendations for future observations and data processing}

Building on these findings, we propose an optimized observing and data processing strategy for molecular mapping with FastCurves:

\begin{itemize}
    \item {Performance assessment:} We recommend using FastCurves in advance to evaluate whether the chosen instrument, considering its known systematic effects, can achieve the scientific objectives within a reasonable exposure time.

    \item {Exposure time optimization:} We emphasize the need to carefully plan exposure time, as beyond a certain threshold the S/N no longer increases linearly, leading to diminishing returns and reducing the instrument's efficiency. This aspect is crucial when optimizing telescope time allocation.

    \item {Post-processing parameter optimization:} After data acquisition, the optimal cutoff resolution should be selected to enhance planetary signal detection while efficiently suppressing stellar speckle noise and systematic modulations. Applying PCA to the filtered data cube significantly improves the S/N, but the number of principal components subtracted must be carefully tuned to maximize systematic noise suppression while minimizing signal loss, following the same principle as the cutoff resolution selection.

    \item {Systematic noise mitigation in instrument design:} Future instrument development must incorporate constraints on systematic noise to ensure the continued effectiveness of cross-correlation techniques for exoplanet detection and characterization. Additionally, we stress the importance of assessing potential systematics and anticipating precise calibrations to minimize their impact on overall performance.
    
\end{itemize}

\subsection{Prospects for future instruments}

While this study focuses on MIRI/MRS, the methodology and insights gained are broadly applicable to other instruments, including ground-based IFSs such as ELT/HARMONI, ELT/METIS, ELT/ANDES, or ELT/PCS, as well as future space-based observatories. The refined noise models and systematic noise mitigation techniques presented here will be essential for defining noise budgets, optimizing observational strategies, and improving exoplanet characterization capabilities across a range of instruments.

\subsection{Final remarks}

\begin{figure}
    \centering
    \includegraphics[width=8.5cm]{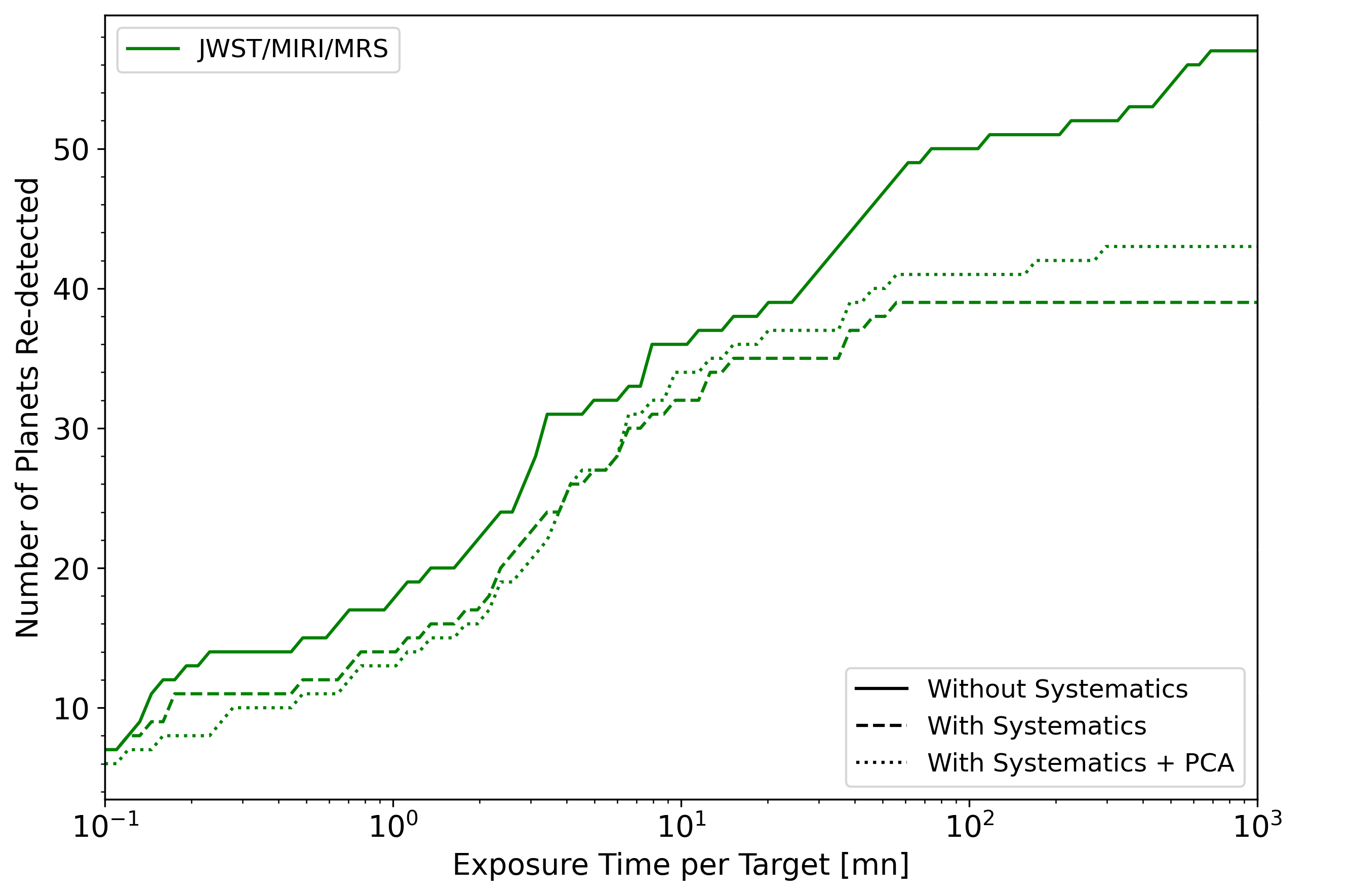}
    \caption{\label{Known exoplanets detection yield as a function of exposure time MIRIMRS systematics} MIRI/MRS known exoplanets detection yields with molecular mapping. Yields were estimated by applying the FastCurves S/N calculation (i.e., Fig. \ref{CT Cha b SNR estimation}) to each of the 5600 planets in the Planetary Systems and Planetary Systems Composite Parameters Table from the NASA exoplanet archive \citep{Akeson_2013}. A planet is considered detected when the S/N on one of the MIRI/MRS bands is greater than 5 for the exposure time considered. It can be observed that PCA would allow MIRI/MRS to approach the performance levels it could achieve without systematics, though not fully reach them, due to the signal loss caused by the subtraction of the principal components.}
\end{figure}

This work demonstrates the feasibility of predicting instrument performance when combining spectroscopy with high-contrast imaging while incorporating systematic effects. By verifying that noise and signal levels in real data match model predictions, we have developed an extension of FastCurves\footnote{\url{https://github.com/StevMartos/FastYield}} to assess the detection yield of an instrument across an exoplanet catalog. Applying this tool to an archive of known exoplanets reveals how systematic effects impact instrument performance and potentially limit detections (see Fig. \ref{Known exoplanets detection yield as a function of exposure time MIRIMRS systematics}). This highlights the crucial role of systematics in defining detection limits and shaping the future of exoplanet characterization.

\begin{acknowledgements}
The authors extend their gratitude to Benjamin Charnay for generously sharing the Exo-REM spectra grid with a resolution of $R = 200 000$, which proved more suitable for the MIRI/MRS case compared to the publicly available grid ($R=20 000$ at $1~\mu \mathrm{m}$). Additionally, they express appreciation to Mathilde Mâlin for providing the simulated data cubes used in her article, and to David Law for engaging discussions regarding resampling noise and the projection of noise statistics through the pipeline's 3D cube reconstruction algorithm. This project receives funding from the European Research Council (ERC) under the European Union's Horizon 2020 research and innovation programme (grant agreement n°866001 - EXACT).
\end{acknowledgements}

\bibliography{bibliography}

\begin{thebibliography}{57}
\expandafter\ifx\csname natexlab\endcsname\relax\def\natexlab#1{#1}\fi

\bibitem[{Akeson {et~al.}(2013)Akeson, Chen, Ciardi, Crane, Good, Harbut,
  Jackson, Kane, Laity, Leifer, Lynn, McElroy, Papin, Plavchan, Ramírez, Rey,
  von Braun, Wittman, Abajian, Ali, Beichman, Beekley, Berriman, Berukoff,
  Bryden, Chan, Groom, Lau, Payne, Regelson, Saucedo, Schmitz, Stauffer, Wyatt,
  \& Zhang}]{Akeson_2013}
Akeson, R.~L., Chen, X., Ciardi, D., {et~al.} 2013, PASP, 125, 989–999

\bibitem[{{Allard} {et~al.}(2012){Allard}, {Homeier}, \&
  {Freytag}}]{Allard_2012}
{Allard}, F., {Homeier}, D., \& {Freytag}, B. 2012, Philos. Trans. R. Soc.
  Lond. A., 370, 2765

\bibitem[{Argyriou {et~al.}(2023)Argyriou, Glasse, Law, Labiano, Álvarez
  Márquez, Patapis, Kavanagh, Gasman, Mueller, Larson, Vandenbussche, Glauser,
  Royer, Dicken, Harkett, Sargent, Engesser, Jones, Kendrew, Noriega-Crespo,
  Brandl, Rieke, Wright, Lee, \& Wells}]{Argyriou_2023}
Argyriou, I., Glasse, A., Law, D.~R., {et~al.} 2023, A\&A, 675, A111

\bibitem[{Argyriou {et~al.}(2020)Argyriou, Wells, Glasse, Lee, Royer,
  Vandenbussche, Malumuth, Glauser, Kavanagh, Labiano, Lahuis, Mueller, \&
  Patapis}]{Argyriou_2020}
Argyriou, I., Wells, M., Glasse, A., {et~al.} 2020, A\&A, 641, A150

\bibitem[{Batalha {et~al.}(2019)Batalha, Marley, Lewis, \&
  Fortney}]{Batalha_2019}
Batalha, N.~E., Marley, M.~S., Lewis, N.~K., \& Fortney, J.~J. 2019, ApJ, 878,
  70

\bibitem[{Beuzit {et~al.}(2019)Beuzit, Vigan, Mouillet, Dohlen, Gratton,
  Boccaletti, Sauvage, Schmid, Langlois, Petit, Baruffolo, Feldt, Milli,
  Wahhaj, Abe, Anselmi, Antichi, Barette, Baudrand, Baudoz, Bazzon, Bernardi,
  Blanchard, Brast, Bruno, Buey, Carbillet, Carle, Cascone, Chapron, Charton,
  Chauvin, Claudi, Costille, De~Caprio, de~Boer, Delboulbé, Desidera, Dominik,
  Downing, Dupuis, Fabron, Fantinel, Farisato, Feautrier, Fedrigo, Fusco,
  Gigan, Ginski, Girard, Giro, Gisler, Gluck, Gry, Henning, Hubin, Hugot,
  Incorvaia, Jaquet, Kasper, Lagadec, Lagrange, Le~Coroller, Le~Mignant,
  Le~Ruyet, Lessio, Lizon, Llored, Lundin, Madec, Magnard, Marteaud, Martinez,
  Maurel, Ménard, Mesa, Möller-Nilsson, Moulin, Moutou, Origné, Parisot,
  Pavlov, Perret, Pragt, Puget, Rabou, Ramos, Reess, Rigal, Rochat, Roelfsema,
  Rousset, Roux, Saisse, Salasnich, Santambrogio, Scuderi, Segransan, Sevin,
  Siebenmorgen, Soenke, Stadler, Suarez, Tiphène, Turatto, Udry, Vakili,
  Waters, Weber, Wildi, Zins, \& Zurlo}]{Beuzit_2019}
Beuzit, J.-L., Vigan, A., Mouillet, D., {et~al.} 2019, A\&A, 631, A155

\bibitem[{Bidot {et~al.}(2024)Bidot, Mouillet, \& Carlotti}]{Bidot_2024}
Bidot, A., Mouillet, D., \& Carlotti, A. 2024, A\&A, 682, A10

\bibitem[{Boccaletti {et~al.}(2024)Boccaletti, Mâlin, Baudoz, Tremblin,
  Perrot, Rouan, Lagage, Whiteford, Mollière, Waters, Henning, Decin, Güdel,
  Vandenbussche, Absil, Argyriou, Bouwman, Cossou, Coulais, Gastaud, Glasse,
  Glauser, Kamp, Kendrew, Krause, Lahuis, Mueller, Olofsson, Patapis, Pye,
  Royer, Serabyn, Scheithauer, Colina, van Dishoeck, Ostlin, Ray, \&
  Wright}]{Boccaletti_2024}
Boccaletti, A., Mâlin, M., Baudoz, P., {et~al.} 2024, A\&A, 686, A33

\bibitem[{Bonnefoy {et~al.}(2014)Bonnefoy, Chauvin, Lagrange, Rojo, Allard,
  Pinte, Dumas, \& Homeier}]{Bonnefoy_2014}
Bonnefoy, M., Chauvin, G., Lagrange, A.-M., {et~al.} 2014, A\&A, 562, A127

\bibitem[{Charnay {et~al.}(2018)Charnay, Bézard, Baudino, Bonnefoy,
  Boccaletti, \& Galicher}]{Charnay_2018}
Charnay, B., Bézard, B., Baudino, J.-L., {et~al.} 2018, ApJ, 854, 172

\bibitem[{Cugno {et~al.}(2024)Cugno, Patapis, Banzatti, Meyer, Dannert,
  Stolker, MacDonald, \& Pontoppidan}]{Cugno_2024}
Cugno, G., Patapis, P., Banzatti, A., {et~al.} 2024, ApJ Letters, 966, L21

\bibitem[{Delorme {et~al.}(2021)Delorme, Jovanovic, Echeverri, Mawet, Wallace,
  Bartos, Cetre, Wizinowich, Ragland, Lilley, Wetherell, Doppmann, Wang,
  Morris, Ruffio, Martin, Fitzgerald, Ruane, Schofield, Suominen, Calvin, Wang,
  Magnone, Johnson, Sohn, Lopez, Bond, Pezzato, Sayson, Chun, \&
  Skemer}]{Delorme_2021}
Delorme, J.-R., Jovanovic, N., Echeverri, D., {et~al.} 2021, The Keck Planet
  Imager and Characterizer: A dedicated single-mode fiber injection unit for
  high resolution exoplanet spectroscopy

\bibitem[{Deming {et~al.}(2024)Deming, Fu, Bouwman, Dicken, Espinoza, Glasse,
  Greene, Kendrew, Law, Lustig-Yaeger, Marin, \& Schlawin}]{Deming_2024}
Deming, D., Fu, G., Bouwman, J., {et~al.} 2024, Toward Exoplanet Transit
  Spectroscopy Using JWST/MIRI's Medium Resolution Spectrometer

\bibitem[{Fruchter \& Hook(2002)}]{Fruchter_2002}
Fruchter, A.~S. \& Hook, R.~N. 2002, PASP, 114, 144

\bibitem[{Garvin {et~al.}(2024)Garvin, Bonse, Hayoz, Cugno, Spiller, Patapis,
  de~la Roche, Nath-Ranga, Absil, Meinshausen, \& Quanz}]{Garvin_2024}
Garvin, E.~O., Bonse, M.~J., Hayoz, J., {et~al.} 2024, Machine Learning for
  Exoplanet Detection in High-Contrast Spectroscopy: Revealing Exoplanets by
  Leveraging Hidden Molecular Signatures in Cross-Correlated Spectra with
  Convolutional Neural Networks

\bibitem[{{Gasman} {et~al.}(2024){Gasman}, {Argyriou}, {Morrison}, {Law},
  {Glasse}, {Gordon}, {Kavanagh}, {Lage}, {Patapis}, \& {Sloan}}]{Gasman_2024}
{Gasman}, D., {Argyriou}, I., {Morrison}, J.~E., {et~al.} 2024, \aap, 688, A226

\bibitem[{Gasman {et~al.}(2023)Gasman, Argyriou, Sloan, Aringer, Álvarez
  Márquez, Fox, Glasse, Glauser, Jones, Justtanont, Kavanagh, Klaassen,
  Labiano, Larson, Law, Mueller, Nayak, Noriega-Crespo, Patapis, Royer, \&
  Vandenbussche}]{Gasman_2023}
Gasman, D., Argyriou, I., Sloan, G.~C., {et~al.} 2023, A\&A, 673, A102

\bibitem[{Henning {et~al.}(2024)Henning, Kamp, Samland, Arabhavi, Kanwar, van
  Dishoeck, Guedel, Lagage, Waelkens, Abergel, Absil, Barrado, Boccaletti,
  Bouwman, o~Garatti, Geers, Glauser, Lahuis, Nehme, Olofsson, Pantin, Ray,
  Vandenbussche, Waters, Wright, Christiaens, Franceschi, Gasman, Guadarrama,
  Jang, Morales-Calderon, Pawellek, Perotti, Rodgers-Lee, Schreiber, Schwarz,
  Tabone, Temmink, Vlasblom, Colina, Greve, \& Oestlin}]{Henning_2024}
Henning, T., Kamp, I., Samland, M., {et~al.} 2024, MINDS: The JWST MIRI
  Mid-INfrared Disk Survey

\bibitem[{Hoeijmakers {et~al.}(2018)Hoeijmakers, Schwarz, Snellen, de~Kok,
  Bonnefoy, Chauvin, Lagrange, \& Girard}]{Hoeijmakers_2018}
Hoeijmakers, H.~J., Schwarz, H., Snellen, I. A.~G., {et~al.} 2018, A\&A, 617,
  A144

\bibitem[{Jorquera {et~al.}(2024)Jorquera, Bonnefoy, Pérez, Chauvin, Aguinaga,
  Dougados, Julo, Demars, Andrews, Ricci, Zhu, kurtovic, Cuello, ning Bai,
  Birnstiel, Dullemond, \& Guzmán}]{Jorquera_2024}
Jorquera, S., Bonnefoy, M., Pérez, L.~M., {et~al.} 2024, VLT/MUSE detection of
  accretion-ejection associated with the close stellar companion in the HT Lup
  system

\bibitem[{Jovanovic {et~al.}(2015)Jovanovic, Martinache, Guyon, Clergeon,
  Singh, Kudo, Garrel, Newman, Doughty, Lozi, Males, Minowa, Hayano, Takato,
  Morino, Kuhn, Serabyn, Norris, Tuthill, Schworer, Stewart, Close, Huby,
  Perrin, Lacour, Gauchet, Vievard, Murakami, Oshiyama, Baba, Matsuo,
  Nishikawa, Tamura, Lai, Marchis, Duchene, Kotani, \&
  Woillez}]{Jovanovic_2015}
Jovanovic, N., Martinache, F., Guyon, O., {et~al.} 2015, PASP, 127, 890–910

\bibitem[{Klaassen {et~al.}(2020)Klaassen, Geers, Beard, O’Brien, Cossou,
  Gastaud, Coulais, Schreiber, Kavanagh, Topinka, Azzollini, De~Meester,
  Bouwman, Glasse, Glauser, Law, Cracraft, Murray, Sargent, Jones, \&
  Wright}]{Klaassen_2020}
Klaassen, P.~D., Geers, V.~C., Beard, S.~M., {et~al.} 2020, MNRAS, 500,
  2813–2821

\bibitem[{Konopacky {et~al.}(2013)Konopacky, Barman, Macintosh, \&
  Marois}]{Konopacky_2013}
Konopacky, Q.~M., Barman, T.~S., Macintosh, B.~A., \& Marois, C. 2013, Science,
  339, 1398–1401

\bibitem[{Kravchenko {et~al.}(2022)Kravchenko, Dallilar, Absil, Berbel,
  Baruffolo, Bonse, Buron, Cao, Cortes, Dannert, Davies, De~Rosa, Deysenroth,
  Doelman, Eisenhauer, Esposito, Feuchtgruber, Förster~Schreiber, Gao,
  Gemperlein, Genzel, Gillessen, Ginski, Glauser, Glindemann, Grani,
  Haguenauer, Hartwig, Hayoz, Heida, Kenworthy, Kolb, Kuntschner, Lutz, Liu,
  MacIntosh, Marsset, Orban~de Xivry, Özdemir, Puglisi, Quanz, Rau, Riccardi,
  Schuppe, Snik, Sturm, Tacconi, Taylor, \& Wiezorrek}]{Kravchenko_2022}
Kravchenko, K., Dallilar, Y., Absil, O., {et~al.} 2022, in Ground-based and
  Airborne Instrumentation for Astronomy IX, ed. C.~J. Evans, J.~J. Bryant, \&
  K.~Motohara, Vol. 9908 (SPIE), 211

\bibitem[{Labiano-Ortega {et~al.}(2016)Labiano-Ortega, Dicken, Vandenbussche,
  Lahuis, Muller, Beard, Justtanont, Azzollini, Law, Gordon, Glasse, Wright,
  Rieke, Klaassen, Glauser, Morrison, Geers, Bailey, \&
  Garcia-Marin}]{Labiano_Ortega_2016}
Labiano-Ortega, A., Dicken, D., Vandenbussche, B., {et~al.} 2016, in
  Observatory Operations: Strategies, Processes, and Systems VI, ed. A.~B.
  Peck, C.~R. Benn, \& R.~L. Seaman (SPIE)

\bibitem[{Landman {et~al.}(2023{\natexlab{a}})Landman, Snellen, Keller,
  N’Diaye, Fagginger-Auer, \& Desgrange}]{Landman_2023a}
Landman, R., Snellen, I. A.~G., Keller, C.~U., {et~al.} 2023{\natexlab{a}},
  A\&A, 675, A157

\bibitem[{Landman {et~al.}(2023{\natexlab{b}})Landman, Stolker, Snellen,
  Costes, de~Regt, Zhang, Gandhi, Mollière, Kesseli, Vigan, \&
  Sánchez-López}]{Landman_2023b}
Landman, R., Stolker, T., Snellen, I., {et~al.} 2023{\natexlab{b}}, $\beta$
  Pictoris b through the eyes of the upgraded CRIRES+

\bibitem[{Law {et~al.}(2023)Law, E.~Morrison, Argyriou, Patapis, Álvarez
  Márquez, Labiano, \& Vandenbussche}]{Law_2023}
Law, D.~R., E.~Morrison, J., Argyriou, I., {et~al.} 2023, AJ, 166, 45

\bibitem[{Lissauer {et~al.}(2011)Lissauer, Fabrycky, Ford, Borucki, Fressin,
  Marcy, Orosz, Rowe, Torres, Welsh, Batalha, Bryson, Buchhave, Caldwell,
  Carter, Charbonneau, Christiansen, Cochran, Desert, Dunham, Fanelli, Fortney,
  Gautier~III, Geary, Gilliland, Haas, Hall, Holman, Koch, Latham, Lopez,
  McCauliff, Miller, Morehead, Quintana, Ragozzine, Sasselov, Short, \&
  Steffen}]{Lissauer_2011}
Lissauer, J.~J., Fabrycky, D.~C., Ford, E.~B., {et~al.} 2011, Nature, 470,
  53–58

\bibitem[{Macintosh {et~al.}(2014)Macintosh, Graham, Ingraham, Konopacky,
  Marois, Perrin, Poyneer, Bauman, Barman, Burrows, Cardwell, Chilcote,
  De~Rosa, Dillon, Doyon, Dunn, Erikson, Fitzgerald, Gavel, Goodsell, Hartung,
  Hibon, Kalas, Larkin, Maire, Marchis, Marley, McBride, Millar-Blanchaer,
  Morzinski, Norton, Oppenheimer, Palmer, Patience, Pueyo, Rantakyro, Sadakuni,
  Saddlemyer, Savransky, Serio, Soummer, Sivaramakrishnan, Song, Thomas,
  Wallace, Wiktorowicz, \& Wolff}]{Macintosh_2014}
Macintosh, B., Graham, J.~R., Ingraham, P., {et~al.} 2014, PNAS, 111,
  12661–12666

\bibitem[{{M{\^a}lin} {et~al.}(2023){M{\^a}lin}, {Boccaletti}, {Charnay},
  {Kiefer}, \& {B{\'e}zard}}]{Malin_2023}
{M{\^a}lin}, M., {Boccaletti}, A., {Charnay}, B., {Kiefer}, F., \&
  {B{\'e}zard}, B. 2023, \aap, 671, A109

\bibitem[{Marley {et~al.}(2021)Marley, Saumon, Visscher, Lupu, Freedman,
  Morley, Fortney, Seay, Smith, Teal, \& Wang}]{Marley_2021}
Marley, M.~S., Saumon, D., Visscher, C., {et~al.} 2021, ApJ, 920, 85

\bibitem[{Miles {et~al.}(2023)Miles, Biller, Patapis, Worthen, Rickman, Hoch,
  Skemer, Perrin, Whiteford, Chen, Sargent, Mukherjee, Morley, Moran, Bonnefoy,
  Petrus, Carter, Choquet, Hinkley, Ward-Duong, Leisenring, Millar-Blanchaer,
  Pueyo, Ray, Sallum, Stapelfeldt, Stone, Wang, Absil, Balmer, Boccaletti,
  Bonavita, Booth, Bowler, Chauvin, Christiaens, Currie, Danielski, Fortney,
  Girard, Grady, Greenbaum, Henning, Hines, Janson, Kalas, Kammerer, Kennedy,
  Kenworthy, Kervella, Lagage, Lew, Liu, Macintosh, Marino, Marley, Marois,
  Matthews, Matthews, Mawet, McElwain, Metchev, Meyer, Molliere, Pantin,
  Quirrenbach, Rebollido, Ren, Schneider, Vasist, Wyatt, Zhou, Briesemeister,
  Bryan, Calissendorff, Cantalloube, Cugno, De~Furio, Dupuy, Factor, Faherty,
  Fitzgerald, Franson, Gonzales, Hood, Howe, Kraus, Kuzuhara, Lagrange, Lawson,
  Lazzoni, Liu, Llop-Sayson, Lloyd, Martinez, Mazoyer, Quanz, Redai, Samland,
  Schlieder, Tamura, Tan, Uyama, Vigan, Vos, Wagner, Wolff, Ygouf, Zhang,
  Zhang, \& Zhang}]{Miles_2023}
Miles, B.~E., Biller, B.~A., Patapis, P., {et~al.} 2023, ApJ Letters, 946, L6

\bibitem[{Mollière {et~al.}(2022)Mollière, Molyarova, Bitsch, Henning,
  Schneider, Kreidberg, Eistrup, Burn, Nasedkin, Semenov, Mordasini, Schlecker,
  Schwarz, Lacour, Nowak, \& Schulik}]{Molliere_2022}
Mollière, P., Molyarova, T., Bitsch, B., {et~al.} 2022, ApJ, 934, 74

\bibitem[{Morley {et~al.}(2012)Morley, Fortney, Marley, Visscher, Saumon, \&
  Leggett}]{Morley_2012}
Morley, C.~V., Fortney, J.~J., Marley, M.~S., {et~al.} 2012, ApJ, 756, 172

\bibitem[{Mâlin {et~al.}(2024)Mâlin, Boccaletti, Perrot, Baudoz, Rouan,
  Lagage, Waters, Güdel, Henning, Vandenbussche, Absil, Barrado, Bouwman,
  Cossou, Decin, Glauser, Pye, Olofsson, Glasse, Lahuis, Patapis, Royer,
  Scheithauer, Whiteford, Serabyn, Choquet, Colina, Ostlin, van Dishoeck, Ray,
  \& Wright}]{Malin_2024}
Mâlin, M., Boccaletti, A., Perrot, C., {et~al.} 2024, Unveiling the HD 95086
  system at mid-infrared wavelengths with JWST/MIRI

\bibitem[{Oberg \& Bergin(2016)}]{Oberg_2016}
Oberg, K.~I. \& Bergin, E.~A. 2016, ApJ Letters, 831, L19

\bibitem[{Parker {et~al.}(2024)Parker, Birkby, Landman, Wardenier, Young,
  Vaughan, van Sluijs, Brogi, Parmentier, \& Line}]{Parker_2024}
Parker, L.~T., Birkby, J.~L., Landman, R., {et~al.} 2024, Into the red: an
  M-band study of the chemistry and rotation of $\beta$ Pictoris b at high
  spectral resolution

\bibitem[{Patapis {et~al.}(2022)Patapis, Nasedkin, Cugno, Glauser, Argyriou,
  Whiteford, Mollière, Glasse, \& Quanz}]{Patapis_2022}
Patapis, P., Nasedkin, E., Cugno, G., {et~al.} 2022, A\&A, 658, A72

\bibitem[{Patience {et~al.}(2012)Patience, King, De~Rosa, Vigan, Witte, Rice,
  Helling, \& Hauschildt}]{Patience_2012}
Patience, J., King, R.~R., De~Rosa, R.~J., {et~al.} 2012, A\&A, 540, A85

\bibitem[{Perrin {et~al.}(2003)Perrin, Sivaramakrishnan, Makidon, Oppenheimer,
  \& Graham}]{Perrin_2003}
Perrin, M., Sivaramakrishnan, A., Makidon, R., Oppenheimer, B., \& Graham, J.
  2003, ApJ, 596

\bibitem[{Petit dit de~la Roche {et~al.}(2018)Petit dit de~la Roche,
  Hoeijmakers, \& Snellen}]{Petit_dit_de_la_Roche_2018}
Petit dit de~la Roche, D. J.~M., Hoeijmakers, H.~J., \& Snellen, I. A.~G. 2018,
  A\&A, 616, A146

\bibitem[{{Petrus} {et~al.}(2021){Petrus}, {Bonnefoy, M.}, {Chauvin, G.},
  {Charnay, B.}, {Marleau, G.-D.}, {Gratton, R.}, {Lagrange, A.-M.}, {Rameau,
  J.}, {Mordasini, C.}, {Nowak, M.}, {Delorme, P.}, {Boccaletti, A.},
  {Carlotti, A.}, {Houll\'e, M.}, {Vigan, A.}, {Allard, F.}, {Desidera, S.},
  {D\'{}Orazi, V.}, {Hoeijmakers, H. J.}, {Wyttenbach, A.}, \& {Lavie,
  B.}}]{Petrus_2021}
{Petrus}, S., {Bonnefoy, M.}, {Chauvin, G.}, {et~al.} 2021, A\&A, 648, A59

\bibitem[{Pontoppidan {et~al.}(2024)Pontoppidan, Salyk, Banzatti, Zhang,
  Pascucci, Oberg, Long, Munoz-Romero, Carr, Najita, Blake, Arulanantham,
  Andrews, Ballering, Bergin, Calahan, Cobb, Colmenares, Dickson-Vandervelde,
  Dignan, Green, Heretz, Herczeg, Kalyaan, Krijt, Pauly, Pinilla, Trapman, \&
  Xie}]{Pontoppidan_2024}
Pontoppidan, K.~M., Salyk, C., Banzatti, A., {et~al.} 2024, High-contrast
  JWST-MIRI spectroscopy of planet-forming disks for the JDISC Survey

\bibitem[{Rieke {et~al.}(2015)Rieke, Ressler, Morrison, Bergeron, Bouchet,
  García-Marín, Greene, Regan, Sukhatme, \& Walker}]{Rieke_2015}
Rieke, G.~H., Ressler, M.~E., Morrison, J.~E., {et~al.} 2015, PASP, 127,
  665–674

\bibitem[{Rigby {et~al.}(2023)Rigby, Lightsey, García~Marín, Bowers, Smith,
  Glasse, McElwain, Rieke, Chary, Liu, Clampin, Kimble, Kinzel, Laidler,
  Mehalick, Noriega-Crespo, Shivaei, Skelton, Stark, Temim, Wei, \&
  Willott}]{Rigby_2023}
Rigby, J.~R., Lightsey, P.~A., García~Marín, M., {et~al.} 2023, PASP, 135,
  048002

\bibitem[{Ruffio {et~al.}(2019)Ruffio, Macintosh, Konopacky, Barman, De~Rosa,
  Wang, Hoch, Czekala, \& Marois}]{Ruffio_2019}
Ruffio, J.-B., Macintosh, B., Konopacky, Q.~M., {et~al.} 2019, AJ, 158, 200

\bibitem[{Ruffio {et~al.}(2024)Ruffio, Perrin, Hoch, Kammerer, Konopacky,
  Pueyo, Madurowicz, Rickman, Theissen, Agrawal, Greenbaum, Miles, Barman,
  Balmer, Llop-Sayson, Girard, Rebollido, Soummer, Allen, Anderson, Beichman,
  Bellini, Bryden, Espinoza, Glidden, Huang, Lewis, Libralato, Louie, Sohn,
  Seager, van~der Marel, Wakeford, Watkins, Ygouf, \& Mountai}]{Ruffio_2024}
Ruffio, J.-B., Perrin, M.~D., Hoch, K. K.~W., {et~al.} 2024, JWST-TST High
  Contrast: Achieving direct spectroscopy of faint substellar companions next
  to bright stars with the NIRSpec IFU

\bibitem[{Schmidt {et~al.}(2008)Schmidt, Neuhäuser, Seifahrt, Vogt, Bedalov,
  Helling, Witte, \& Hauschildt}]{Schmidt_2008}
Schmidt, T. O.~B., Neuhäuser, R., Seifahrt, A., {et~al.} 2008, A\&A, 491,
  311–320

\bibitem[{Smith {et~al.}(2007)Smith, Armus, Dale, Roussel, Sheth, Buckalew,
  Jarrett, Helou, \& Kennicutt}]{Smith_2007}
Smith, J. D.~T., Armus, L., Dale, D.~A., {et~al.} 2007, PASP, 119, 1133–1144

\bibitem[{{Snellen} {et~al.}(2015){Snellen}, {de Kok, R.}, {Birkby, J. L.},
  {Brandl, B.}, {Brogi, M.}, {Keller, C.}, {Kenworthy, M.}, {Schwarz, H.}, \&
  {Stuik, R.}}]{Snellen2015}
{Snellen}, I., {de Kok, R.}, {Birkby, J. L.}, {et~al.} 2015, A\&A, 576, A59

\bibitem[{Turbet(2018)}]{Turbet_2018}
Turbet, M. 2018, Theses, {Sorbonne Universit{\'e}}

\bibitem[{Vigan {et~al.}(2023)Vigan, Morsy, Lopez, Otten, Garcia, Costes,
  Muslimov, Viret, Charles, Zins, Murray, Costille, Paufique, Seemann, Houllé,
  Anwand-Heerwart, Phillips, Abinanti, Balard, Baraffe, Benedetti, Blanchard,
  Blanco, Beuzit, Choquet, Cristofari, Desidera, Dohlen, Dorn, Ely, Fuenteseca,
  Garcia, Jaquet, Jaubert, Kasper, Merrer, Maire, N'Diaye, Pallanca, Popovic,
  Pourcelot, Reiners, Rochat, Sehim, Schmutzer, Smette, Tchoubaklian,
  Tomlinson, \& Soto}]{Vigan_2023}
Vigan, A., Morsy, M.~E., Lopez, M., {et~al.} 2023, First light of VLT/HiRISE:
  High-resolution spectroscopy of young giant exoplanets

\bibitem[{Wang {et~al.}(2018)Wang, Mawet, Hu, Ruane, Delorme, \&
  Klimovich}]{J_Wang_2018}
Wang, J., Mawet, D., Hu, R., {et~al.} 2018, JATIS, 4, 035001

\bibitem[{Wells {et~al.}(2015)Wells, Pel, Glasse, Wright, Aitink-Kroes,
  Azzollini, Beard, Brandl, Gallie, Geers, Glauser, Hastings, Henning, Jager,
  Justtanont, Kruizinga, Lahuis, Lee, Martinez-Delgado, Martínez-Galarza,
  Meijers, Morrison, Müller, Nakos, O’Sullivan, Oudenhuysen, Parr-Burman,
  Pauwels, Rohloff, Schmalzl, Sykes, Thelen, van Dishoeck, Vandenbussche,
  Venema, Visser, Waters, \& Wright}]{Wells_2015}
Wells, M., Pel, J.-W., Glasse, A., {et~al.} 2015, PASP, 127, 646–664

\bibitem[{Worthen {et~al.}(2024)Worthen, Chen, Law, Lu, Hoch, Chai, Sloan,
  Sargent, Kammerer, Hines, Rebollido, Balmer, Perrin, Watson, Pueyo, Girard,
  Lisse, \& Stark}]{Worthen_2024}
Worthen, K., Chen, C.~H., Law, D.~R., {et~al.} 2024, ApJ, 964, 168

\bibitem[{Wu {et~al.}(2015)Wu, Close, Males, Barman, Morzinski, Follette,
  Bailey, Rodigas, Hinz, Puglisi, Xompero, \& Briguglio}]{Wu_2015}
Wu, Y.-L., Close, L.~M., Males, J.~R., {et~al.} 2015, ApJ, 801, 4

\end{thebibliography}
\bibliographystyle{aa}

\begin{appendix}

\section{Used filter}\label{appendixA}

The high-pass filtering operation was calculated according to
\begin{equation}
     \\{[S(\lambda)]_{\mathrm{HF}} = S(\lambda) - [S(\lambda)]_{\mathrm{LF}}}
.\end{equation}
Here, scipy's Gaussian filtering (gaussian_filter\footnote{https://github.com/scipy/scipy/blob/v1.12.0/scipy/ndimage/_filters.py}) was used. The low-pass filtering operation is a convolution of the considered spectrum and a Gaussian kernel of an arbitrarily defined width, $\sigma_c$. This width determines the cutoff frequency, $f_c$, and the cutoff resolution, $R_c$:
\begin{equation}
     \\{[S(\lambda)]_{\mathrm{LF}} = S(\lambda) \ast G_{\sigma_c}(\lambda) \text{, with } G_{\sigma_c}(\lambda) = \frac{1}{\sqrt{2\pi\sigma_c^2}} \e^{-\frac{\lambda^2}{2\sigma_c^2}}}
.\end{equation}
If the focus is on the PSD of the spectrum, it can be written as
\begin{equation}
     \\{PSD(f) = | \mathrm{TF} \lbrace S(\lambda) \rbrace (f) \times \mathrm{TF} \lbrace G_{\sigma_c}(\lambda) \rbrace (f) | ^2}
.\end{equation}
The cutoff frequency, $f_c$, is defined as
\begin{equation}
     \\{\mathrm{TF} \lbrace G_{\sigma_c}(\lambda) \rbrace (f_c) = \frac{1}{2}}
.\end{equation}
The Fourier transform of the Gaussian kernel is given by
\begin{equation}
     \\{\mathrm{TF} \lbrace G_{\sigma_c}(\lambda) \rbrace (f) = \e^{-2\pi^2 f^2 \sigma_c^2}}
.\end{equation}
So,
\begin{equation}
     \\{\e^{-2\pi^2 f_c^2 \sigma_c^2} = \frac{1}{2} \rightarrow f_c = \frac{1}{\pi \sigma_c} \sqrt{\frac{ln(2)}{2}}}
.\end{equation}
The cutoff frequency, $f_c$, is then converted into the cutoff resolution, $R_c$, by
\begin{equation}
     \\{R_c = R_{\mathrm{sampling}} \times f_c = 2 R_{\mathrm{instru}} \times f_c  }
,\end{equation}
where $R_{\mathrm{sampling}}$ is the sampling frequency and $R_{instru}$ is the instrument resolution (assuming Nyquist sampling), with
\begin{equation}
     \\{ R_{\mathrm{instru}} = \frac{\lambda}{2 \Delta \lambda} \text{ and } R_{\mathrm{sampling}} = \frac{\lambda}{\Delta \lambda}  }
.\end{equation}
Finally,
\begin{equation}
     \\{ \sigma_c = \frac{2 R_{\mathrm{instru}}}{\pi  R_c} \sqrt{\frac{ln(2)}{2}} }
.\end{equation}

\section{Molecular mapping formalism (with systematics)} \label{appendixB}

To isolate the planetary spectrum from the stellar spectrum, we considered the TexTris pre-processing routine \citep{Petrus_2021}. First, the spectrum of the star was empirically estimated from the data:
\begin{equation}
     \\{\hat{\gamma}(\lambda) \hat{S}_{\mathrm{*}}(\lambda) = \sum_{i,j} S(\lambda,x_i,y_j)
     \label{eq1}}
.\end{equation}
Then, the stellar modulation function was estimated with 
\begin{equation}
     \\{\hat{M}(\lambda,x,y) = \Bigg[\frac{S(\lambda,x,y)}{\hat{\gamma}(\lambda) \hat{S}_{\mathrm{*}}(\lambda)}\Bigg]_{\mathrm{LF}}}
.\end{equation}
Next, the stellar component was subtracted from the above-estimated quantities:
\begin{equation}
     \\{S_{\mathrm{res}}(\lambda,x,y) = S(\lambda,x,y) - \hat{M}(\lambda,x,y) \hat{\gamma}(\lambda) \hat{S}_{\mathrm{*}}(\lambda)}
.\end{equation}
It was assumed that $\hat{\gamma} \hat{S}_* \approx \gamma S_*$ and that $ [S \times S']_{\mathrm{LF}} \approx [S]_{\mathrm{LF}} \times [S']_{\mathrm{LF}} $. Thus, the estimated stellar modulation function can be written as
\begin{dmath}
     \hat M(\lambda,x,y) = \Bigg{[} \frac{S(\lambda,x,y)}{\hat{\gamma}(\lambda) \hat{S}_*(\lambda)} \Bigg{]}_{\mathrm{LF}} = [M(\lambda,x,y)]_{\mathrm{LF}} + \Bigg{[} \frac{M_{\mathrm{p}}(\lambda,x,y) S_{\mathrm{p}}(\lambda)}{{S}_*(\lambda)} \Bigg{]}_{\mathrm{LF}} + \Bigg{[} \frac{n(\lambda,x,y)}{\gamma (\lambda) S_*(\lambda)} \Bigg{]}_{\mathrm{LF}}
.\end{dmath}
So:
\begin{dmath}
     S_{\mathrm{res}}(\lambda,x,y) = S(\lambda,x,y) - \hat{M}(\lambda,x,y) \hat{\gamma}(\lambda) \hat{S}_*(\lambda)
     = [M(\lambda,x,y)]_{\mathrm{HF}} \gamma(\lambda) S_*(\lambda) + \gamma(\lambda) M_{\mathrm{p}}(\lambda,x,y) S_{\mathrm{p}}(\lambda) + n(\lambda,x,y) -  \Bigg{[} \frac{M_{\mathrm{p}}(\lambda,x,y) S_{\mathrm{p}}(\lambda)}{{S}_*(\lambda)} \Bigg{]}_{\mathrm{LF}} \gamma(\lambda) S_*(\lambda) -  \Bigg{[} \frac{n(\lambda,x,y)}{\gamma (\lambda) S_*(\lambda)} \Bigg{]}_{\mathrm{LF}} \gamma(\lambda) S_*(\lambda)
.\end{dmath}
\cite{Bidot_2024} have shown that the last low-frequency filtered noise term $[n / \gamma S_* ]_{\mathrm{LF}} \gamma S_*$ is negligible compared with the noise term $n$. Then:
\begin{dmath}
     S_{\mathrm{res}}(\lambda,x,y) = \gamma(\lambda) [M_{\mathrm{p}}(\lambda,x,y) S_{\mathrm{p}}(\lambda)]_{\mathrm{HF}} - \Bigg{[} \frac{M_{\mathrm{p}}(\lambda,x,y) S_{\mathrm{p}}(\lambda)}{{S}_*(\lambda)} \Bigg{]}_{\mathrm{LF}} \gamma(\lambda) [S_*(\lambda)]_{\mathrm{HF}} 
     \newline{+ n(\lambda,x,y) + n_{\mathrm{syst}}(\lambda,x,y)}
     \label{eq_S_res}
,\end{dmath}
with
\begin{equation}
     \\{n_{\mathrm{syst}}(\lambda,x,y) = [M(\lambda,x,y)]_{\mathrm{HF}} \gamma(\lambda) S_*(\lambda) }
.\end{equation}
This way, a cube $S_{\mathrm{res}}$ is obtained in which the stellar contribution modulated by $M$ was estimated and subtracted.

If $\mathbb{S}_{\mathrm{p}}$ is the planetary spectrum model considered, the template used is this same spectrum but normalized, high-pass filtered, and shifted with a radial velocity, $rv$; that is,
\begin{equation}
     \\{ \hat t (\lambda,rv) = \frac{\gamma(\lambda)[\mathbb{S}_{\mathrm{p}}(\lambda,rv)]_{\mathrm{HF}}}{\|\gamma(\lambda)[\mathbb{S}_{\mathrm{p}}(\lambda,rv)]_{\mathrm{HF}}\|} }
.\end{equation}
To simplify the notation, the radial velocity dependence, $rv$, is omitted, but it should be kept in mind that the observed stellar and planetary spectra depend on their respective radial velocities. The 2D CCF is simply given by the scalar product between the residual signal and the template, such as
\begin{dmath}
     \mathrm{CCF}(x,y) = \langle S_{\mathrm{res}}(\lambda,x,y) , \hat t(\lambda) \rangle 
     = { \left\| \gamma(\lambda) [M_{\mathrm{p}}(\lambda,x,y) S_{\mathrm{p}}(\lambda)]_{\mathrm{HF}} \right\| \cos \theta_{\mathrm{p}}} 
     - { \left\|  \Bigg{[} \frac{M_{\mathrm{p}}(\lambda,x,y) S_{\mathrm{p}}(\lambda)}{{S}_*(\lambda)} \Bigg{]}_{\mathrm{LF}} \gamma(\lambda) [S_*(\lambda)]_{\mathrm{HF}} \right\| \cos \theta_*} 
     + \langle n(\lambda,x,y)+n_{\mathrm{syst}}(\lambda,x,y) , \hat t(\lambda) \rangle
\end{dmath}

\section{Corrective factor $R_{\mathrm{corr}}$ in dithered data} \label{appendixC}

\begin{figure*}
    \centering
    \includegraphics[width=18cm]{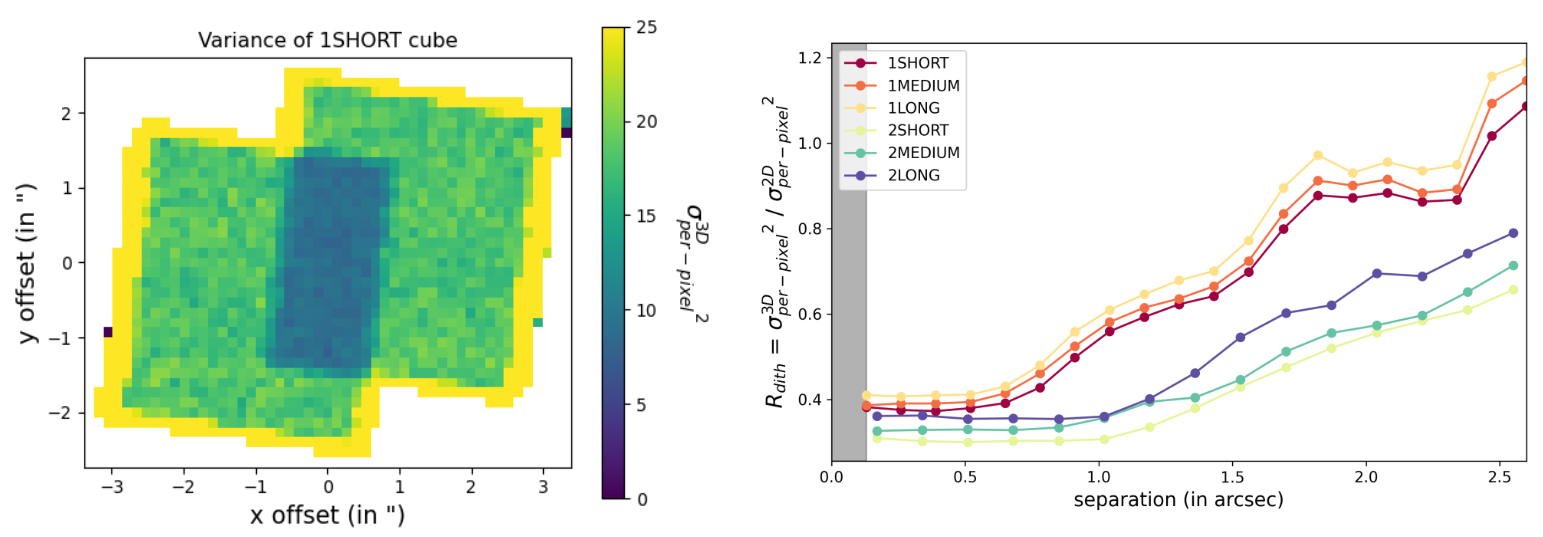}
    \caption{\label{Variance dithering + Ratio_dith per-pixel} Left: Variance (in the spectral dimension) of a cube on the 1SHORT-band, while injecting an arbitrarily defined variance. Right: Ratios between the per-pixel variances in the 3D cubes and the per-pixel variances in the 2D detector images used to reconstruct the cubes.}
\end{figure*}

\begin{figure}
    \centering
    \includegraphics[width=9cm]{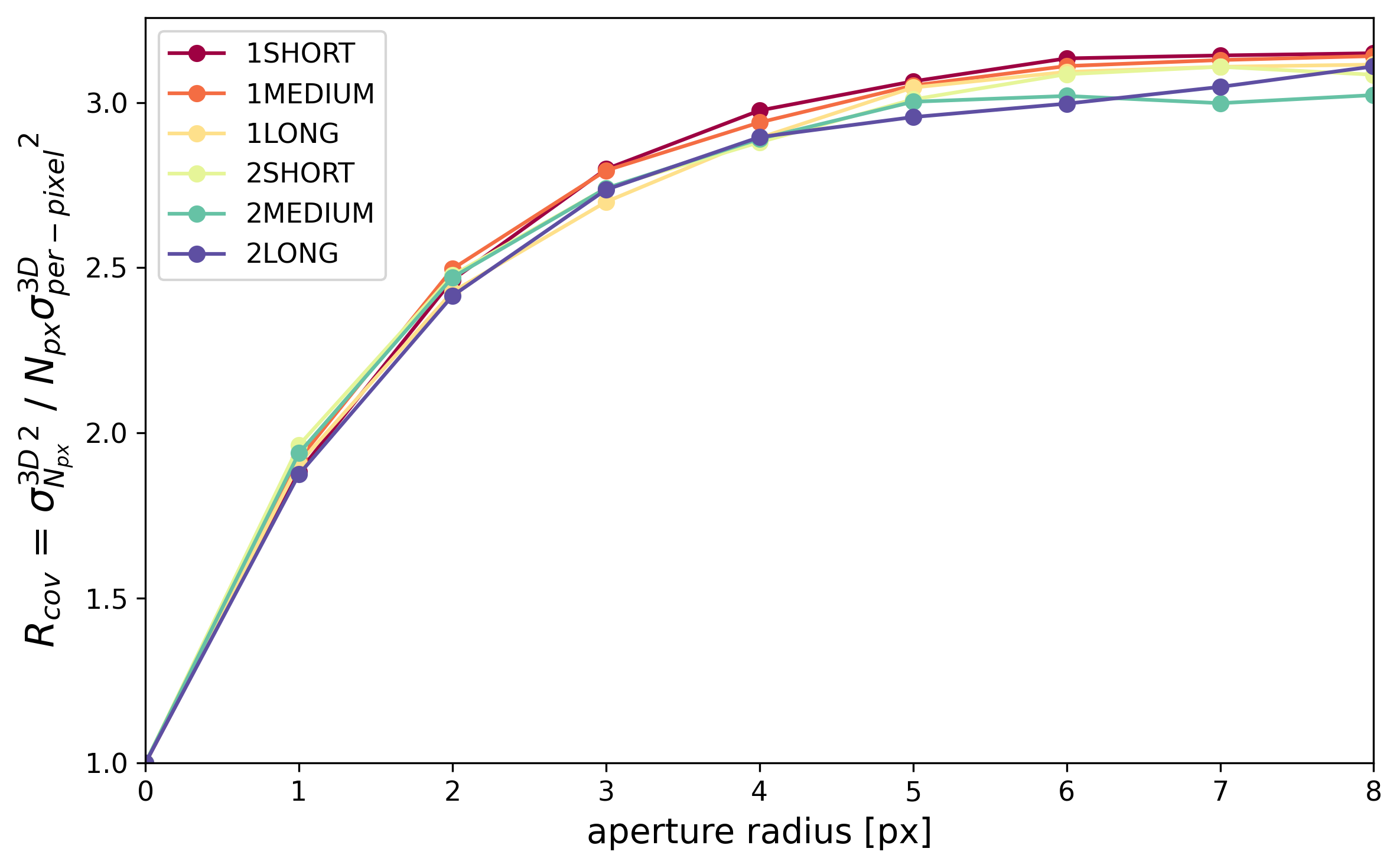}
    \caption{\label{Covariance impact due to the dithering} Ratio between the measured variance in a given aperture radius and the variance that would be measured without covariance between spaxels.}
\end{figure}

\begin{figure}
    \centering
    \includegraphics[width=9cm]{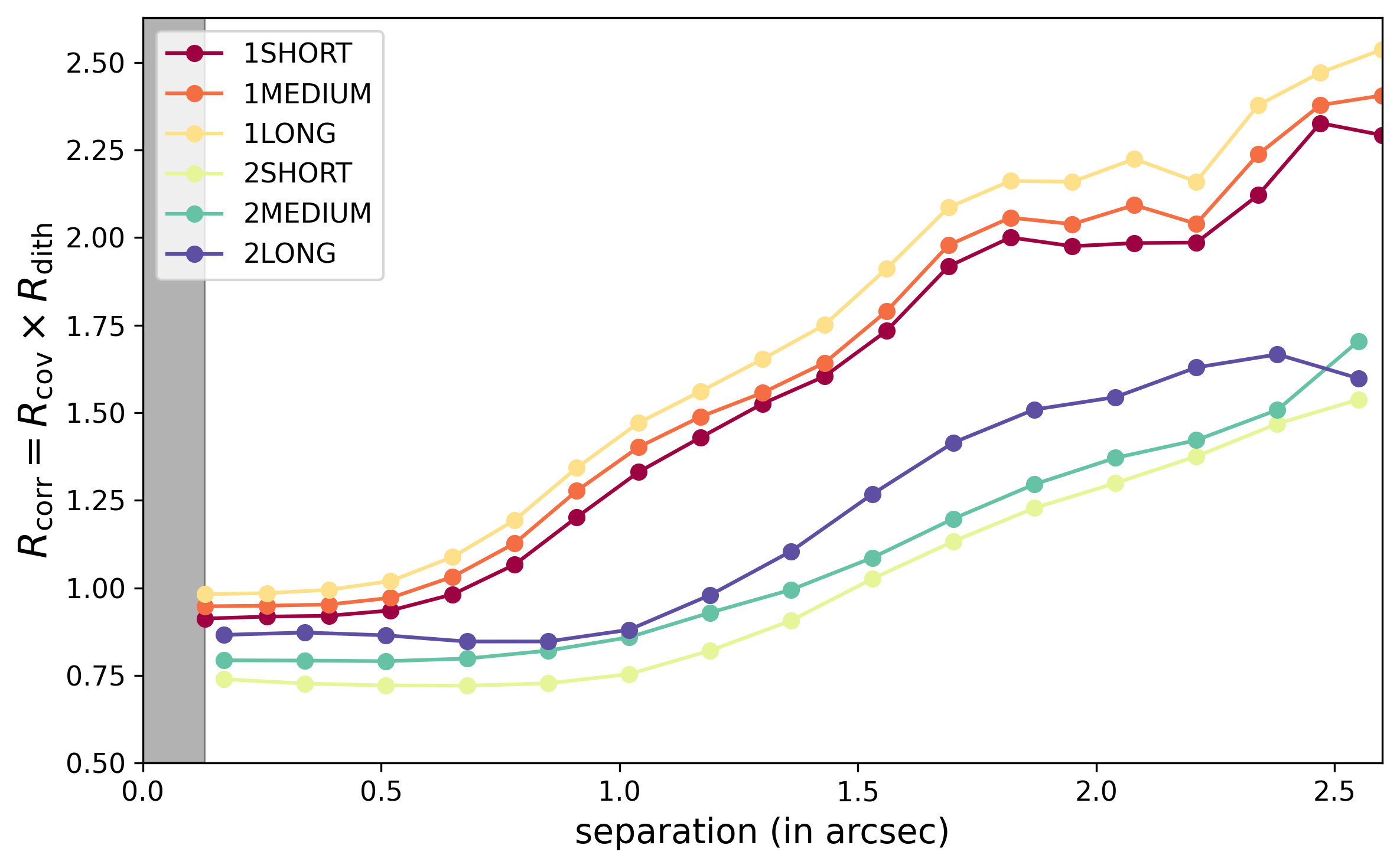}
    \caption{\label{Ratio between the measured and input per FWHM noise} Ratios between the per-FWHM variances in the 3D cubes and the per-FWHM variances in the 2D detector images used to reconstruct the cubes.}
\end{figure}

It is fairly straightforward to estimate the noise that 2D detector images of MIRI/MRS should theoretically have, given the (Gaussian) photon and detector noises. However, the drizzle method of reconstruction to a 3D cube \citep{Law_2023} modifies the statistics of the noise.

We let ${\sigma_{\mathrm{per-pixel}}^{\mathrm{2D}}}^2$ be the variance per-pixel of the 2D detector image considered (this is the variance calculated with FastCurves for photon and detector noises) and ${\sigma_{\mathrm{per-pixel}}^{\mathrm{3D}}}^2$ be the variance actually found in the reconstructed cube. The ratio between these two quantities is called $R_{\mathrm{dith}}$. To measure this quantity empirically, the values of four 2D detector images corresponding to four different dithering points are replaced by values for which the variance is arbitrarily defined. Then, the data cubes are reconstructed with the pipeline and the effective variances measured. But, as is shown on the left panel of the Fig. \ref{Variance dithering + Ratio_dith per-pixel}, the dithering is not spatially homogeneous. In particular, the center is a region where all four integrations of the four-point dithering are being stacked, while at the edges only two out of four will be stacked, giving a variance two times greater than at the center. This implies the need to calculate $R_{\mathrm{dith}}$ for each separation (see right panel of Fig. \ref{Variance dithering + Ratio_dith per-pixel}). If there is no obligation to place the star in the center and the planet's position is known, it is then advisable to place the planet in the central region of the FoV, where all the dithering positions will be stacked. 

In the method considered, the planet's signal is integrated over a box of $A_{\mathrm{fwhm}}$ spaxels and the dithering implies spatial covariance between adjacent spaxels, so the variance in this box will not be equal to $A_{\mathrm{fwhm}} \times {\sigma_{\mathrm{per-pixel}}^{\mathrm{3D}}}^2$. We defined
\begin{equation}
     \\{ R_{\mathrm{cov}} = \frac{{\sigma_{N_{px}}^{\mathrm{3D}}}^2}{{N_{px} \sigma_{\mathrm{per-pixel}}^{\mathrm{3D}}}^2} }
,\end{equation}
where ${\sigma_{N_{px}}^{\mathrm{3D}}}^2$ is the variance measured when summing the signal over an aperture of $N_{px}$ pixels. This calculation is similar to the one performed in \cite{Law_2023}. Figure \ref{Covariance impact due to the dithering} shows that this ratio increases asymptotically with the aperture radius due to the covariance. Nevertheless, the interest is on an aperture of $A_{\mathrm{fwhm}}$ spaxels, so: $R_{\mathrm{cov}} = {\sigma_{A_{\mathrm{fwhm}}}^{\mathrm{3D}}}^2 / {A_{\mathrm{fwhm}} \sigma_{\mathrm{per-pixel}}^{\mathrm{3D}}}^2$, where ${\sigma_{A_{\mathrm{fwhm}}}^{\mathrm{3D}}}^2$ is the effective variance when summing the signal over an aperture of $A_{\mathrm{fwhm}}$ spaxels (this is the quantity that must be expressed in terms of ${\sigma_{\mathrm{per-pixel}}^{\mathrm{2D}}}^2$). Finally: 
\begin{equation}
     \\{ {\sigma_{A_{\mathrm{fwhm}}}^{\mathrm{3D}}}^2 = A_{\mathrm{fwhm}} R_{\mathrm{corr}} {\sigma_{\mathrm{per-pixel}}^{\mathrm{2D}}}^2}
,\end{equation}
where $R_{\mathrm{corr}} = R_{\mathrm{cov}} \times R_{\mathrm{dith}}$ is the corrective ratio introduced in the core of the paper. This quantity is plotted for each band in Fig. \ref{Ratio between the measured and input per FWHM noise}.

\section{Structure and spatial variations in the modulations through PSD} \label{appendixD}

To assess the impact of various effects, the previous method used to obtain the simulated stellar modulation function $M_{\mathrm{sim}}$ is repeated, but with a choice to include or exclude specific corrections. In particular, fringes can be corrected or not, since MIRISim simulates the fringes with the same extended-source fringe pattern as the 2D flat used in the first pipeline correction step. This is why only the first fringe subtraction step is necessary to perfectly eliminate all fringes from the data cubes simulated by MIRISim. In addition, bad pixels, hot pixels, and cosmic rays (outliers) are absent from the noiseless 2D detector images simulated with MIRISim. Therefore, applying the outliers subtraction step may yield to a modulation function assuming bad outliers subtraction. Similarly, as MIRISim does not simulate straylight, applying the straylight subtraction step will result in a modulation function with bad straylight subtraction. Finally, an Exponential Modified Shepard Method (EMSM) can be used to reconstruct the cube, also detailed in \cite{Law_2023}. This enables a comparison between the two cube reconstruction methods, drizzle and EMSM, to identify any discrepancies. Although the focus here is on the 1SHORT band, similar outcomes can be derived for other bands as well.

\begin{figure*}
    \centering
    \includegraphics[width=18cm]{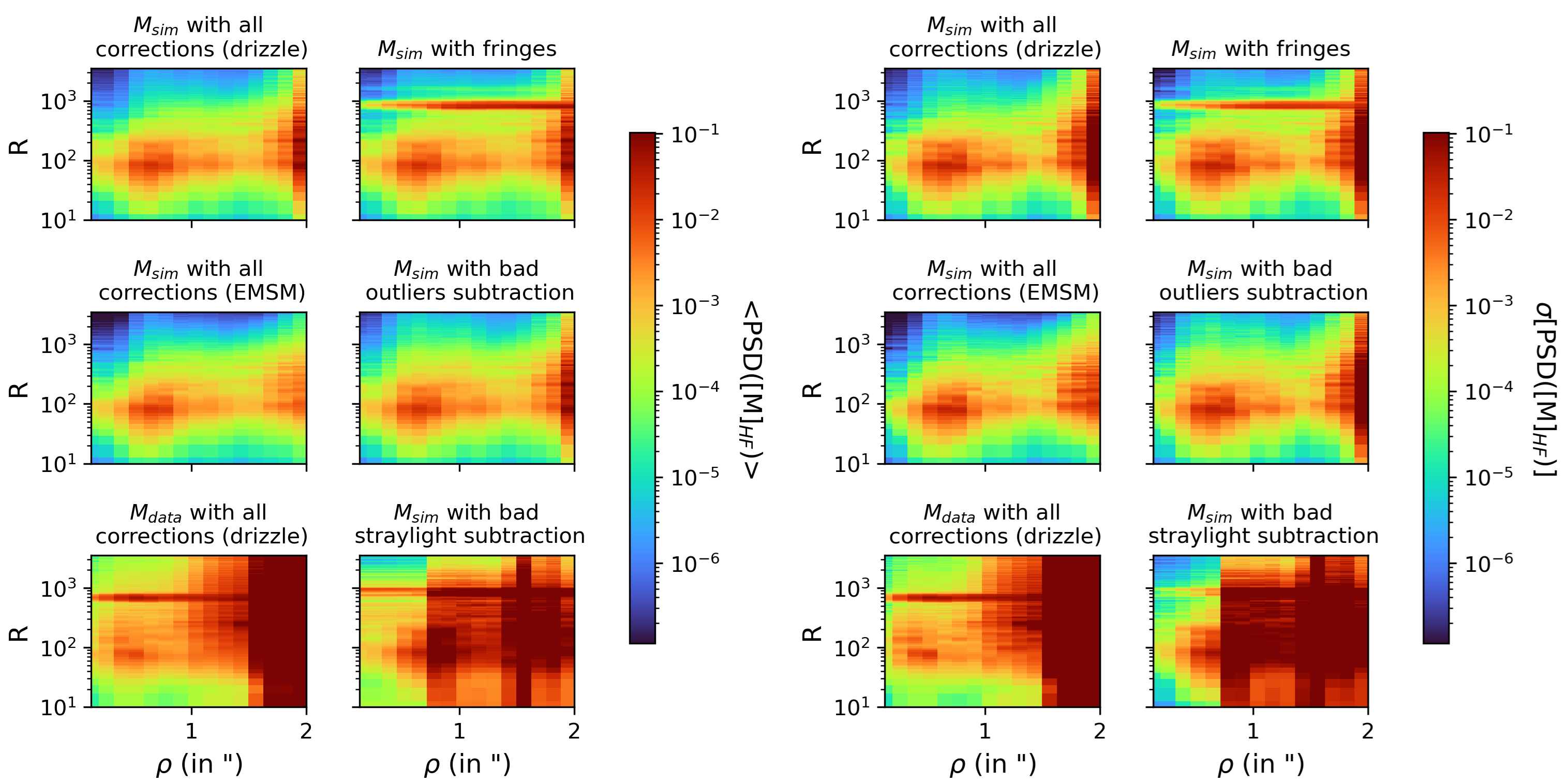}
    \caption{\label{mean and std PSD modulations with differents effects} Left: Average of PSDs for each separation of the high-filtered stellar modulation functions on 1SHORT with $R_c = 100$. Right: Spatial variations in PSDs for each separation of the high-filtered stellar modulation functions (same as left panel but with standard deviation). The simulated stellar modulation functions $M_{\mathrm{sim}}$ were estimated by injecting a BT-NextGen stellar spectrum at 6000K into MIRISim. The stellar modulation functions at each point are renormalized so as not to have the flux dependency of the PSF (which would otherwise only show a PSD concentrated at $\rho\approx0$). $M_{\mathrm{data}}$ was estimated from HD159222 data. Edge effects, arising from both the detector and the cube reconstruction, begin to manifest at around 2 arcseconds.}
\end{figure*}

To get an idea of the impact of these different effects and corrections on the modulation structure, the PSD of $[M_{\mathrm{sim}}]_{\mathrm{HF}}$ is computed for each spaxel and averaged for each separation (see left panel of Fig. \ref{mean and std PSD modulations with differents effects}). Even with all the corrections applied (top left), residual modulations remain after high-pass filtering, peaking at $R\approx100$.  This peak primarily arises from residual aliasing artifacts, and shows why it makes sense to filter at a higher cutoff resolution to mitigate systematics (Sect. \ref{subsection4_3}). In addition, a different reconstruction method (middle left) or a bad outliers subtraction (middle right) does not seem to imply any differences in modulations. If fringes are left uncorrected (top right), an additional peak emerges in the PSD at $R\approx800$, corresponding to the frequency of simulated fringes. Conversely, bad straylight subtraction (bottom right) leads to a drastic alteration in the PSD, affecting nearly all resolutions and compromising fringe suppression, even with the correct flat applied. Furthermore, a significant difference is found between the PSDs of stellar modulation functions estimated from simulations (top left) and data (bottom left) when all corrections are applied. While fringes persist in the data, as expected with only the first fringe correction step applied, they occur at a different resolution ($R\approx650$) compared to simulated fringes. There are also additional modulations in the data at high-resolution, stemming from bad straylight subtraction (directly visible in the data), but to a lesser extent than simulated (bottom right). Lastly, it can be seen from Fig. \ref{mean and std PSD modulations with differents effects} that the spatial standard deviations of the PSDs are found to be equivalent to their means.

\section{Simulated systematic effects on noise} \label{appendixE}

\subsection{Photometric calibration errors} \label{appendixE_1}

\begin{figure}
    \centering
    \includegraphics[width=9cm]{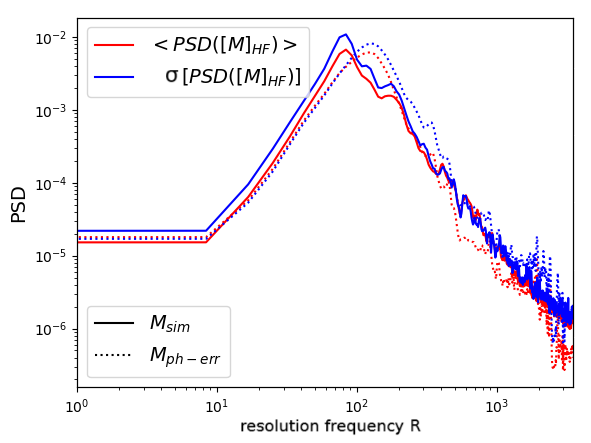}
    \caption{\label{PSD M photometric effect} Comparison of PSDs of stellar modulation functions when the amplitude of simulated photometric errors is equivalent to that of aliasing artifacts in the data. The modulation amplitude of the resampling noise on 1SHORT measured by \cite{Law_2023} implies an error of $\sigma_{\mathrm{ph-err}}\approx10\%/\sqrt{2}\approx7\%$. The mean and standard deviation are calculated all over the cubes.}
\end{figure}

The impact of photometric calibration errors, akin to resampling noise, on contrast is quantified using the following approach. A low-pass filter with a strong cutoff resolution ($R_c=10$) is applied on $M_{\mathrm{sim}}$ (with all corrections) to retrieve a stellar modulation function without any systematic high-frequency modulations (only speckles). On the latter ($[M_{\mathrm{sim}}]_{\mathrm{LF}}$), photometric errors are simulated following the methodology outlined in \cite{Patapis_2022}. \footnote{In that study, photometric and wavelength calibration errors are simulated to assess their impact on correlation. A similar approach is adopted in Sect. \ref{subsection5_4}, but with the formalism detailed here and revised assumptions.} Photometric errors of the same nature as resampling noise are therefore modeled by a sine function: 
\begin{equation}
      \\{ \epsilon(\lambda,x,y) = \sqrt{2} \sigma_{\mathrm{ph-err}} \sin( 2 \pi f(x,y) \lambda) }
,\end{equation}
and the stellar modulation function with simulated photometric errors is given by
\begin{equation}
      \\{ M_{\mathrm{ph-err}} (\lambda,x,y) = [M_{\mathrm{sim}}(\lambda,x,y)]_{\mathrm{LF}} \times (1 + \epsilon(\lambda,x,y)) }
,\end{equation}
where $\sigma_{\mathrm{ph-err}}$ is the photometric calibration error and $f(x,y)$ is a frequency randomly drawn for each spaxel. The probability density function of aliasing artifact frequencies is presumed to be dictated by the mean PSD (normalized) of the stellar modulation function, $M_{\mathrm{sim}}$ (with all corrections). This way, when the simulated photometric errors match the amplitude of those induced by aliasing artifacts in the data \citep[measured at $7\%$ by][]{Law_2023}, both the mean PSD of the modulations and its spatial variation are recovered (see Fig. \ref{PSD M photometric effect}). This suggests that the employed model for aliasing artifacts is suitable and adequate to recover the structure and variations in the modulations in the simulated data.

\begin{figure}
    \centering
    \includegraphics[width=9cm]{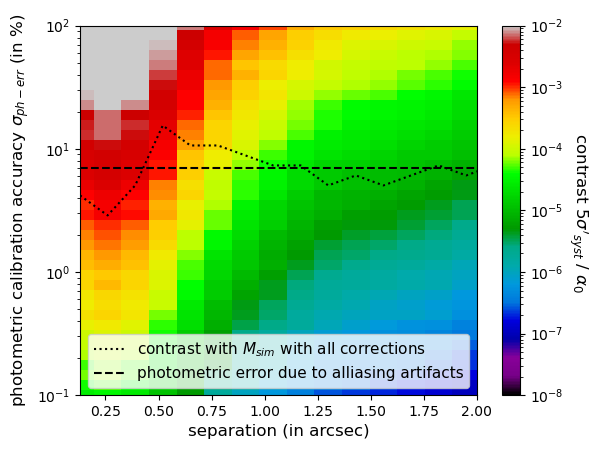}
    \caption{\label{effect of photometric calibration accuracy on systematic noise} Simulated effect of photometric calibration errors of the same nature as aliasing artifacts on systematic noise on 1SHORT with a BT-Settl template at $1000~\mathrm{K}$ and $R_c = 100$.}
\end{figure}

The contrast arising from systematics induced by these simulated photometric errors is computed using Eq. \ref{eq_sigma_syst_prime_est} with $M_{\mathrm{ph-err}}$ for each $\sigma_{\mathrm{ph-err}}$ and depicted in Fig. \ref{effect of photometric calibration accuracy on systematic noise}. Likewise, when these simulated photometric errors match the amplitude observed in the data due to aliasing artifacts, the contrast obtained with $M_{\mathrm{sim}}$ (with all corrections, i.e., solid black line in Fig. \ref{contrast level due to different effects 1SHORT}) is reproduced.

\subsection{Wavelength calibration errors} \label{appendixE_2}

\begin{figure}
    \centering
    \includegraphics[width=9cm]{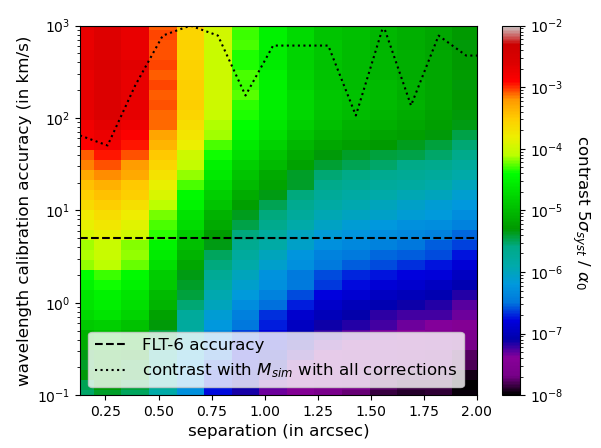}
    \caption{\label{effect of wavelength calibration accuracy on systematic noise} Simulated effect of wavelength calibration errors on systematic noise on 1SHORT with a BT-Settl template at $1000~\mathrm{K}$ and $R_c = 100$. The current wavelength solution (FLT-6) offers a calibration accuracy of a few kilometers per second in channels 1 and 2.}
\end{figure}

Simultaneously, it is crucial to assess the potential influence of wavelength calibration errors on noise. To achieve this, Doppler shifts were applied for each spaxel to the stellar modulation function without systematic high-frequency modulation (including the stellar spectrum):
\begin{equation}
      \\ { M_{\lambda-err} (\lambda,x,y) S_*(\lambda) = [M_{\mathrm{sim}}(\lambda + \Delta\lambda(x,y) ,x,y)]_{\mathrm{LF}} S_*(\lambda+ \Delta\lambda(x,y)) }
,\end{equation}
where the offsets, $\Delta\lambda(x,y)$, were drawn randomly according to a normal distribution centered at 0 and with a standard deviation of $\sigma_{\lambda-err}$. As previously, the contrast due to systematics induced by these simulated wavelength errors was calculated according to Eq. \ref{eq_sigma_syst_prime_est} for each $\sigma_{\lambda-err}$ and plot in Fig. \ref{effect of wavelength calibration accuracy on systematic noise}. It emerges that a calibration error of $100\text{--}1000~\mathrm{km/s}$ would be required to explain the systematic noise observed with $M_{\mathrm{sim}}$ (with all corrections, i.e., solid black line in Fig. \ref{contrast level due to different effects 1SHORT}). In other words, when the simulated error matches the expected FLT-6 wavelength solution's amplitude error (few kilometers per second), the contrast is underestimated by a factor of roughly 30. Especially as the errors were simulated under extreme conditions: assuming no spatial correlation in the errors ($\Delta\lambda(x,y)$), thus maximizing the systematic effect and the induced contrast. In fact, it is highly probable that there exists covariance in the wavelength error between adjacent spaxels. For instance, assuming a covariance radius of 4 spaxels (deduced from Fig. \ref{Covariance impact due to the dithering}, since a plateau is reached after 4 pixels), the contrast is now underestimated by approximately a factor of 200. Hence, the impact of wavelength calibration errors on noise seems negligible. This also suggests that if only wavelength calibration errors (on the order of a few kilometers per second) were present as systematic noise, a contrast of $10^{-4}$ could be achieved at $0.5~"$ and $10^{-6}$ at $1.0~"$ for a planet at $1000~\mathrm{K}$.

\section{Simulated systematic effects on correlation} \label{appendixF}

\subsection{Photometric calibration errors} \label{appendixF_1}

\begin{figure}
    \centering
    \includegraphics[width=9cm]{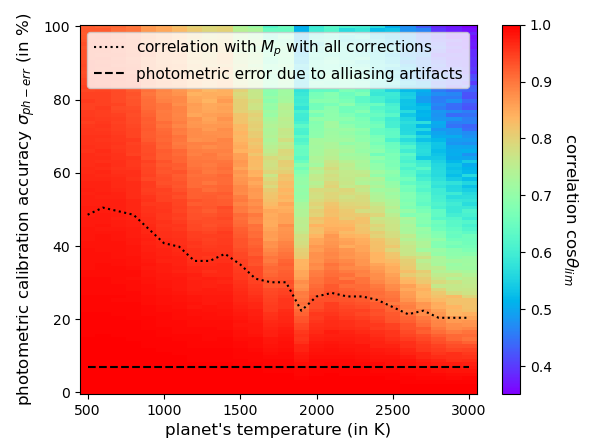}
    \caption{\label{effect of photometric calibration accuracy on correlation} Simulated effect of photometric calibration errors of the same nature as aliasing artifacts on correlation on 1SHORT with $R_c=100$. Photometric error due to resampling noise on 1SHORT: $\sigma_{\mathrm{ph-err}}\approx7\%$.}
\end{figure}

The idea is to replace $M_{\mathrm{p}}$ by $M_{\mathrm{ph-err}}$ in Eq. \ref{eq_systematic_effect_on_correlation} and estimate the induced correlation drop for various photometric error levels $\sigma_{\mathrm{ph-err}}$ (see Fig. \ref{effect of photometric calibration accuracy on correlation}).
\begin{figure}
    \centering
    \includegraphics[width=9cm]{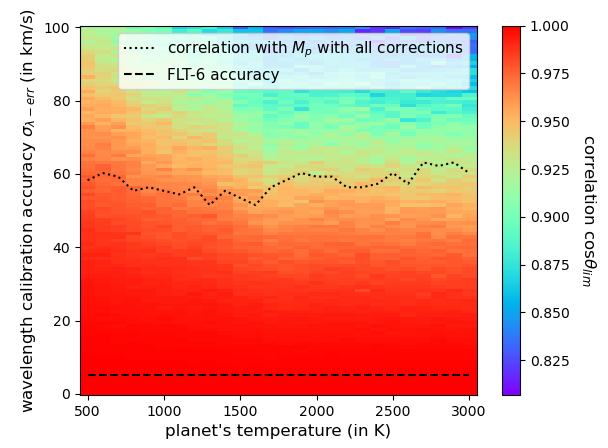}
    \caption{\label{effect of wavelength calibration accuracy on correlation} Simulated effect of wavelength calibration errors on correlation on 1SHORT with $R_c=100$. The current wavelength solution (FLT-6) offers a calibration accuracy of a few kilometers per second in channels 1 and 2.}
\end{figure}

\subsection{Wavelength calibration errors} \label{appendixF_2}

In the same way, replacing $M_{\mathrm{p}}$ with $M_{\lambda-err}$ in Eq. \ref{eq_systematic_effect_on_correlation} allows one to quantify the impact of wavelength calibration errors on the correlation (see Fig. \ref{effect of wavelength calibration accuracy on correlation}).

\end{appendix}
\end{document}